\PassOptionsToPackage{varg}{newtxmath}
\documentclass{optica-article}

\hypersetup{linkcolor=blue,citecolor=blue}

\journal{opticajournal} 
\articletype{Research Article}

\usepackage{xspace, xr-hyper, textcomp, amsmath, bm, booktabs, makecell}

\makeatletter
\newcommand*{\addFileDependency}[1]{ \typeout{(#1)}
  \@addtofilelist{#1}
  \IfFileExists{#1}{}{\typeout{No file #1.}}
}
\makeatother

\newcommand*{\myexternaldocument}[3][SI]{\externaldocument[#1]{#2/#3}\addFileDependency{#3.tex}\addFileDependency{#2/#3.aux}}

\myexternaldocument[S-]{build}{supplemental-document}

\graphicspath{{./figs/}}

\newcommand{\Invivo}{\textit{In vivo}\xspace}
\newcommand{\invivo}{\textit{in vivo}\xspace}

\newcommand{\invitro}{\textit{in vitro}\xspace}
\newcommand{\enface}{\textit{en face}\xspace}
\newcommand{\Enface}{\textit{En face}\xspace}
\newcommand{\um}{\textmu m\xspace}

\newcommand{\figs}{Figs.}

\begin{document}

\title{Image formation theory of optical coherence tomography with optical aberrations and its application to computational aberration correction}

\author{
    Shuichi Makita,\authormark{1,*}
    Naoki Fukutake,\authormark{1, 2}
    Lida Zhu,\authormark{1}
    and Yoshiaki Yasuno\authormark{1}
}

\address{
    \authormark{1}Computational Optics Group, University of Tsukuba, 1-1-1 Tennodai, Tsukuba, Ibaraki 305-8573, Japan\\
    \authormark{2}Nikon Corporation, 1-5-20 Nishioi, Shinagawa-ku, Tokyo, 140-8601, Japan\\
}

\email{\authormark{*}shuichi.makita@cog-labs.org} 

\begin{abstract*}
    Computational corrections of defocus and aberrations in optical coherence tomography (OCT) offers a promising approach to realize high-resolution imaging with deep imaging depth, but without additional high hardware costs.
    However, these techniques are not well understood owing to a lack of accurate theoretical models and investigation tools.
    The image formation theory for OCT with optical aberrations is thus formulated here.
    Based on this theory, a numerical simulation method is developed, and computational refocusing and computational aberration correction (CAC) methods are designed.
    The CAC method based on the image formation theory is applied to simulated OCT signals and OCT images of a microparticle phantom and an \invivo human retina for simultaneous multi-depth correction of systematic aberration.
    The numerical simulation under the effective numerical aperture of 0.2 and 1.05 \um central wavelength shows that the proposed method can obtain the Strehl ratios of more than 0.8 over a $\pm$ 100 \um defocus range, while the conventional method cannot achieve this under the simulated conditions.
    Imaging results show that the CAC method designed based on the image formation theory can correct optical aberrations and improve the image quality more than the conventional CAC method.
    The proposed method improved the frequency component corresponding to the density of cone photoreceptors in OCT photoreceptor images by 1.2 to 1.4 times under the multi-depth correction.
    This theoretical model-based approach provides a powerful aid for understanding OCT imaging properties and processing method design.
\end{abstract*}

\section{Introduction}

Optical coherence tomography (OCT) is a non-invasive imaging technique that provides high-resolution cross-sectional images of biological tissues, and it is widely applied in fields including ophthalmology, cardiology, and dermatology \cite{huang_optical_1991,drexler_optical_2015}.
As the development of several medical fields continues in areas including tissue diagnostics, disease screening, evaluation of novel treatment methods, optimization of existing treatments, and drug development, there is an increasing demand for higher imaging resolutions to capture more detailed images of thick biological tissues, including \invitro tissues such as spheroids, organoids, and grafts, as well as \invivo human tissues.

In OCT, high depth resolution is achieved by using a short temporal coherence gate.
However, the lateral spatial resolution is limited in OCT imaging.
This is because high-speed OCT resolves a depth profile by scanning the temporal coherence gate while the focal spot along the depth remains fixed.
Techniques have been proposed to overcome this limitation, e.g., dynamic focus OCT\cite{schmitt_optical_1997,lexer_dynamic_1999,pircher_dynamic_2006}.
This method is, however, not compatible with high-speed Fourier-domain OCT (FD-OCT)\cite{fercher_measurement_1995,hausler_coherence_1998}.
FD-OCT acquires signals with several temporal delays (A-line) simultaneously, and thus moving the focus during A-line acquisition causes the image quality to deteriorate.
In this case, focus fusion\cite{drexler_vivo_1999,rolland_gabor-based_2010} offers another solution.
However, this fusion technique requires multiple image acquisitions and thus involves longer imaging times.

Furthermore, the lateral resolution of OCT imaging is limited by optical aberrations in some cases.
Ocular aberrations represent the most significant factor affecting high-resolution OCT imaging of the retina.
Use of the adaptive optics (AO) technique has been proposed to correct the ocular aberrations for high-resolution retinal OCT imaging\cite{pircher_combining_2007,kurokawa_adaptive_2017}.
This technique enables cellular-level imaging of the retina \invivo, but it is realized at the cost of the high complexity of the system and the AO devices.
Furthermore, the AO technique does not solve the limited depth of focus problem in OCT imaging, and thus focus fusion is still required to achieve all-depth in-focus imaging\cite{jian_adaptive_2013,kurokawa_multi-reference_2021}.

In contrast to these hardware-based techniques, computational correction\cite{yasuno_-focus_2005,ralston_inverse_2006,adie_computational_2012,kumar_numerical_2014,kumar_subaperture_2013,shemonski_computational_2015,ruiz-lopera_computational_2020,ginner_noniterative_2017} offers a promising approach to overcome all the issues discussed above.
However, the performance and the limitations of the proposed computational correction techniques are not well investigated.
For computational refocusing (CR), the comparison of several methods has been done \cite{kumar_numerical_2014}, however, it is only for 2D \enface imaging.
The rigorous OCT image formation theories\cite{davis_nonparaxial_2007,villiger_image_2010,fukutake_four-dimensional_2025-1} suggest that the elongation of the OCT's PSF along the depth (delay) direction occurs with the strong defocus in a high numerical aperture system.
The investigation of the performance of computational correction methods in detail should require a 3D PSF assessment.
For aberration correction, it is known that there is an issue in aberration estimation from OCT signals with a point-scanning OCT configuration because of the convolution of the illumination and collection pupils\cite{kumar_-vivo_2017,south_wavefront_2018}.
Previously existing computational aberration correction (CAC) or computational adaptive optics (CAO) methods were used to estimate the phase errors in the spatial frequency components of OCT signals\cite{adie_computational_2012,hillmann_aberration-free_2016,auksorius_vivo_2020,ruiz-lopera_computational_2020,kumar_-vivo_2017,kumar_digital_2021}.
This approach is perhaps suitable for full-field SS-OCT\cite{hillmann_aberration-free_2016,auksorius_vivo_2020} because the illumination is in the form of a plane wave.
Therefore, the illumination pupil is a delta function, and then the \enface spatial frequency of OCT signals can be treated as the collection pupil.
The impact of the convolution of pupils in aberration estimation is reported \cite{kumar_-vivo_2017,south_wavefront_2018,liu_closed-loop_2021} but only for 2D (single depth) cases.
In the case of point-scanning FD-OCT (PSFD-OCT), the A-line profiles are acquired simultaneously, the depth-(defocus-)dependent phase errors must be considered for volumetric CAC\@.

The image formation process of OCT is complicated because the 4D space \cite{fukutake_four-dimensional_2025-1} is required to be represented accurately.
To investigate the influence of the optical aberrations and performance of computational methods in detail, the properly formulating an accurate model based on rigorous image formation theory will thus be an important aid to the understanding of the imaging properties and prediction of new processing approaches.

In this paper, we reformulate the OCT image formation theory and include optical aberrations in it for interpreting the definition of aberrations, which is corrected by CAC, and for developing an accurate OCT signal simulator.
The utility of this theoretical framework is demonstrated by investigating CR and CAC methods with numerically simulated OCT signals that have been generated using the rigorous theory.
A volumetric CAC filter for use in fiber-optic-based PSFD-OCT has been designed that, to our knowledge, is new.
Computational corrections for the OCT images of numerically simulated signals, of a microparticle phantom, and of an \invivo human retina are performed by applying the newly designed method, and the results are compared with those of previous approaches.

\section{OCT image formation theory}
\label{sec:OCT_image_formation_theory}

Several previous works have formulated the image formation theory for OCT\cite{davis_nonparaxial_2007,villiger_image_2010,sheppard_reconstruction_2012,zhou_unified_2021}.
In this paper, we reformulate the image formation theory of OCT to include optical aberrations and to ease the interpretation of computational methods applied to OCT signals.
Through the following formulation, the reference plane of the wavefront error that is corrected by the computational methods using \enface Fourier transformation is explicitly defined, and the numerical model for simulating the performance of CAC methods is established.

We began by formulating the imaging theory for the PSFD-OCT system because it is the most common type among the OCT systems in current use.
Other OCT system types can be obtained by modifying the system illumination and collection configurations, as shown in Supplement 1 Section~\ref{S-sec:CR_other_OCT} and Ref.~\cite{fukutake_four-dimensional_2025-1}.

The interference signal detected by PSFD-OCT can be described from the rigorous OCT image formation theory (Supplement 1, Section~\ref{S-sec:OCT_theory}) as:
\begin{equation}\label{eq:3}
    \begin{split}
        I^{'} (\mathbf{r}_{0\parallel}, \omega; z_0)
        =& \sqrt{p p_\mathrm{r}} S(\omega)
            \mathrm{e}^{\mathrm{i} 2 [k_\mathrm{s}(\omega) z_0 - k_\mathrm{r}(\omega) z_\mathrm{r}]} \\
        & \times U_\mathrm{r}^*
        \iiint
            h_\mathrm{RCI} [\mathbf{r}_{0\parallel} - \mathbf{r}_\parallel, z_0 - z, k_{\mathbf{s}}(\omega)]
            \frac{k_\mathrm{s}^2 (\omega)}{4\uppi} \psi_\mathrm{p} (\omega) N(\mathbf{r}_\parallel, z)
        \mathrm{d}\mathbf{r}_\parallel \mathrm{d} z,
    \end{split}
\end{equation}
where $h_{\mathrm{RCI}}$ is the 3D complex point spread function (cPSF) for reflection confocal imaging (RCI), $\mathbf{r}_{0\parallel} = (x_0, y_0)$ is the transversal focal scanning location, $z_0$ is the axial location of the focus, $k_\mathrm{s} = n_\mathrm{BG} k$ is the wavenumber in the sample with background refractive index $n_\mathrm{BG}$, and $S$ is the normalized spectral density of the light source [Hz$^{-1}$].
In addition, $p$ and $p_\mathrm{r}$ are the light powers of the sample and reference arms [W], respectively, and $z_\mathrm{r}$ is the single-trip path length of the reference arm.
$U_\mathrm{r}$ is the monochromatic wave function of the reference light at the detector plane [m$^{-1}$], where a wave function is a solution of the Helmholtz equation.
$\psi_{\mathrm{p}}$ is the molar relative electric susceptibility [m$^3$], $N$ is the density distribution of the molecules that contribute to the backscattering [m$^{-3}$], and they are related as $\psi = \psi_\mathrm{p} N$.
Here, $I^{'}$ is the complex-valued signal.

\subsection{Complex point-spread function for reflection confocal imaging}
\label{sec:PSF_RCI}

\begin{figure}
    \centering
    \includegraphics[width=0.6\linewidth]{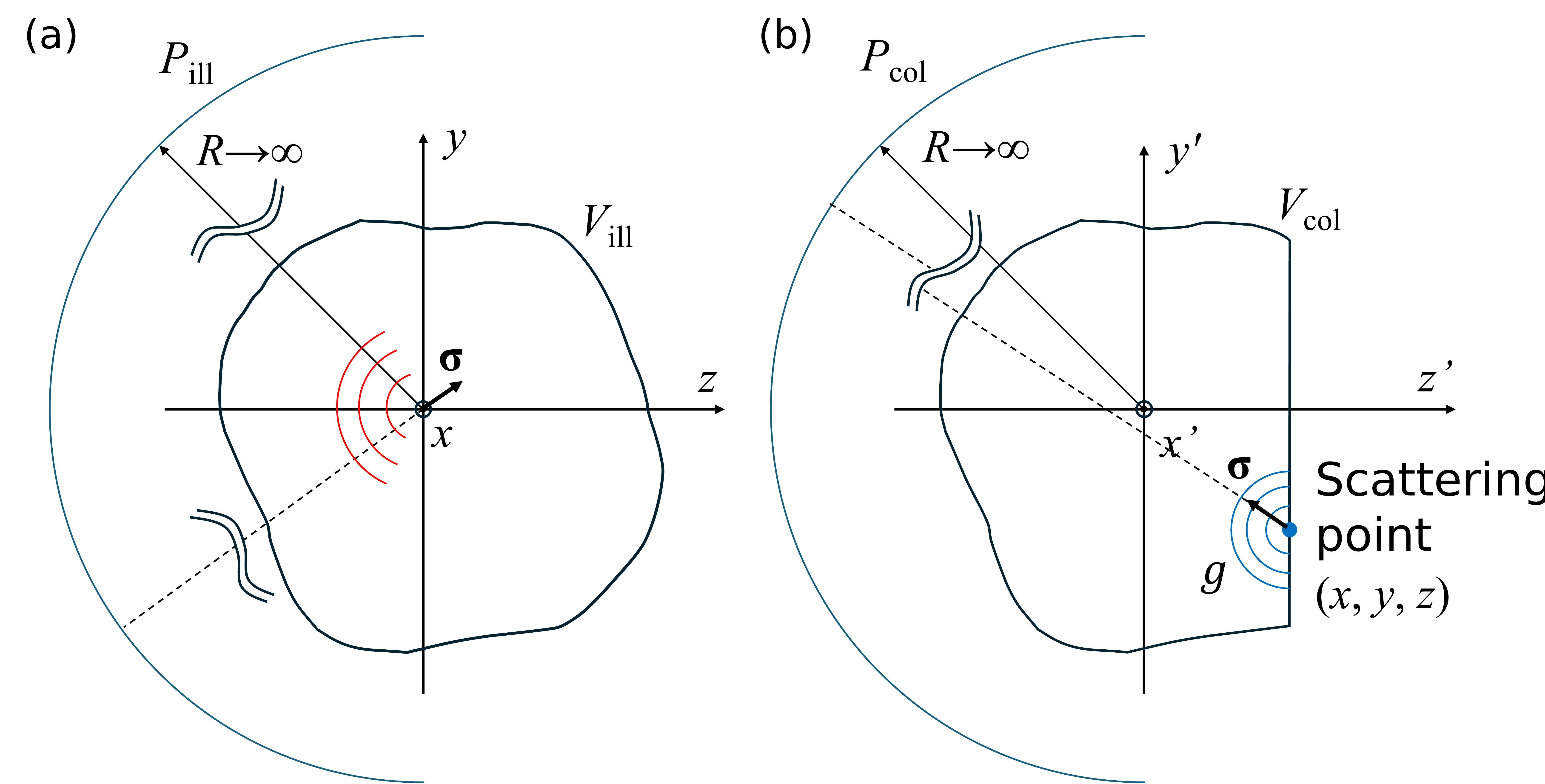}
    \caption{Definition of the spaces of the (a) illumination and (b) collection fields.}\label{fig:fieldspace}
\end{figure}

The cPSF $h_\mathrm{RCI}$ for RCI are can be expressed as the product of the illumination field and collection mode, $f_\mathrm{ill}$ and $f_\mathrm{col}$, respectively, as follows\cite{sheppard_three-dimensional_1994,davis_nonparaxial_2007,villiger_image_2010}:

\begin{equation}\label{eq:9}
    \begin{split}
        h_\mathrm{RCI}(\mathbf{r}_\parallel, z, k) =& f_\mathrm{col}(-\mathbf{r}_\parallel, -z, k) f_\mathrm{ill}(-\mathbf{r}_\parallel, -z, k) \\
        =&
        \frac{-k^2}{4\uppi^2}
        \iint_{|\bm{\upsigma}_\parallel| \le 1}
            P_\mathrm{col} (\bm{\upsigma}_\parallel) \mathrm{e}^{\mathrm{i} k(\mathbf{r}_\parallel\cdot\bm{\upsigma}_\parallel - z \sigma_z)}
        \mathrm{d}\Omega \\
        & \times
        \iint_{|\bm{\upsigma}_\parallel| \le 1}
            P_\mathrm{ill}(-\bm{\upsigma}_\parallel) \mathrm{e}^{-\mathrm{i} k(\mathbf{r}_\parallel\cdot\bm{\upsigma}_\parallel + z \sigma_z)}
        \mathrm{d}\Omega.
    \end{split}
\end{equation}
Here we derive $f_\mathrm{ill}$ and $f_\mathrm{col}$ as the Debye diffraction integrals (The details of the derivation are available in the Supplement 1, Section~\ref{S-sec:PSF_RCI}).
$P_\mathrm{ill}$ and $P_\mathrm{col}$ are the illumination and collection field distributions, respectively, on a spherical plane centered at the geometrical focus according to the definition of the Debye diffraction integral [\figurename~\ref{fig:fieldspace}], and they are assumed to be independent of the optical frequency $\omega$; i.e., the angular distribution of both the illumination and collection fields are the same for all optical frequencies.
$P_{\mathrm{ill}/\mathrm{col}}/R$ is the amplitude of the wave function for a monochromatic wave.
Therefore, $P_{\mathrm{ill}/\mathrm{col}}$ is unitless.
Thus, the unit of $h_\mathrm{RCI}$ is [m$^{-2}$].
Here, $f_\mathrm{ill}$ and $f_\mathrm{col}$ are also wave functions, and they should satisfy $\iint_{-\infty}^{\infty} |f_{\mathrm{ill}/\mathrm{col}}|^2 \mathrm{d}x \mathrm{d}y = 1$ to meet their physical meanings in Eq.~(\ref{eq:3}).

Note that the Debye diffraction integral is based on the approximation that the volume of interest is small when compared with the radius of the sphere, $R$.
When we consider $R$ to be infinitely large, as illustrated in \figurename~\ref{fig:fieldspace}, Eqs.~(\ref{S-eq:4}) and (\ref{S-eq:8}) are then probably accurate \cite{braat_imaging_2019} within the entire object spaces of both illumination $V_\mathrm{ill}$ and collection $V_\mathrm{col}$.
Here, the space $V_\mathrm{col}$ is bounded at the axial location of each emitter to define the backscattered field alone.
In this case, $P_\mathrm{ill}$ and $P_\mathrm{col}$ correspond to infinitely distant distributions.
If we consider the objective to be a thin lens that has no chromatic aberrations, $P_\mathrm{ill}$ and $P_\mathrm{col}$ may then be equivalent to the illumination distribution and collection modes when expressed at the front focal plane of the objective, respectively.

\subsubsection{Spatial frequency representation}

Because the CR and CAC methods are applied as a phase filter in the spatial frequency domain, it is suitable to express the cPSF $h_\mathrm{RCI}$ and also OCT's PSF in the spatial frequency domain.
The lateral Fourier transform of the RCI cPSF $\tilde{h}_\mathrm{RCI}$ can now be expressed as a 2D convolution within the lateral spatial frequency space by using Eq.~(\ref{S-eq:10}) as follows:
\begin{equation}\label{eq:12}
    \begin{split}
        \tilde{h}_\mathrm{RCI}(\bm{\upnu}_\parallel, z, k_\mathrm{s})
        =& \mathcal{F}_{\mathbf{r}_\parallel}[h_\mathrm{RCI}(\mathbf{r}_\parallel, z, k_\mathrm{s})](\bm{\upnu}_\parallel)\\
        =&
\left[
            \tilde{f}_\mathrm{ill}(-\bm{\upnu}_\parallel, -z, k_\mathrm{s})
            \otimes_{\bm{\upnu}_\parallel}^2
            \tilde{f}_\mathrm{col}(-\bm{\upnu}_\parallel, -z, k_\mathrm{s})
        \right](\bm{\upnu}_\parallel, z, k_\mathrm{s}),
    \end{split}
\end{equation}
where $[f \otimes_{\mathbf{x}}^n g]$ is the $n$-dimensional convolution operation between $f$ and $g$ with respect to $\mathbf{x}$, $\mathcal{F}_{x}[f](\nu)$ is the Fourier transform of the function $f$ from the $x$-axis to the $\nu$-axis, and $\tilde{f}_\mathrm{ill}$ and $\tilde{f}_\mathrm{col}$ are the lateral 2D Fourier transform of the illumination and collection fields.
Then, the 1D Fourier transform of Eq.~(\ref{eq:12}) along $z$ is given by:
\begin{equation}\label{eq:13}
    H_\mathrm{RCI}(\bm{\upnu}_\parallel, \nu_z, k_\mathrm{s})
    =
\left[
        F_\mathrm{ill}(-\bm{\upnu}_\parallel, -\nu_z, k_\mathrm{s})
        \otimes_{\bm{\upnu}}^3
        F_\mathrm{col}(-\bm{\upnu}_\parallel, -\nu_z, k_\mathrm{s})
    \right](\bm{\upnu}_\parallel, \nu_z, k_\mathrm{s}),
\end{equation}
where $F_\mathrm{ill}$ and $F_\mathrm{col}$ are the spatial 3D Fourier transforms of the illumination field $f_\mathrm{ill}$ and the collection field $f_\mathrm{col}$, respectively.
This three-dimensional spatial Fourier transform of $h_\mathrm{RCI}$ is called the coherent transfer function (cTF) of RCI\cite{sheppard_three-dimensional_1994,gu_image_1991}.
Because the cPSF $h_\mathrm{RCI}$ has the unit of [m$^{-2}$], the cTF $H_\mathrm{RCI}$ has the unit of [m].
For each $k$, Eq.~(\ref{eq:13}) represents a 3D convolution of spherical shell caps [see Eq.~(\ref{S-eq:S2})].
This image formation approach is used in a wide range of microscopy techniques\cite{sheppard_three-dimensional_1994,gu_image_1991,fukutake_general_2020,fukutake_unified_2025-1}.
As shown in examples from previous studies, the RCI cTF has a thickness along the axial frequency direction of $\nu_z$.
This nature makes it difficult to reconstruct accurate 3D object structure in PSFD-OCT because some axial frequency components of the object have already been integrated within the axial thickness of the cTF during the OCT detection process.
Additionally, this is the source of the signal loss that occurs at defocus\cite{villiger_image_2010}.
However, in the case where the system is a fiber-optic system, the thickness is mitigated by the use of the apodized detection\cite{gu_image_1991}.
Because PSFD-OCT is usually based on use of a fiber-optic system and low numerical aperture (NA) illumination/collection, the cTF is not thick along $\nu_z$, as illustrated in \figurename~\ref{S-fig:Validation-sim}.

\subsubsection{Representation of systematic aberrations}

In the case where high-order systematic aberrations exist, the cPSF $h_\mathrm{RCI}$ can be expressed by substituting $P_\mathrm{ill}(\bm{\upsigma}_\parallel) \rightarrow P_\mathrm{ill} (\bm{\upsigma}_\parallel) \mathrm{e}^{\mathrm{i} k W_\mathrm{ill}(\bm{\upsigma}_\parallel, k)}$ and $P_\mathrm{col}(\bm{\upsigma}_\parallel) \rightarrow P_\mathrm{col} \mathrm{e}^{\mathrm{i} k W_\mathrm{col}(\bm{\upsigma}_\parallel, k)}$, where $W_\mathrm{ill}$ and $W_\mathrm{col}$ are the wavefront errors, except the defocus, for illumination and collection, respectively, at the infinitesimally distant spherical reference plane.
Using the spatial frequency $\bm{\upnu}_\parallel$, wavefront errors can be written as
\begin{equation}
    W(\bm{\upnu}_\parallel, k) \equiv \sum_{j} w_j(k) Z_j \left(\frac{\frac{2\uppi}{k}\bm{\upnu}_\parallel}{\sigma_{\parallel, \mathrm{c}}}\right),
\end{equation}
where $Z_j$ is a Zernike polynomial, $w_j$ is the wavefront error coefficient when expanded using the Zernike polynomial, and $j$ is the index of the Zernike coefficients in the single indexing scheme\cite{thibos_standards_2002}.
$\sin^{-1}\sigma_{\parallel, \mathrm{c}}$ is the cut-off angle for illumination or collection, and $n_\mathrm{BG} \sigma_{\parallel, \mathrm{c}}$ is the cut-off NA\@.

Strictly speaking, the pupil $P$ and wavefront error $W$ will be dependent on the scanning location $\mathbf{r}_{0\parallel}$ according to the beam scan configurations.
In the case of translational beam scanning [\figurename~\ref{S-fig:Interpret_Aberrations}(a)], the displacement of the scanning beam on the physical aperture is equivalent to the displacement of the aperture of the entire optical system.
The same plane wave component will suffer different wavefront aberrations.
Therefore, $W$ becomes a function of $\mathbf{r}_{0\parallel}$ as $W(\mathbf{r}_{0\parallel})$.

In contrast, the fan-beam scanning procedure [\figurename~\ref{S-fig:Interpret_Aberrations}(b)] tilts the beam wavefront at the physical aperture.
Therefore, $P$ becomes a function of $\mathbf{r}_{0\parallel}$ as $P(\mathbf{r}_{0\parallel})$.

In both cases, the cTF becomes a function of $\mathbf{r}_{0\parallel}$ as $h_\mathrm{RCI} (\mathbf{r}_\parallel, z, k, \mathbf{r}_{0\parallel})$.
If we assume that the image is constructed via convolution of the object structure and $h_\mathrm{RCI}$, the scanning range must then be small enough for $h_\mathrm{RCI}$ to be regarded as being constant over the entire scanning range.
The maximum of this range may then define the acceptable scanning range for CAC\@.

Note here that the ocular aberrations $W_\mathrm{ocular}$ are usually defined as wavefront errors at the anterior part of the eye\cite{thibos_standards_2002}.
This corresponds to the plane immediately before the objective.
This definition differs from that of $W$, as illustrated in [\figurename~\ref{S-fig:Interpret_Aberrations}(c)].
Therefore, many studies on retinal imaging methods, including on adaptive optics imaging methods and aberrometers for the eye, reported the aberrations that were defined at this plane.
These different definitions of $W$ may make it difficult to compare the aberrations estimated using CAC with PSFD-OCT directly with the ocular aberrations reported in other studies.

\subsection{Complex OCT signal reconstruction}

Here, the reconstructed OCT signal, i.e., the complex OCT signal in the spatial domain, is described using the derivations presented above.
The 1D Fourier transform of Eq.~(\ref{eq:3}) along $\omega$ becomes the complex OCT signal:
\begin{equation}\label{eq:18}
    \begin{split}
        s(\mathbf{r}_{0\parallel}, \tau; z_0)
        =&
        \int
            I^{'} (\mathbf{r}_{0\parallel}, \omega; z_0) \mathrm{e}^{- \mathrm{i} \tau\omega}
        \mathrm{d}\omega\\
        =& \frac{\sqrt{pp_\mathrm{r}} U_\mathrm{r}^*}{4\uppi}
        \iiint
            h_\mathrm{OCT} (\mathbf{r}_{0\parallel} - \mathbf{r}_\parallel, z_0 - z, \tau)
            N(\mathbf{r}_\parallel, z)
        \mathrm{d}\mathbf{r}_\parallel
        \mathrm{d}z,
    \end{split}
\end{equation}
where $\tau$ represents the optical delay.
Here, $h_\mathrm{OCT}$ is the cPSF of the OCT signal and can be described 
as follows:
\begin{equation}\label{eq:19}
h_\mathrm{OCT}(\mathbf{r}_\parallel, z, \tau)
    = \int
        \psi_\mathrm{p}(\omega)
        S(\omega)
        k_\mathrm{s}^2(\omega)
\mathrm{e}^{\mathrm{i} 2 [k_\mathrm{s}(\omega) z_0 - k_\mathrm{r}(\omega) z_\mathrm{r}]}
        h_\mathrm{RCI}(\mathbf{r}_\parallel, z, k_\mathrm{s}(\omega))
        \mathrm{e}^{- \mathrm{i} \tau \omega}
    \mathrm{d}\omega.
\end{equation}
The unit of $h_\mathrm{OCT}$ is [m$^{-1}$].

\section{Numerical simulation method}\label{sec:numerical_simulation_method}

Based on the OCT image formation theory, the numerical simulator is developed to efficiently compute the OCT PSF and cTF\@.
The OCT image formulation theory requires treatment of a 4D function [Eq.~(\ref{eq:3})] and thus the PSF of OCT is also a 4D function [Eq.~(\ref{eq:19})].
However, the simulation of these functions requires large amounts of memory and computation time.
For simulation of the PSFs, one or a few depth locations of the sample $z_{\mathrm{s}}$ are selected.

Because the 3D spatial frequency spectra of illumination and collection fields are delta functions [Eq.~(\ref{S-eq:S2})], it is not practical to directly start the numerical simulation from the 3D spatial frequency domain.
The calculations start by defining $P_\mathrm{ill}$ and $P_\mathrm{col}$.
Then, $\tilde{f}_\mathrm{ill}$ and $\tilde{f}_\mathrm{col}$ [Eq.~\ref{S-eq:10}] are determined by multiplying them by the propagation phase term and the inclination factor $1/\sigma_z$.
A 2D convolution is then calculated between $\tilde{f}_\mathrm{ill}$ and $\tilde{f}_\mathrm{col}$ to obtain $\tilde{h}_\mathrm{RCI}$.
This procedure is iterated for the desired depth locations and optical frequencies.

Finally, the 3D inverse Fourier transform for the spatial frequencies $\bm{\upnu}_\parallel$ and the optical frequency $\omega$ is calculated for the PSF simulation.
These last-stage Fourier transforms are calculated using a zoom FFT\cite{leutenegger_fast_2006} to avoid calculation of the blank regions.
For validating the numerical simulation procedure of $\tilde{h}_\mathrm{RCI}$, the cTF is simulated using the same approach (Supplement 1, Section~\ref{S-sec:numerical_simulation-ctf}) and compared with the analytical solution (Supplement 1, Section~\ref{S-sec:ctf_anal}).
Simulated cTFs were found to match analytical cTFs in both distribution and value (Supplement 1, Section~\ref{S-sec:validation-cTFsim}).

An open source simulator of the PSF of OCT and cTF of reflection imaging can be found at \url{https://github.com/ComputationalOpticsGroup/COG-OCTPSF-simulator}\cite{OCTPSF_simulator}.

\section{Design of computational refocusing and aberration correction based on the image formation theory}\label{sec:DRandCAC}

\subsection{Simplified model of the OCT image formation process}\label{sec:simplified_model}

The forward aberration correction and refocusing method is based on \enface OCT signal processing.
To describe them, the approximated OCT signal formation process is desired.
Here, we formulate the simplification of the OCT image formation to clarify the required approximations of the forward-based methods.

\subsubsection{Paraxial approximation}

The axial frequency thickness of the RCI cTF becomes thicker as the NA values of the illumination and collection fields increase.
However, this change also means that the axial imaging range of PSFD-OCT (i.e., the depth of focus) becomes shallower.
Therefore, PSFD-OCT is commonly used under low NA conditions.
In this case, the paraxial approximation $|\bm{\upsigma}_\parallel| \ll 1$ may be valid.
Then, $\sigma_z \approx 1 - \frac{|\bm{\upsigma}_\parallel|^2}{2}$.

The lateral Fourier transform of the RCI cPSF $\tilde{h}_\mathrm{RCI}$ [Eq.~(\ref{eq:12})] with the paraxial approximation can now be expressed
as:
\begin{equation}\label{eq:23}
    \tilde{h}_\mathrm{RCI}(\bm{\upnu}_\parallel, z, k_\mathrm{s}) \approx \mathrm{e}^{- 2\mathrm{i} k_\mathrm{s} z} \tilde{\Gamma}(\bm{\upnu}_\parallel, z, k_\mathrm{s}),
\end{equation}
where
\begin{eqnarray}
    \tilde{\Gamma} (\bm{\upnu}_\parallel, z, k_\mathrm{s}) &\equiv&
    - \frac{4\uppi^{2}}{k_\mathrm{s}^{2}}
    \left[
        A_\mathrm{ill} \mathrm{e}^{\mathrm{i} \phi_\mathrm{ill}}
        \otimes_{\bm{\upnu}_\parallel}^2
        A_\mathrm{col} \mathrm{e}^{\mathrm{i} \phi_\mathrm{col}}
    \right](\bm{\upnu}_\parallel, z, k_\mathrm{s}) \label{eq:24}\\
    A(\bm{\upnu}_\parallel, k_\mathrm{s}) &\equiv& \frac{P\left(\frac{2\uppi}{k_\mathrm{s}} \bm{\upnu}_\parallel\right)}{1 - \alpha(k_\mathrm{s}) |\bm{\upnu}_\parallel|^2}\label{eq:25}\\
    \phi(\bm{\upnu}_\parallel, z, k_\mathrm{s}) &\equiv& k_\mathrm{s}z \alpha(k_\mathrm{s}) \left|\bm{\upnu}_\parallel\right|^2.\label{eq:26}
\end{eqnarray}
$A$ represents $A_\mathrm{ill}$ or $A_\mathrm{col}$, and $\phi$ represents $\phi_\mathrm{ill}$ or $\phi_\mathrm{col}$.
$\alpha(k) = \frac{1}{2}\left(\frac{2\uppi}{k} \right)^2$.

\subsubsection{Paraxial 2D cTF with aberrations}

The case with high-order aberrations can be expressed by substituting $P_\mathrm{ill} \rightarrow P_\mathrm{ill} \mathrm{e}^{\mathrm{i} k W_\mathrm{ill}}$ and $P_\mathrm{col} \rightarrow P_\mathrm{col} \mathrm{e}^{\mathrm{i} kW_\mathrm{col}}$, where $W_\mathrm{ill}$ and $W_\mathrm{col}$ are the wavefront errors with exclusion of the defocus.
$W_\mathrm{ill}$ and $W_\mathrm{col}$ may be conjugate to the wavefront aberrations occurring at the front focal plane of the objective.
The phase term Eq.~(\ref{eq:26}) can be treated as:
\begin{equation}
    \phi(\bm{\upnu}_\parallel, z, k_\mathrm{s}) \rightarrow k_\mathrm{s} z \alpha(k_\mathrm{s}) \left|\bm{\upnu}_\parallel\right|^2 + k_\mathrm{s} W(\bm{\upnu}_\parallel, k_\mathrm{s}).
\end{equation}
Here, $W$ is $W_\mathrm{ill}$ or $W_\mathrm{col}$.
When the illumination optics and the collection optics are identical, e.g., in the case of a conventional fiber-optics-based PSFD-OCT, it can be considered that $A_\mathrm{ill} = A_\mathrm{col}$ and $W_\mathrm{ill} = W_\mathrm{col}$.

\subsubsection{OCT imaging with paraxial and narrow-band approximations}

By substituting Eq.~(\ref{eq:23}) into Eq.~(\ref{eq:19}), the cPSF of OCT under the paraxial approximation is obtained as follows:

\begin{equation}\label{eq:28}
h_\mathrm{OCT} (\mathbf{r}_\parallel, z, \tau)
    \approx
    \int
        \psi_\mathrm{p} (\omega)
        S(\omega)
        k_\mathrm{s}^2(\omega)
        \mathrm{e}^{\mathrm{i} 2 [(z_0 - z) k_\mathrm{s}(\omega) - z_\mathrm{r} k_\mathrm{r}(\omega)] }
        \Gamma\left(\mathbf{r}_\parallel, z, k_\mathrm{s}(\omega)\right)
        \mathrm{e}^{- \mathrm{i} \tau\omega}
    \mathrm{d}\omega,
\end{equation}
where $\Gamma(\mathbf{r}_\parallel, z, k) = \mathcal{F}^{-1}_{\bm{\upnu}_\parallel} \left[\tilde{\Gamma} (\bm{\upnu}_\parallel, z, k)\right](\mathbf{r}_\parallel)$.

The OCT cPSF with narrow-band approximation and without high-order dispersion mismatch can now be described as [Eq.~\ref{S-eq:parapsf}]:
\begin{equation}\label{eq:parapsf}
h_\mathrm{OCT} (\mathbf{r}_\parallel, z, \tau)
        \propto \psi_\mathrm{p} (\omega_\mathrm{c})
\Gamma^{'}\left(\mathbf{r}_\parallel, z; k_\mathrm{s}(\omega_\mathrm{c} )\right)
        \gamma\left\{\frac{2}{c} \left[
            l + n_\mathrm{g,BG} (\omega_\mathrm{c}) (z - z_0) + n_\mathrm{g,r} (\omega_\mathrm{c}) z_\mathrm{r}
        \right]\right\},
\end{equation}
where $\gamma$ is the complex temporal coherence function, $l = \frac{c\tau}{2}$ is the single-pass optical path length (OPL) corresponding to the optical delay $\tau$, $c$ is the speed of light in vacuum, $C$ is a phase term [Eq.~\ref{S-eq:C}], and $n_{g}$ is the group index.

The forward-based refocusing and aberration correction methods assume that the OCT's PSF is described by Eq.~(\ref{eq:parapsf}).
The following design procedure of the forward-based computational methods is based on this PSF model.

\subsection{Computational refocusing}\label{sec:OCTrefocusing}

Several methods for digital correction of defocus have been proposed previously.
Interferometric synthetic aperture microscopy (ISAM)\cite{ralston_interferometric_2007} involves resampling of the data $\tilde{I}^{'}$ in the $(\bm{\upnu}_{\parallel}, \omega)$ space to retrieve the information in the $(\bm{\upnu}_{\parallel}, \nu_z)$ space.
It is known that this process requires the use of an approximation in the case of PSFD-OCT \cite{sheppard_reconstruction_2012}.
ISAM assumes that the $(\bm{\upnu}_{\parallel}, \omega)$ and $(\bm{\upnu}_{\parallel}, \nu_z)$ spaces are bijective.
Because the cTF for RCI is not infinitesimally thin along $\nu_z$, this assumption is not valid, and thus, ISAM does not perform a rigorously perfect correction.
Theoretically, a low NA condition is required for ISAM to be a good approximation.
Lee et al\cite{lee_wide-field_2023} showed that ISAM then becomes a phase-filtering method with a narrow-band approximation.
Therefore, under both the narrow-band and paraxial approximations, ISAM becomes a forward-phase-filtering method\cite{yasuno_non-iterative_2006}.

Here, we derive this forward-phase-filtering method from the image formation theory.
Because the paraxial approximation and the narrow-band approximation are both required for this derivation, their descriptions are based on Eqs.~(\ref{eq:24})-(\ref{eq:26}).
The OCT signal of a scatterer at a location of $\mathbf{r}_\mathrm{s} = (\mathbf{r}_{\mathrm{s}\parallel}, z_\mathrm{s})$ can be expressed using Eqs.~(\ref{eq:18}) and (\ref{eq:parapsf}) as:
\begin{equation}\label{eq:paraoctsig}
s \left(\mathbf{r}_{0\parallel}, \frac{2}{c}l; z_0, \mathbf{r}_\mathrm{s}\right)
        \propto
\psi_\mathrm{p} (\omega_\mathrm{c})
\Gamma^{'}\left(\mathbf{r}_{0\parallel} - \mathbf{r}_{\mathrm{s}\parallel}, z_0 - z_\mathrm{s}; k_\mathrm{s}(\omega_\mathrm{c} )\right)\\
\gamma\left[\frac{2}{c} \left\{
            l - n_\mathrm{g,BG} (\omega_\mathrm{c}) z_\mathrm{s} + n_\mathrm{g,r} (\omega_\mathrm{c}) z_\mathrm{r}
        \right\}\right].
\end{equation}

In the case of a single-mode fiber-based system, the same optical fiber is used both in the illumination and collection paths and thus limits the propagation mode.
Therefore, $P_\mathrm{ill}$ and $P_\mathrm{col}$ can be assumed to have the same Gaussian distribution.
With a low NA and the narrow-band approximation, the magnitude distributions may be assumed to be 2D Gaussian functions that are independent of the wavenumber, i.e., $A = A_\mathrm{ill} = A_\mathrm{col} = \mathrm{e}^{-|\bm{\upnu}_\parallel|^2 / \Delta f^2}$.
If there is only the defocus, then the phase term $\phi(\bm{\upnu}_\parallel, z, k_\mathrm{s}(\omega_\mathrm{c})) \rightarrow \frac{2\uppi^2}{n_{\mathrm{BG}} k_\mathrm{c}} z |\bm{\upnu}_\parallel|^2$, where $k_\mathrm{c}$ is the central wavenumber of the light source in a vacuum.
The lateral frequency spectrum of an OCT signal when a scatterer is located at $(\mathbf{0}, z_\mathrm{s})$ can be expressed using Eq.~(\ref{eq:24}) as:
\begin{equation}\label{eq:36}
    \tilde{\Gamma}(\bm{\upnu}_\parallel, z_0 - z_\mathrm{s}; k_\mathrm{s}(\omega_\mathrm{c}))
    =
    - \frac{4\uppi^{2}}{k_\mathrm{s}^{2}}
    \frac{\uppi}{2}
    \frac{A^{1/2} (\bm{\upnu}_\parallel, k_\mathrm{s}(\omega_\mathrm{c}))}
    {1/\Delta f^2 - \mathrm{i} \phi_\mathrm{def} (z_0 - z_\mathrm{s})}
    \mathrm{e}^{\mathrm{i} \frac{\phi_\mathrm{def}}{2} (z_0 - z_\mathrm{s}) |\bm{\upnu}_\parallel|^2 },
\end{equation}
where
\begin{equation}\label{eq:37}
    \phi_\mathrm{def} (\omega_\mathrm{c})
    = k_\mathrm{s}(\omega_\mathrm{c}) \alpha(k_\mathrm{s}(\omega_\mathrm{c}))
\end{equation}
represents the defocus phase factor.
The phase-only refocus filter is thus given by
\begin{equation}\label{eq:38}
    \tilde{\Gamma}^{-1}_{\mathrm{CR},\mathrm{PSFD}} (\bm{\upnu}_\parallel, l_\mathrm{s}; k_\mathrm{s}(\omega_\mathrm{c}))
    = \mathrm{e}^{\mathrm{i} \frac{\phi_\mathrm{def}}{2} \delta z_\mathrm{s}(l_\mathrm{s}) |\bm{\upnu}_\parallel |^2},
\end{equation}
where $\delta z_\mathrm{s}(l_\mathrm{s}) \equiv \frac{l_\mathrm{s} - l_0}{n_\mathrm{g,BG}}$, and $l_\mathrm{s} - l_0 \approx n_\mathrm{g,BG}(\omega_\mathrm{c}) (z_s - z_0)$ is the OPL in the sample from the focus at $z = z_0$.
This is identical to the corresponding expression from Ref.~\cite{yasuno_non-iterative_2006}.
Furthermore, this method is equivalent to assuming that the cTF forms a parabola in the spatial frequency domain and that it is independent of the wavenumber.
The discrepancy between the real cTF and this shape should be the source of the correction error.

Note here that the phase term of $\tilde{\Gamma}$ depends on the magnitude distributions of the pupils, i.e., $P_\mathrm{ill}$ and $P_\mathrm{col}$, because of the convolution performed in Eq.~(\ref{eq:24}).
The defocus phase in Ref.~\cite{yasuno_non-iterative_2006} is based on the fact that the illumination and collection paths are identical, and $A_\mathrm{ill} = A_\mathrm{col}$ is assumed to form a Gaussian distribution.
When the system condition differs from this assumption, the optimal refocusing filter should then be different from that of Eq.~(\ref{eq:38}).
The impact of the amplitude distribution of the pupils is discussed in Section~\ref{sec:discuss-apodization}.

\subsection{Computational aberration correction}
\label{sec:CAC}

\begin{figure}
    \centering
    \includegraphics[width=0.8\linewidth]{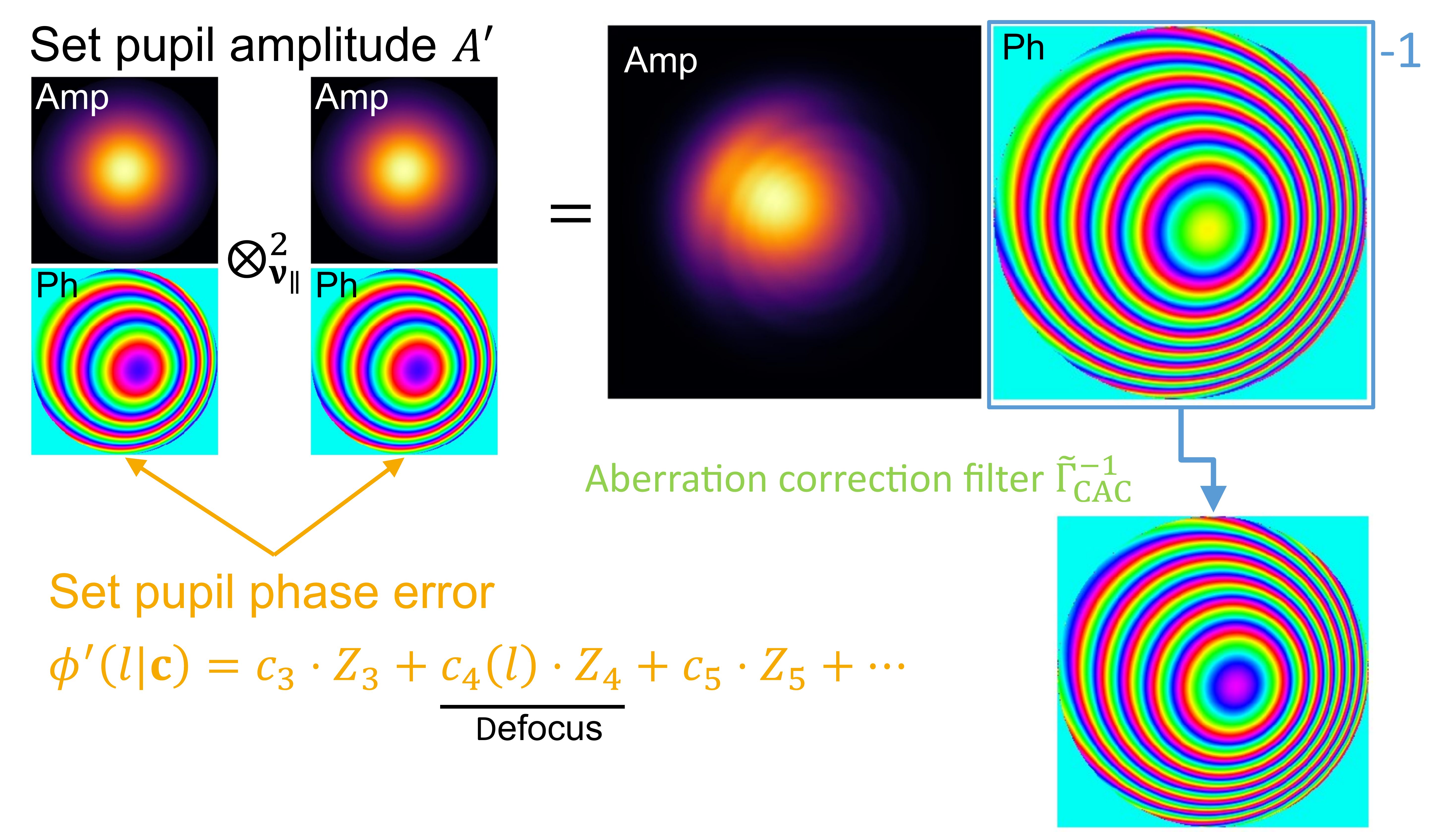}
    \caption{Schematic diagram of new CAC filter calculations.}
    \label{fig:3}
\end{figure}

In the case with the paraxial and narrow-band approximations, we can define the aberration correction filter for volumetric aberration correction using Eq.~(\ref{eq:24}).
The schematic diagram used to calculate the CAC filter is shown in \figurename~\ref{fig:3}.
The filter response in the spatial frequency domain can be defined using the phase coefficients of the Zernike polynomials, i.e., $\mathbf{c} = \{..., c_j, ...\}$, for wavefront aberrations in the pupil plane as follows:
\begin{equation}\label{eq:42}
    \tilde{\Gamma}^{-1}_{\mathrm{CAC}} (\bm{\upnu}_\parallel, l, \mathbf{c}) = \mathrm{e}^{- \mathrm{i} \arg [P' \otimes_{\bm{\upnu}_\parallel}^2 P'](\bm{\upnu}_\parallel; l, \mathbf{c})},
\end{equation}
where
\begin{equation}\label{eq:43}
    P'(\bm{\upnu}_\parallel; l, \mathbf{c}) = A' (\bm{\upnu}_\parallel) \mathrm{e}^{\mathrm{i} \phi'(\bm{\upnu}_\parallel, l, \mathbf{c})}.
\end{equation}
Here, $A'$ and $\phi'$ are the amplitude and phase of the expected aberrated pupil $P'$, respectively.
The inverse Fourier transform of Eq.~(\ref{eq:42}), i.e., $\Gamma^{-1}$, corresponds to the inverse filter used to cancel the broadening caused by the aberrations.

The same or similar models to Eq.~(\ref{eq:42}) have been previously reported\cite{kumar_-vivo_2017,south_wavefront_2018,liu_closed-loop_2021}, but they have only been used for 2D cases.
Here we extend the model to 3D cases for volumetric aberration and defocus correction.
The pupil phase can be modeled as follows:
\begin{equation}\label{eq:44}
    \phi'(\bm{\upnu}_\parallel, l, \mathbf{c})
    =
    c_4 (l) Z_4 \left(\frac{\bm{\upnu}_\parallel}{f_\mathrm{co}/2}\right)
    +
    \sum_{j \neq 4} c_j Z_j \left(\frac{\bm{\upnu}_\parallel}{f_\mathrm{co}/2}\right),
\end{equation}
where $f_\mathrm{co}$ is the OCT signal's cut-off frequency.
According to Eq.~(\ref{eq:26}), the coefficients $\mathbf{c}$ are expressions of the wavefront aberrations with respect to the pupil plane, and thus only the defocus $c_4$ depends on the optical path.
By comparing the 4th Zernike polynomial in Eq.~(\ref{eq:44}) with the defocus phase [Eq.~(\ref{eq:26})] and its influence on the OCT signal [Eq.~\ref{eq:paraoctsig}], we find that
\begin{equation}\label{eq:45}
    c_4(l) Z_4 \left(\frac{\bm{\upnu}_\parallel}{f_\mathrm{co}/2}\right)
    =
    - \phi_\mathrm{def}(\omega_\mathrm{c})
    \delta z_\mathrm{s}(l_\mathrm{s})
    \left|\bm{\upnu}_\parallel \right|^2
    + C(l),
\end{equation}
where $C(l)$ is a constant phase in the spatial frequency domain.
Therefore, it can be omitted from the analysis.
Because $Z_4 (\boldsymbol{\uprho}) \equiv \sqrt{3} (2|\boldsymbol{\uprho}|^2 - 1)$, we can define $c_4$ as:
\begin{equation}\label{eq:46}
    \begin{split}
        c_4(l)
        =&
        - \frac{\phi_\mathrm{def} (\omega_\mathrm{c}) (f_\mathrm{co}/2)^2}{2\sqrt{3}}
        \delta z_\mathrm{s}(l_\mathrm{s})\\
        \approx&
        - \frac{\uppi \lambda_{0}(\omega_\mathrm{c}) f_\mathrm{co}^2}{8 \sqrt{3} n_\mathrm{BG} (\omega_\mathrm{c}) n_\mathrm{g,BG} (\omega_\mathrm{c})}
        (l - l_0),
    \end{split}
\end{equation}
where $\lambda_0(\omega)$ is the wavelength of light with a frequency of $\omega$ in a vacuum.
Here, we assume that the OPL is proportional to the distance along the axial direction because of the use of the paraxial and narrow-band approximations.
This filter will be convolved with the complex OCT signal to cancel the aberration effects.
Full details of the implementation of this filter are presented in the Supplement 1 (Section~\ref{S-sec:CAC_implementation}).

Use of this filter design means that only a single set of the polynomial coefficients, which describe the wavefront error with respect to the pupil plane, can be used for correction over the entire OCT imaging depth range.

The conventional CAC filter is based on phase error estimation in the spatial frequency domain by using the Zernike polynomial directly.
\begin{equation}\label{eq:47}
    \tilde{\Gamma}^{-1}_{\mathrm{CAC}, \mathrm{conv}} (\bm{\upnu}_\parallel, l, \mathbf{a})
    =
    \mathrm{e}^{- \mathrm{i} a_4 (l) Z_4 \left(\frac{\bm{\upnu}_\parallel}{f_\mathrm{co}}\right)}
    \mathrm{e}^{- \mathrm{i} \sum_{j \neq 4} a_j Z_j \left(\frac{\bm{\upnu}_\parallel}{f_\mathrm{co}}\right)},
\end{equation}
where $a_j (l)$ represents the depth-dependent coefficients of the Zernike polynomial.
Only the defocus term $a_4 (l)$ is assumed to be depth-dependent, and the other terms are all depth-independent.
By comparing the 4th-order Zernike polynomial in Eq.~(\ref{eq:47}) with the defocus-only phase error of OCT signals in Eq.~(\ref{eq:36}),
\begin{equation}
    a_4(l) Z_4 \left(\frac{\bm{\upnu}_\parallel}{f_\mathrm{co}}\right)
    =
    - \frac{\phi_\mathrm{def}(\omega_\mathrm{c})}{2}
    \delta z_\mathrm{s}(l_\mathrm{s})
    \left|\bm{\upnu}_\parallel \right|^2
    + C'(l).
\end{equation}
We can define $a_4 (l)$ as:
\begin{equation}\label{eq:48}
    \begin{split}
        a_4(l)
=&
        - \frac{k_\mathrm{s} (\omega_\mathrm{c}) \alpha(k_\mathrm{s}(\omega_\mathrm{c})) f_\mathrm{co}^2}
        {2 \times 2\sqrt{3}}
        \delta z_\mathrm{s}(l_\mathrm{s})\\
        \approx&
        - \frac{\uppi \lambda_0 (\omega_\mathrm{c}) f_\mathrm{co}^2}{4 \sqrt{3} n_\mathrm{BG} (\omega_\mathrm{c}) n_\mathrm{g,BG} (\omega_\mathrm{c})}
        (l - l_0).
    \end{split}
\end{equation}
This conventional method does not take into account the wavefront error interactions between the illumination and collection paths in the model.
The Zernike coefficients, which are obtained by optimizing image quality using this method, do not accurately represent the aberration\cite{south_wavefront_2018}.
The crosstalk in phase errors due to aberration interaction will contain higher-order components compared with the aberrations.
When the number of Zernike modes is sufficient to represent the phase error, this method can estimate and correct the phase errors.
However, the model does not account for the depth-dependent crosstalk.
Corrections using coefficients estimated at different depths will result in degraded performance.

\section{Numerical simulation results}
\label{sec:simulations}

In the following numerical simulations (Section~\ref{sec:CR_simulation} and \ref{sec:CAC_simulation}), OCT signals and PSFs are calculated by the OCT signal simulator (Section~\ref{sec:numerical_simulation_method}), and then, the CR and CAC filters are applied to the generated OCT signals.

The simulation conditions are: (1) the all wavelengths are focused with the same illumination NA and the backscattered light is collected with the same collection NA ($P_\mathrm{ill}$ and $P_\mathrm{col}$ are $\omega$-independent.), (2) the relative electric susceptibility $\psi_\mathrm{p}(\omega)$ is a constant, and (3) the zero-delay point of the interferometer arms coincides with the focal plane $z_\mathrm{0} = z_\mathrm{r}$, and (4) the origin of the coordinate is on the focal plane $z_0 = 0$.

\subsection{Numerical simulation of computational refocusing}\label{sec:CR_simulation}

\begin{figure}
    \centering
    \includegraphics[width=8cm]{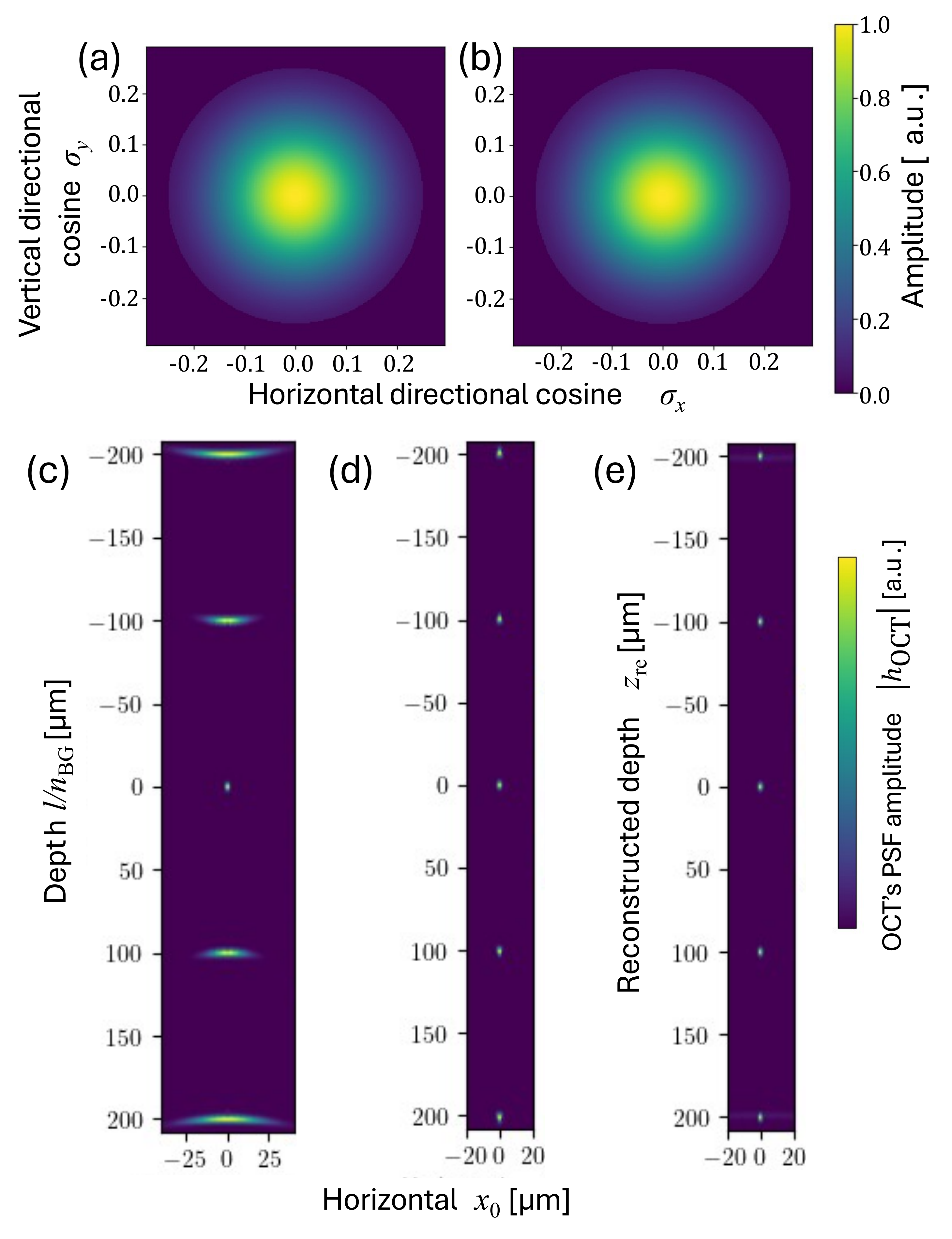}
    \caption{
        Numerical simulations of PSFD-OCT PSFs with and without computational defocus corrections.
        The cut-off NA is 0.25 $n_\mathrm{BG}$, and the effective NA is 0.15 $n_\mathrm{BG}$, where $n_\mathrm{BG}$ = 1.34.
        The simulated (a) illumination and (b) collection pupil distributions are used to calculate (c) the OCT PSFs.
        (d) The computational refocusing filter [Eq.~(\ref{eq:38})] is applied to the simulated OCT signals.
        (e) The PSFs after the ISAM for PSFD-OCT (Supplement 1, Section~\ref{S-sec:ISAM_PSFD}) applied.
    }\label{fig:PSFD}
\end{figure}

\begin{figure}
    \centering
    \includegraphics[width=13cm]{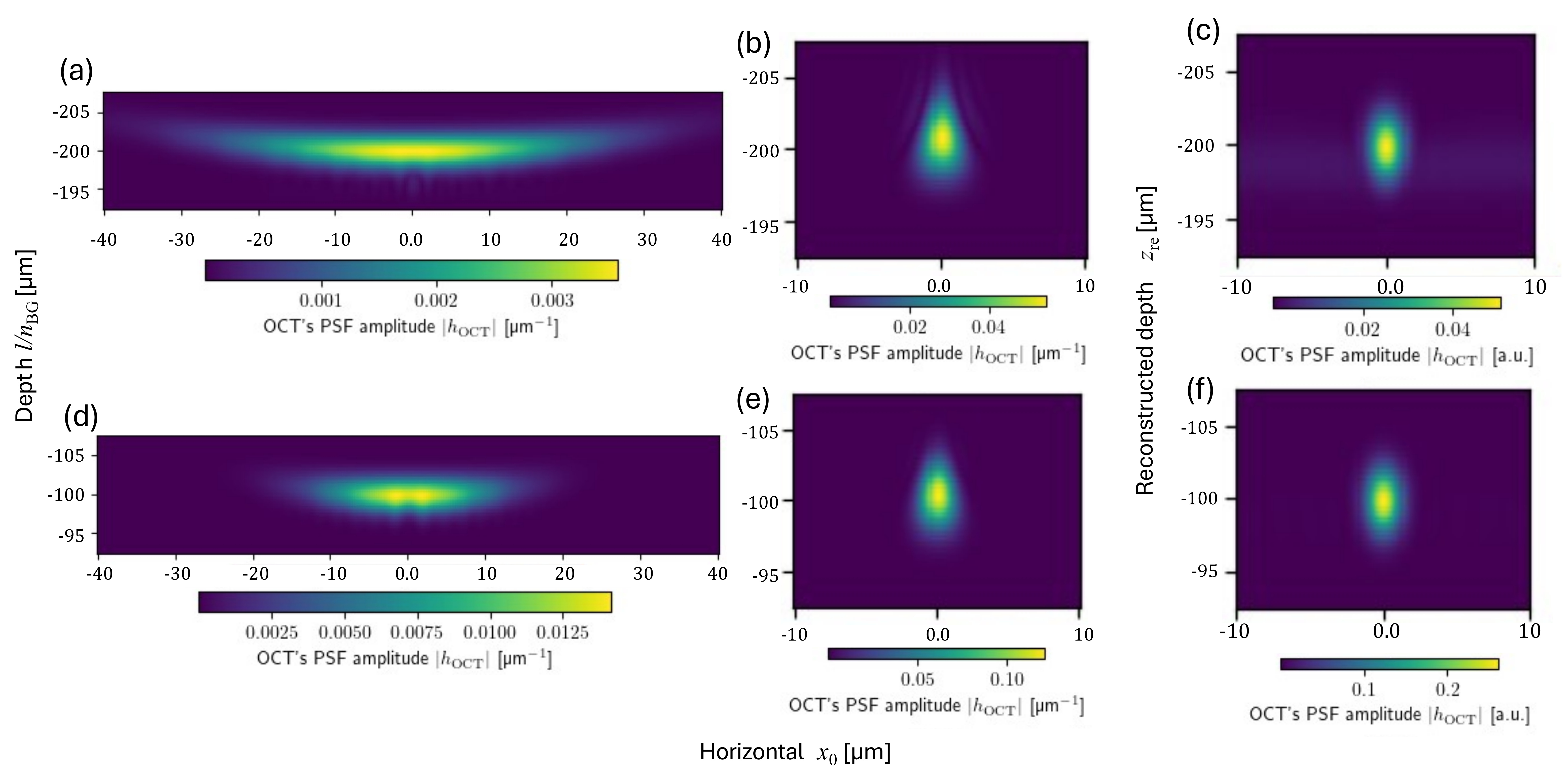}
    \caption{
        Numerically simulated PSFD-OCT PSFs at (a-c) -200 \um defocus and (d-f) -100 \um defocus.
        (b,e) The computational refocusing filter [Eq.~(\ref{eq:38})] is applied.
        (c,f) The PSFs after the ISAM for PSFD-OCT (Supplement 1, Section~\ref{S-sec:ISAM_PSFD}) applied.
    }\label{fig:PSFD_each}
\end{figure}

The simulated pupil amplitudes $P_\mathrm{ill}$ and $P_\mathrm{col}$ at the central wavelength of 1050 nm and PSFs with several defocus values are shown in \figurename~\ref{fig:PSFD}.
$P_\mathrm{ill}$ and $P_\mathrm{col}$ are identical Gaussian distributions with a cut-off NA of 0.25 $n_\mathrm{BG}$ and an effective ($\mathrm{e}^{-1}$ amplitude) NA of 0.15 $n_\mathrm{BG}$.
No aberration where $\mathbf{w} = \mathbf{0}$ is considered in this simulation.
The CR filter [from Eq.~(\ref{eq:38})] is applied to the simulated OCT signals.
In addition, ISAM\cite{ralston_inverse_2006} is also applied to the simulated OCT signals for comparison.
A simple ISAM implementation (Supplement 1, Section~\ref{S-sec:ISAM_PSFD}) is used in this case.
Note that the amplitudes of the PSFs are normalized to enable the various defocused PSFs to be displayed on the same scale.
The enlarged PSFs with the defocus values of 100 and 200 \um are shown in \figurename~\ref{fig:PSFD_each}.

With these NA and defocus conditions, the CR filter and ISAM can both correct the defocus with a similar performance in the lateral direction.
In the axial direction, however, the CR filter cannot correct the axial elongation of the PSF because it is based on a depth-by-depth correction procedure.
Therefore, the slight axial elongation caused by bending of the PSFs with defocus remains.
In contrast, ISAM can correct the axial elongation of the PSF because it is based on resampling of the data within the three-dimensional frequency domain $(\bm{\upnu}_\parallel, k)$.
Fortunately, the axial elongation is not prominent because of the low amplitudes of the PSFs in the peripheral regions.
In the case of a high NA, however, the axial elongation of the PSF becomes more prominent [see \figs~\ref{S-fig:PSFD-HighNA} and \ref{S-fig:PSFD_each-HighNA}].

This numerical simulation confirms that both the CR filter and ISAM can correct the defocus in PSFD-OCT with moderately high NAs well in the case where there are no HOAs.

\subsection{Numerical comparison of computational aberration corrections}\label{sec:CAC_simulation}

\begin{figure}
    \centering
    \includegraphics[width=13cm]{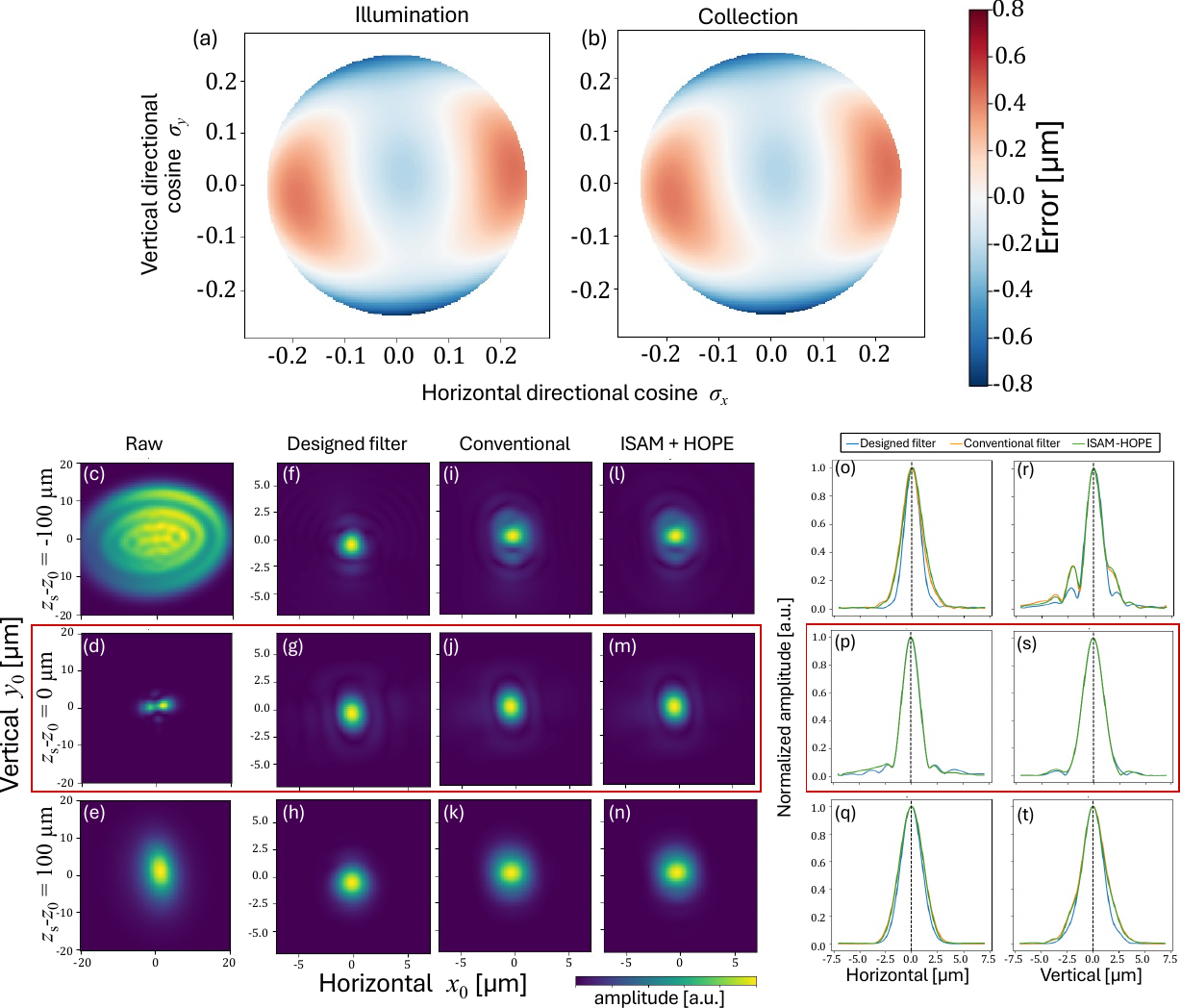}
    \caption{Numerical simulation of aberrated PSFs and corrections.
        (a, b) Numerically simulated wavefront errors on the pupil for PSFD-OCT\@.
        The aberrations for both pupils are assumed to be identical.
        (c,d,e) PSFs with aberrations, (f,g,h) corrected using the designed filter, (i,j,k) corrected using a conventional filter, and (l,m,n) corrected via the ISAM + high-order phase error (HOPE) correction, where its parameters are the estimated ones with the conventional filter (i,j,k) are shown.
        The aberration parameters of each method were estimated at the in-focus signal; hence, the PSFs are well corrected for all collection methods (red box).
        However, the PSFs with correction using the conventional filter exhibit blurring at defocused signals (i,l and k,n).
        (o,p,q) Horizontal and (r,s,t) vertical profiles of corrected PSFs (f--n).
        Profiles are centered at the maximum peak locations and normalized by their maximum values.
        PSF shapes are almost the same at in-focus depth because the correction parameters were optimized with the in-focus signal (red box, p, s).
        However, the PSFs corrected by the conventional filter exhibit increased side lobes and broadening with defocus (o, q, r, t).
    }\label{fig:pupil-CAC-PSFD}
\end{figure}

Numerical simulations of aberrated PSFD-OCT's PSFs and of PSFs corrected using both the designed filter (Section~\ref{sec:CAC}) and the conventional filter were performed.
The pupil size is the same as that in Section~\ref{sec:CR_simulation}.
The HOA coefficients were set as follows: $w_3$: -0.05, $w_5$: 0.2, $w_7$: -0.032, $w_8$: 0.04, and $w_{12}$: -0.1 \um.
The simulated wavefront errors are shown in \figs~\ref{fig:pupil-CAC-PSFD}(a) and \ref{fig:pupil-CAC-PSFD}(b).
The RMS wavefront error is 0.235 \um.
The 12 coefficients (2nd to 4th radial orders) of both conventional [Eq.~(\ref{eq:47})] and designed [Eq.~(\ref{eq:42})] filters were optimized with the in-focus signal ($z_\mathrm{s} - z_0$ = 0 \um).
The correction for other depths was done by putting the according optical pathlength $l-l_0$ into Eqs.~\ref{eq:46} and \ref{eq:48}.
For comparison, a combination of ISAM with a high-order phase error (HOPE) correction [Eq.~(\ref{eq:47})] with the same coefficients except $a_4 = 0$ was also applied.
This process corresponds to the method given in Ref.~\cite{adie_computational_2012}.

\begin{table}
    \centering
    \caption{Strehl ratios of the aberration corrected PSFs [\figurename~\ref{fig:pupil-CAC-PSFD}] with different correction methods and defocus amounts.}\label{tab:strehl-ratio}
    \begin{tabular}{r|ccc}
        Defocus & New CAC & Conventional CAC & ISAM + HOPE correction\\
        \midrule
        - 100 \um & 0.923 & 0.769 & 0.761\\
            0 \um & 0.990 & 0.981 & 0.976\\
          100 \um & 0.981 & 0.845 & 0.838\\
    \end{tabular}
\end{table}

The \enface PSFs of the aberrated PSFD-OCT and the PSFs that were corrected with the designed filter and the conventional filter are shown in \figs~\ref{fig:pupil-CAC-PSFD}(c--n), where the defocus amounts ($z_\mathrm{s} - z_0$) are set to -100, 0, and 100 \um.
Their horizontal and vertical profiles across each maximum peak of the PSFs are shown in \figs~\ref{fig:pupil-CAC-PSFD}(o)-\ref{fig:pupil-CAC-PSFD}(t).
Because the PSFs exhibit different shifts and amplitudes due to different corrections, they are centered at the maximum peak locations and normalized by their maximum values for the comparison.
All filters can correct the defocus and HOAs to some extent.
However, the designed filter shows lower side lobes [\figurename~\ref{fig:pupil-CAC-PSFD}(f)] when compared with both the conventional filter [\figurename~\ref{fig:pupil-CAC-PSFD}(i)] and the ISAM + HOPE correction approach [\figurename~\ref{fig:pupil-CAC-PSFD}(l)].
The difference is pronounced in the profile comparison [\figs~\ref{fig:pupil-CAC-PSFD}(o, r)].

The peak value of each PSF was compared to those of no high-order phase error cases to calculate the Strehl ratio.
The phase errors in \enface spatial frequency components of the simulated OCT signals were removed to obtain reference PSFs.
This metric indicates the degree of success of \enface phase error correction.
The results are shown in \tablename~\ref{tab:strehl-ratio}.
The performance of the designed filter is consistently better than that of the conventional filter across all defocus amounts.
Especially, the Strehl ratios of conventional methods at a defocus of -100 \um are less than 0.8.
According to the definition of the Strehl ratio, the correction with the conventional method at a 100 \um depth discrepancy is not guaranteed to achieve diffraction-limited resolution.

This comparison and other analysis (Supplement 1, Section~\ref{S-sec:other_example}) show that the correction performance of the designed filter is better than that of the conventional filter when the correction is applied at a depth different from that used for coefficient estimation.

It should be noted that the width of the PSF at a defocus of 100 \um is broader than that at 0 \um although the Strehl ratio is high.
This would be because of a narrowing of the signal's spatial frequency bandwidth [\figurename~\ref{S-fig:SystemFunction_Amp}], not a phase error.
The convolution process of aberrated pupils may decrease some spatial components\cite{liu_closed-loop_2021}.

\section{Estimation procedure for the aberration coefficients}
\label{sec:estimation-procedure}

\begin{figure}
    \centering
    \includegraphics[width=10cm]{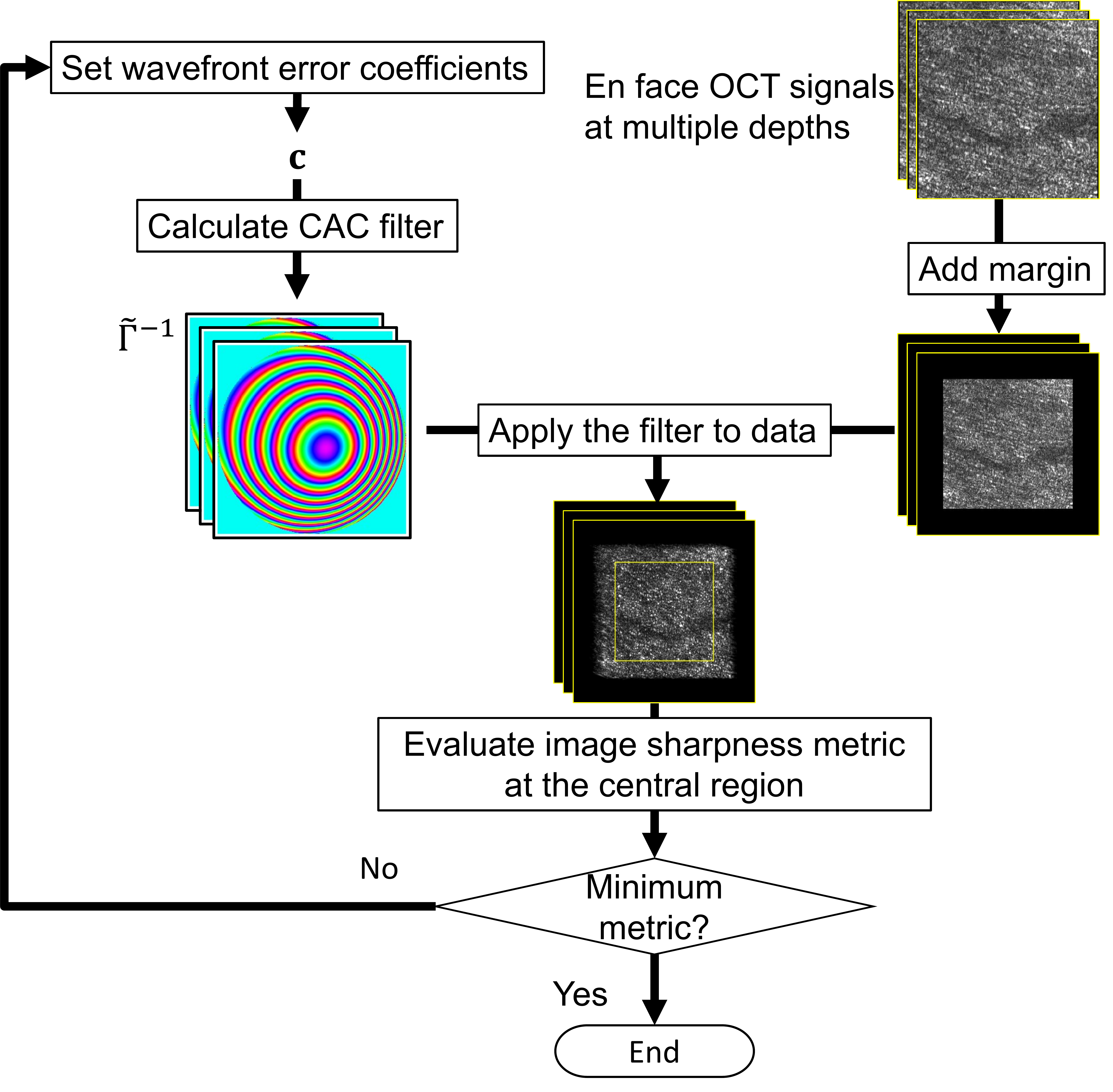}
    \caption{Schematic diagram of the procedure used to estimate the wavefront error coefficients.}\label{fig:CAC-chart}
\end{figure}

Before the CR and CAC filters are applied to estimate or correct the OCT signals, the bulk phase shift correction process for the 2D \enface plane phase error\cite{oikawa_bulk-phase-error_2020} is applied.

First, the CR filter [Eq.~(\ref{eq:38})] is applied to provide a rough estimate of $l_0$.
Then, the CAC filter [Eq.~(\ref{eq:42})] is used to estimate the coefficients of the HOAs, including $l_0$.
The slope of $c_4$ along the depth direction is predefined using the parameters $\lambda_0(\omega_\mathrm{c})$, $f_\mathrm{co}$, and $1/(n_\mathrm{BG} \times n_\mathrm{g,BG}) = 0.55$, which assumes that the surrounding medium is water.

A schematic diagram of the procedure used to estimate the wavefront error coefficients is shown in \figurename~\ref{fig:CAC-chart}.
\Enface OCT signals are extracted from the volumetric data at several depths.
The CAC filter is obtained using the initial coefficients of the HOAs and $l_0$.
The filter is then applied to the selected \enface signals and the image sharpness is assessed.
This procedure is iterated to realize the minimum metric value.
The Nelder-Mead simplex (NMS) algorithm is used to perform the optimization.
This is a robust algorithm and it does not require derivatives\cite{nelder_simplex_1965}.
However, the optimization process is not stable in the case of high-dimensional optimization.
For optimizations including HOAs, the adaptive NMS method is used to improve the multi-dimensional optimization\cite{gao_implementing_2012}.
The parameters used for the NMS algorithm are adapted for the dimensions of the problems.

The image sharpness is evaluated via maximum-intensity projection of the OCT \enface intensity images while using a few slices to suppress speckle patterns \cite{jian_wavefront_2014,cua_coherence-gated_2016,ju_visible_2018}.
The sharpness is assessed using the entropy-like metric\cite{flores_robust_1992}.
This metric is assessed for each selected depth.
Before a single value for the cost can be obtained, the metric values should be standardized at each depth.
The optimization cost can then be calculated as follows:
\[
 m^{'} = \frac{1}{N} \sum_{i}^{N}\frac{m(l_i) - m_0(l_i)}{|m_0(l_i)|},
\]
where $l_i$ is the OPL at the $i$-th extracted axial position, $m$ is the image metric value, and $m_0$ is the image metric value without the correction.

To ensure a stable image sharpness assessment, edge margins are added to the \enface signals and parts of the central regions are then used to calculate the sharpness metric.
Because CR and CAC are signal convolution operations involving use of Fourier transforms, boundary artifacts will occur in both cases.
Zero-padding is applied to all edges of \enface images\cite{bertero_simple_2005,zhu_computational_2022}.
In addition, the performance near the edge region is low because there is no signal outside the boundary that could contain part of the signal; such a signal would be required to recover the signal near the edge.
Therefore, the region of interest (ROI) for the image sharpness metric calculation was reduced to exclude the regions near the edges of image boundaries.
These processes are essential to prevent the effects of the boundary from affecting the image sharpness assessment.

For all optimization of simulated signals, phantom and retinal images, the optimization process has been iterated to reach the condition with a tolerance of $10^{-4}$ for the metric and coefficients.

\section{Computational aberration correction of PSFD-OCT images}
\label{sec:PSFD-CAC}

\subsection{Phantom imaging}

To confirm the aberration correction performance, a phantom experiment was conducted.
The sample used was a scattering phantom, on which polystyrene microspheres were fixed using agar (1 \textmu{}l volume of 1-\um diameter polystyrene microspheres with 6 ml of agar) and sandwiched between two glass plates at the bottom and top.
The phantom imaging should be suffering from the slight systematic aberration caused by the refraction of focused light at the air-glass boundary.
We used a 1.3 \um SS-OCT system\cite{zhu_computational_2022,li_three-dimensional_2017} to perform the phantom imaging process.
In brief, we used a Jones-matrix (SS-OCT) setup operating at a central wavelength of 1.3-\um with a wavelength-swept light source using a 50-kHz sweeping rate.
The objective was replaced with another objective with a short focal length (EFL=18 mm, LSM02, Thorlabs).
The effective NA was approximately 0.1.
The diffraction-limited lateral resolution (1/e\textsuperscript{2} diameter) at the focus was approximately 8.6 \um (in air).
A raster scan with 300 \texttimes\xspace 300 A-lines was applied.
The scanning area was approximately 643 \texttimes\xspace 579 \um which was estimated using a grid pattern sample.
Thus, the sampling density was 2.1 \texttimes\xspace 1.93 \um.

\begin{figure}
    \centering
    \includegraphics[width=10cm]{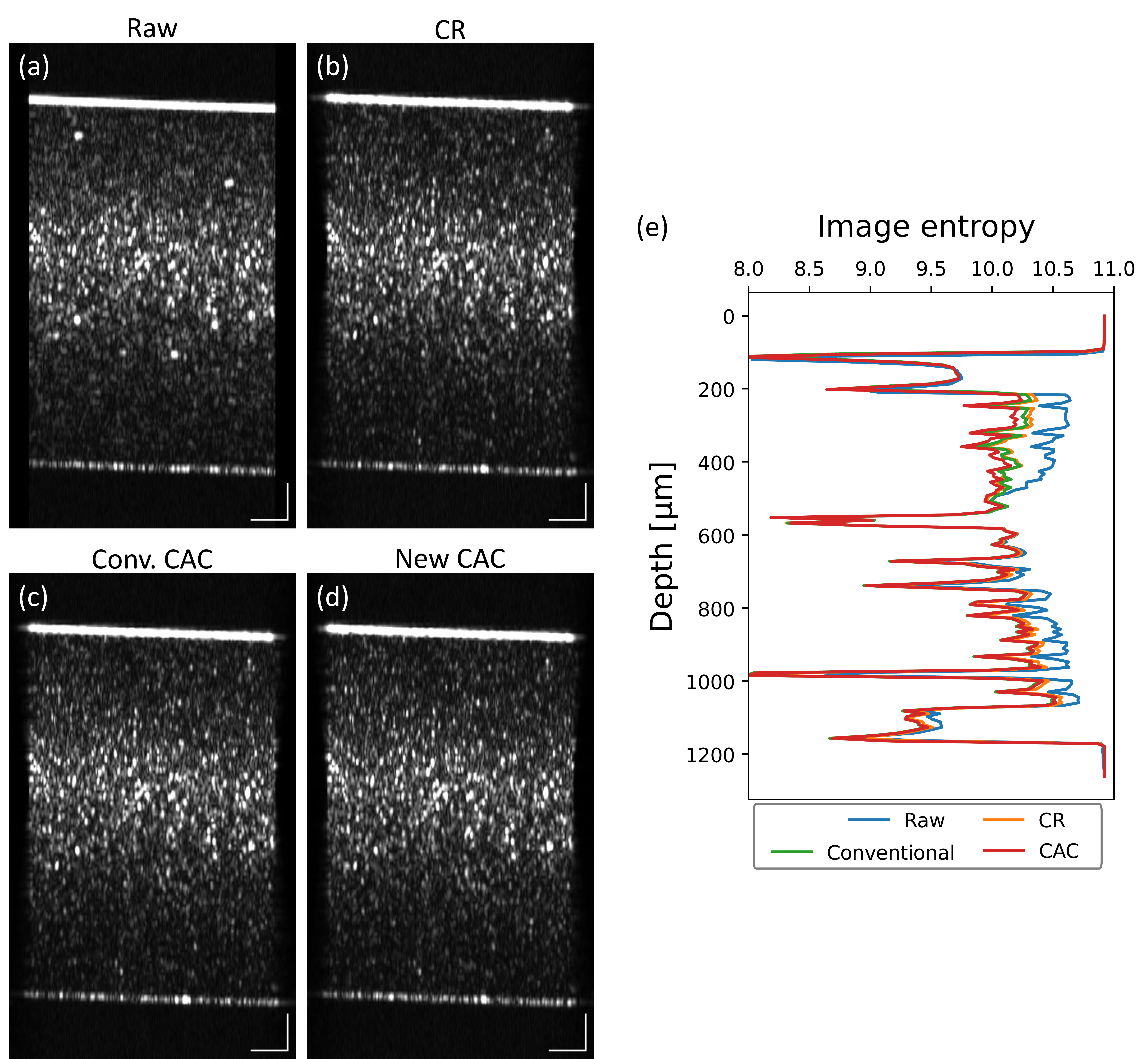}
    \caption{
        Cross-sectional images of microparticle phantom images with computational correction methods: (a) no computational compensation, (b) computational refocusing (CR), (c) conventional computational aberration correction (CAC), (d) new CAC.
        (e) The profiles of the sharpness metric along the depth.}\label{fig:phantom}
\end{figure}

\begin{figure}
    \centering
    \includegraphics[width=13cm]{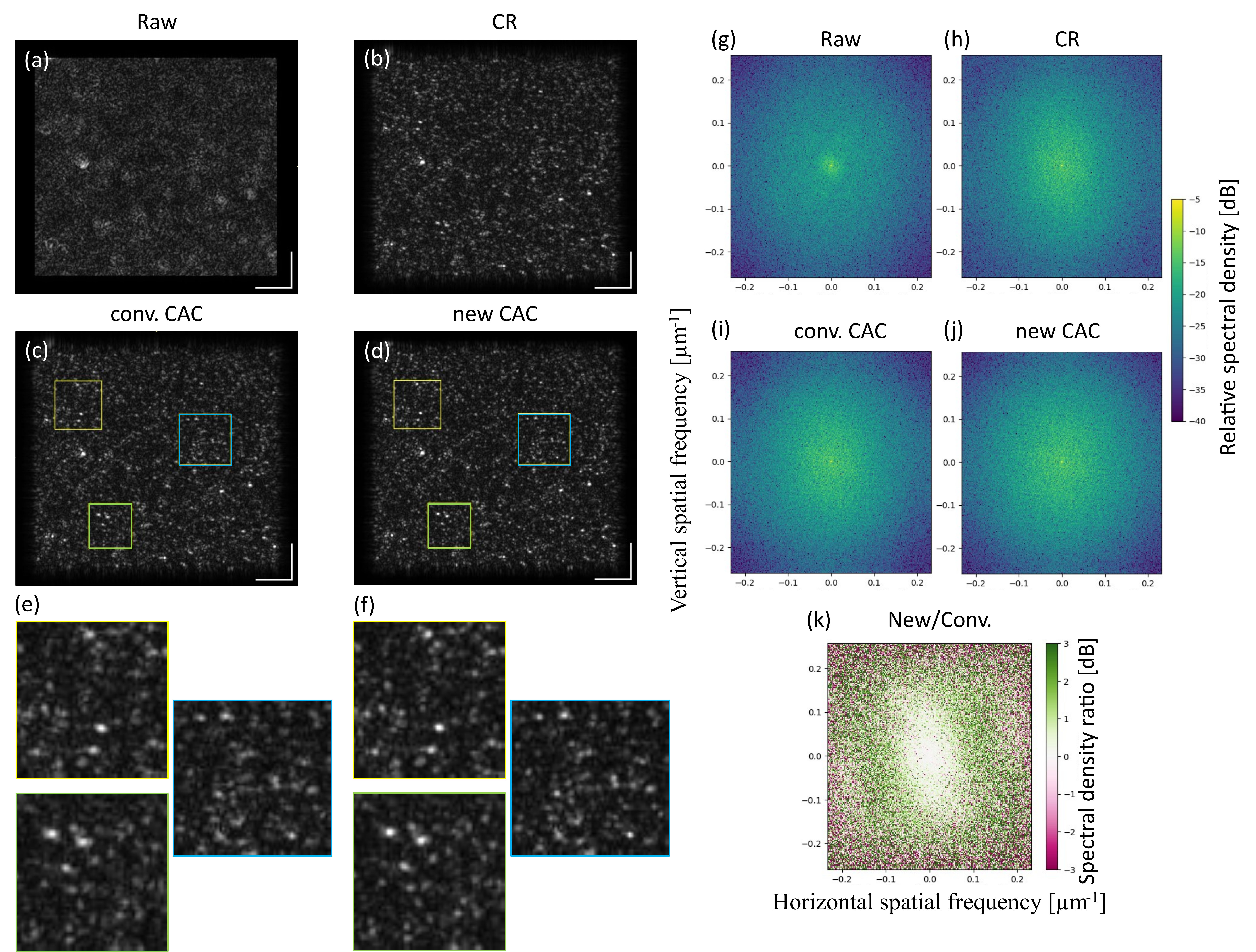}
    \caption{\Enface microparticle phantom images and spatial frequency spectra with computational aberration correction.
    (a,g) no computational compensation (Raw), (b,h) computational refocusing (CR), (c,i) conventional computational aberration correction (CAC), (d,j) new CAC\@.
    (e,f) Enlarged images of some locations and (k) spectral density ratio emphasize the differences between conventional and new CACs.}\label{fig:phantom-enface}
\end{figure}

\begin{figure}
    \centering
    \includegraphics[width=13cm]{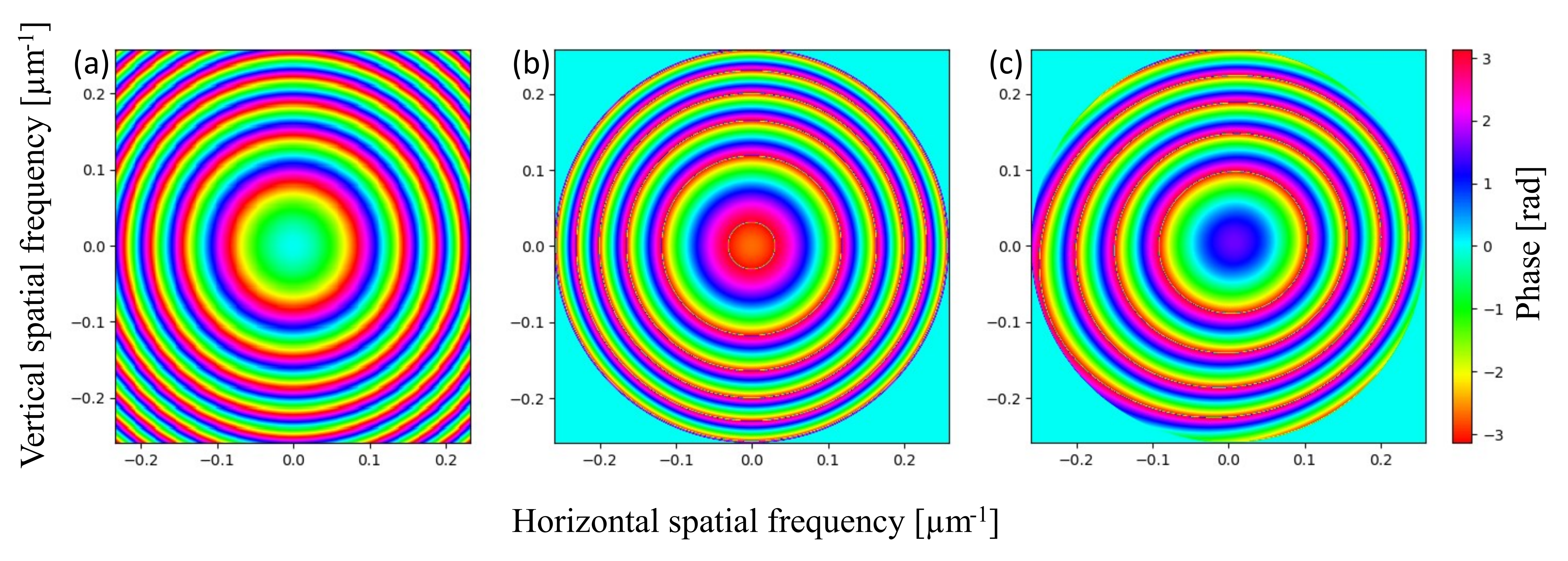}
    \caption{Correction filters used to generate the \enface corrected microparticle phantom images (\figurename~\ref{fig:phantom-enface}) at the depth of 253 \um for (a) computational refocusing, (b) conventional CAC, and (c) the new CAC.}\label{fig:phantom-enface-filt}
\end{figure}

The volumetric correction was performed with both the new and the conventional CAC filters.
The new CAC filter and the conventional CAC filter used the Zernike radial degree from 2 to 4, where $\mathbf{c} = \left\{c_3, c_4(l), c_5, c_7, c_8, c_{12}\right\}$ and $\mathbf{a} = \left\{a_3, a_4(l), a_5, a_7, a_8, a_{12}\right\}$.
Five \enface images were used to estimate the coefficients.
There are six (only offset for $c_4(l)$ and $a_4(l)$) coefficients in total.

The cross-sectional OCT intensity images of the microparticle phantom are shown in \figurename~\ref{fig:phantom}.
The CR-only image with the first estimation step [\figurename~\ref{fig:phantom}(b)] is also shown.
The image sharpness metric at each depth is plotted in \figurename~\ref{fig:phantom}(e).
Among the methods used, the new CAC filter demonstrates the best performance.
In particular, the performance of the new filter is better at shallow depths.
The conventional CAC filter shows slightly lower metric values when compared with the CR filter.

The \enface images of the microparticle phantom near the surface (depth: 253 \um) and their spatial frequency spectra (intensity images rather than complex signals) are shown in \figurename~\ref{fig:phantom-enface}.
The \enface images acquired with the new filter (\figurename~\ref{fig:phantom}d) show slightly sharper particles when compared with those obtained using the conventional filter (\figurename~\ref{fig:phantom}c).
Enlarged panes from the \enface images clearly show this sharpness improvement [see \figs~\ref{fig:phantom-enface}(e) and \ref{fig:phantom-enface}(f)].
The ratio of the spatial frequencies of the images [\figurename~\ref{fig:phantom-enface}(g)] shows that the new filter has higher spatial frequency components when compared with the conventional filter at peripheral frequencies.
The filters used for correction at this depth (253 \um) are shown in \figurename~\ref{fig:phantom-enface-filt}.
The conventional CAC filter almost seems to have corrected the defocus alone
($l_0$ = 579 \um,
 $a_3$ = -4.53 \texttimes\xspace 10$^{-5}$ rad,
 $a_5$ = 1.95 \texttimes\xspace 10$^{-4}$ rad,
 $a_7$ = -8.33 \texttimes\xspace 10$^{-6}$ rad,
 $a_8$ = 1.38 \texttimes\xspace 10$^{-4}$ rad,
 $a_{12}$ = -1.25 \texttimes\xspace 10$^{-4}$ rad).
In contrast, the new CAC filter has substantial HOA coefficients
($l_0$ = 546 \um,
 $c_3$ = -0.319 rad,
 $c_5$ = 0.347 rad,
 $c_7$ = -0.187 rad,
 $c_8$ = -0.114 rad,
 $c_{12}$ = -0.228 rad).
The almost zero high-order correction coefficients with the conventional filter may be due to the fact that the coefficients were estimated with signals from multiple depths simultaneously.
The theory and simulation suggest that the optimum correction coefficients for each depth are not the same.
Hence, the global solution of aberration estimation using multi-depth signals with the conventional filter does not exist or is very difficult to reach.
These results suggest that the conventional filter requires depth-by-depth estimation of the coefficients to provide the best CAC performance.
However, the new filter provides a good aberration correction performance for volumetric data based on a single estimation.
This is advantageous in terms of the time required for estimation of coefficients with volumetric data.

\subsection{\Invivo retinal imaging}

In the case of the retinal imaging, eye optics affect the imaging quality.
We attempted to correct aberrations in the ocular optical system.
A human retina has been imaged  \invivo using a 1-\um SS-OCT system.
Two configurations were used in this work.
The setup and its configurations used for the \invivo retinal imaging are described in detail in the supplementary material (Supplement 1, Section~\ref{S-sec:Optical_setup}).
The study was approved by the Institutional Review Boards of the University of Tsukuba and adhered to the tenets of the Declaration of Helsinki.
The nature of the present study and the implications of participating in this research project were explained to all study participants, and written informed consent was obtained from each participant before any study procedures or examinations were performed.

In the case of the \invivo retinal images, the process of retinal layer segmentation\cite{li_optimal_2006,garvin_automated_2009,abramoff_retinal_2010,antony_automated_2011} and sub-pixel registration among the B-scans were applied to flatten the volume at the retinal pigment epithelium line before the application of the correction procedure described in Section~\ref{sec:estimation-procedure}.

\begin{figure}
    \centering
    \includegraphics[width=10cm]{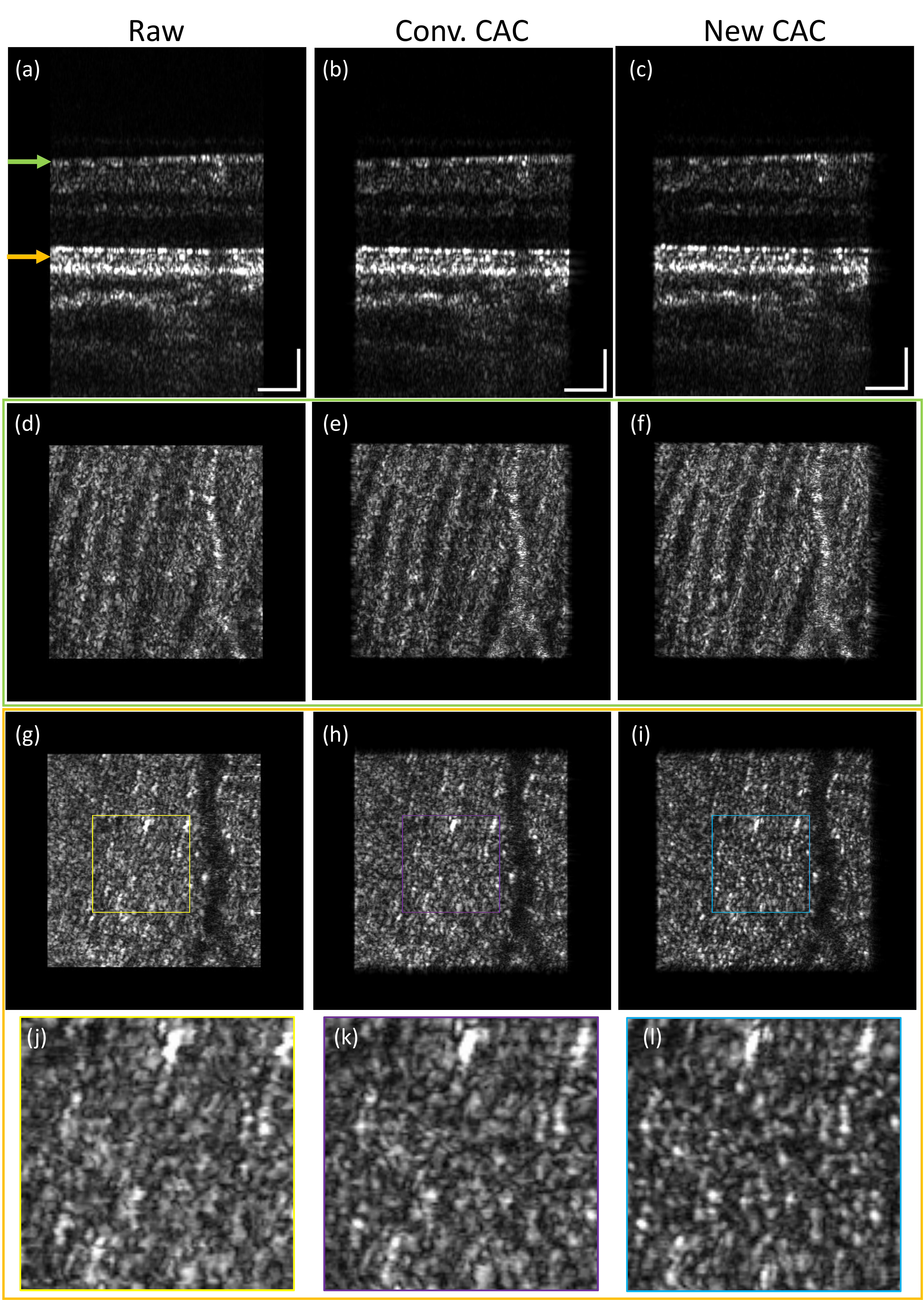}
    \caption{
        \Invivo retinal OCT imaging results obtained by 100 kHz configuration with computational aberration correction.
        (a-c) Cross-sectional and (d-i) \enface slab MIP images at (d-f) the retinal nerve fiber layer and (g-i) the cone outer segment tips (COST) are shown.
        The signal amplitude of OCT are displayed.
        Images are processed with (a,d,g) no computational method (Raw), (b,e,h) conventional CAC, and (c,f,i) new CAC\@.
        (j-l) Enlarged images of the COST region.
        The scale bars are 100 \um.
    }\label{fig:retina}
\end{figure}

\begin{figure}
    \centering
    \includegraphics[width=10cm]{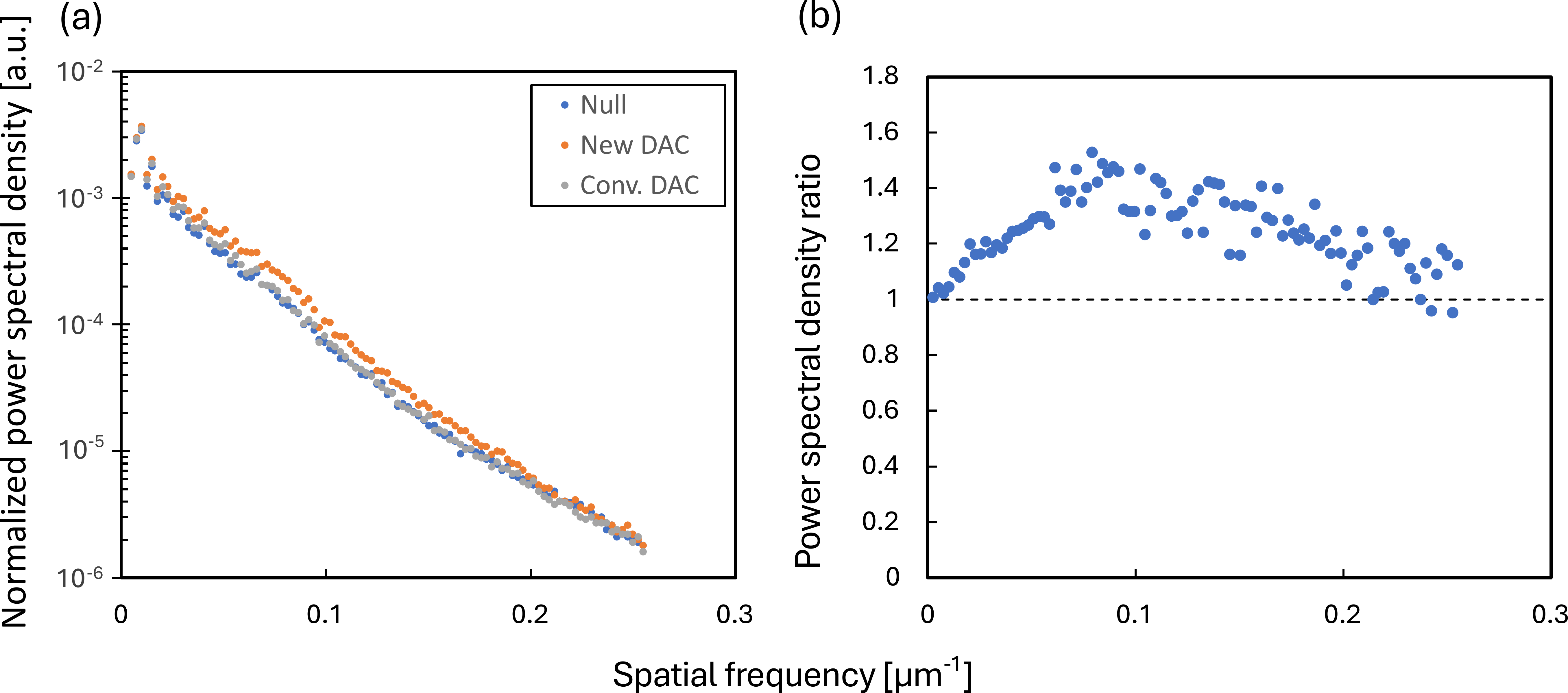}
    \caption{
        Spatial frequency analysis of \invivo retinal OCT imaging obtained with the 100 kHz configuration.
        (a) Spatial frequency spectra of \enface images at the cone outer segment tips (COST) with/without computational aberration correction methods.
        (b) Ratio of the spatial frequency spectra of images obtained by new CAC to that obtained by conventional CAC\@.
    }\label{fig:retina-sp}
\end{figure}

The CAC results for the retinal data obtained by configuration-A are shown in \figurename~\ref{fig:retina}.
The raster scan was performed with 256 $\times$ 256 A-lines over a 0.5 $\times$ 0.5 mm field of view.
The right eye of a healthy volunteer (43 years old) was scanned at approximately 4° temporal from the macula.
The image acquisition speed was 100 kHz and the acquisition time was approximately 0.82 s.
The beam diameter on the cornea was approximately 3.4 mm (1/e\textsuperscript{2}), which corresponds to an effective NA in air of approximately 0.1.
The illumination power on the cornea was approximately 3.3 mW, which is under the limit of the ANSI standard (Z80.36-2016).
The HOAs from the 2nd to 4th radial orders (12 coefficients) were taken into account.
The maximum intensity projection (MIP) of 13-\um slabs at the retinal nerve fiber layer (RNFL) and the cone outer segment tips (COST) are shown in \figurename~\ref{fig:retina}(d-i).
The boundaries of retinal nerve fiber bundles are sharpened up with both the conventional CAC and the new CAC\@.
On the other hand, the COST region shows that the new CAC provides a sharper image when compared with the conventional CAC\@.
This is more apparent in the enlarged images of the COST region shown in \figurename~\ref{fig:retina}(j-l).

The spatial frequency characteristics of the images were also analyzed.
The \enface OCT intensity images were 2D Fourier transformed and 2D power spectra were thus obtained.
These spectra were then normalized by dividing each spectrum by the spectral density at the zero frequency.
Then, the normalized spectra obtained were averaged along the tangential direction to produce the radial spatial frequency profiles.
The radial spatial frequency profiles of the photoreceptor \enface images are shown in \figurename~\ref{fig:retina-sp}(a).
The ratio of the radial frequency spectra obtained for the new CAC to those from the conventional CAC was calculated [\figurename~\ref{fig:retina-sp}(b)].
The results obtained also show that the normalized spectral density at $\le$ 0.2 \um$^{-1}$ was higher for the new CAC when compared with that of the conventional CAC\@.

\begin{figure}
    \centering
    \includegraphics[width=12cm]{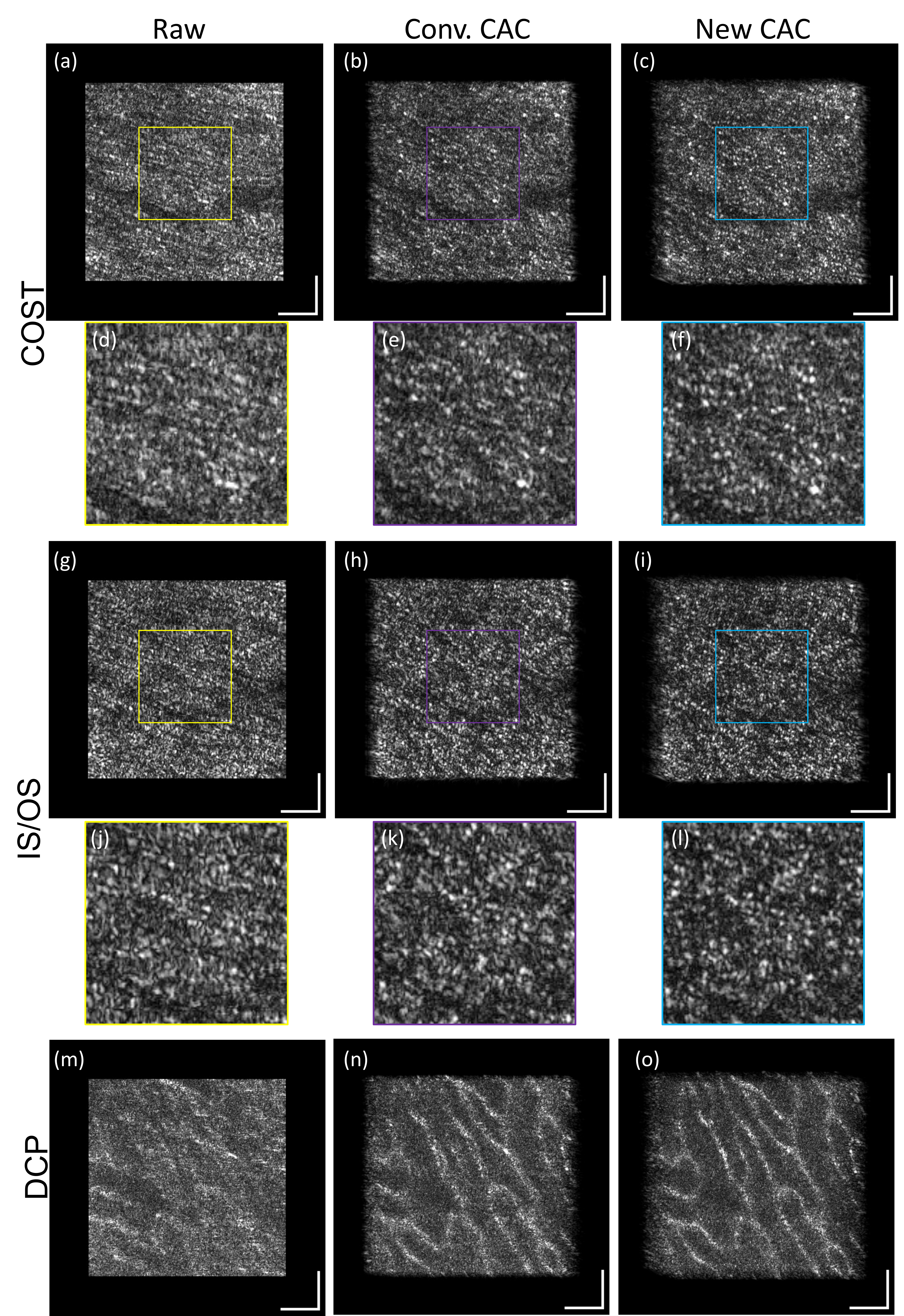}
    \caption{
        \Invivo retinal OCT imaging results obtained by 200 kHz configuration with computational aberration correction.
        \Enface slab MIP images at
        (a-c) the cone outer segment tips (COST),
        (g-i) the inner and outer segment junction (IS/OS), and 
        (m-o) the boundary between the inner nuclear layer and outer plexiform layer, i.e., the deep capillary plexus (DCP) are shown.
        The signal amplitude of OCT are displayed.
        Images are processed with (a,g,m) no computational method (Raw), (b,h,n) conventional CAC, and (c,i,o) new CAC\@.
        (d-f) and (j-l) are enlarged images of the COST and IS/OS regions, respectively.
        The scale bars are 100 \um.
    }\label{fig:retina-200k}
\end{figure}

\begin{figure}
    \centering
    \includegraphics[width=10cm]{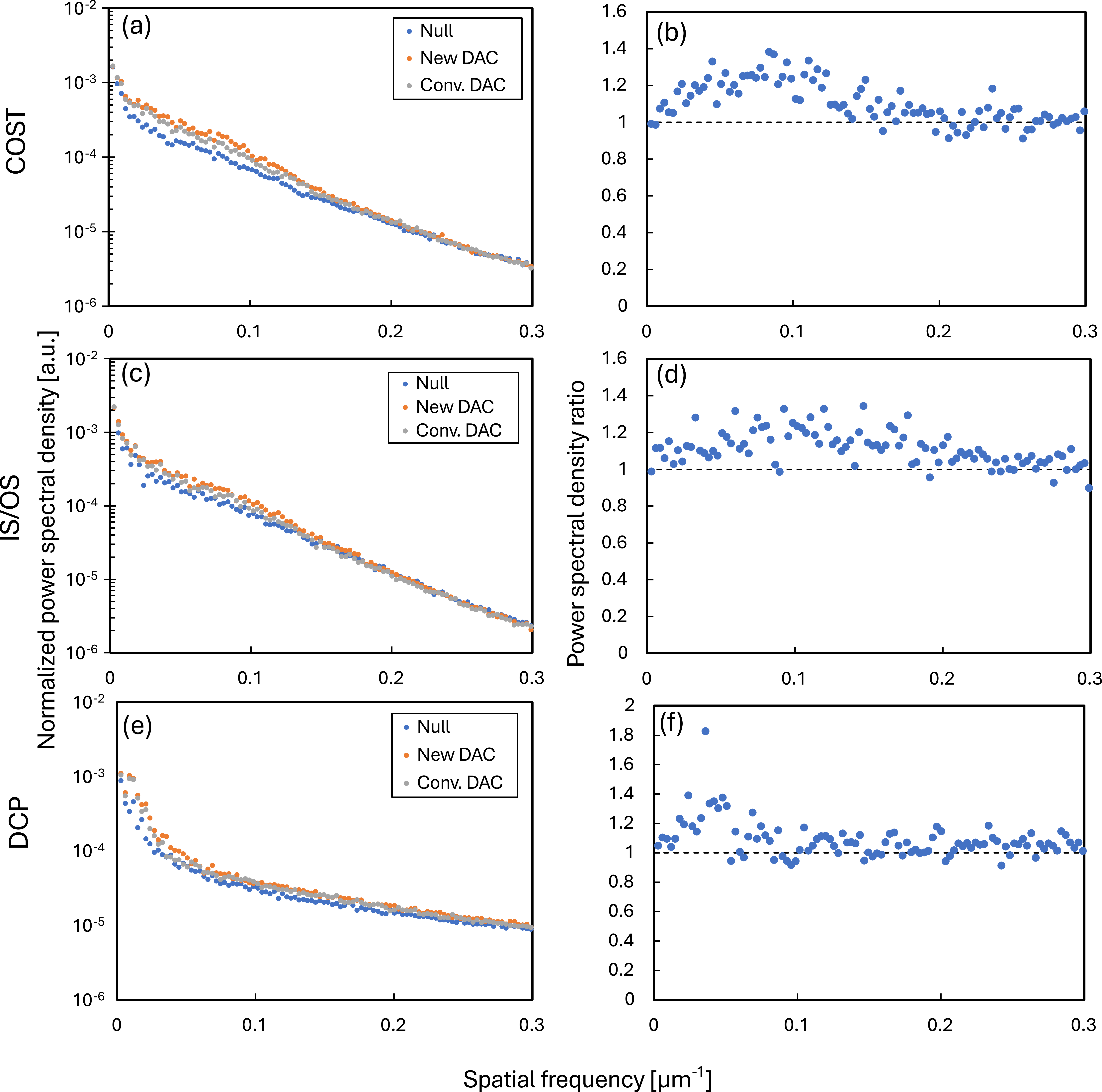}
    \caption{
        Spatial frequency analysis of \invivo retinal OCT imaging obtained with the 200 kHz configuration.
        (a,c,e) Spatial frequency spectra of \enface images with and without computational aberration correction methods are shown.
        (b,d,f) Ratio of the spatial frequency spectra of images obtained by new CAC to that obtained by conventional CAC\@.
        (a,b) the cone outer segment tips (COST),
        (c,d) the inner and outer segment junction (IS/OS), and 
        (e,f) the deep capillary plexus (DCP).
    }\label{fig:retina-sp-cost-200k}
\end{figure}

The other CAC results for the retinal data were obtained by configuration B.
A raster scan with 300 $\times$ 300 A-lines was applied over a 0.5 mm $\times$ 0.5 mm field of view.
The right eye of a healthy volunteer (43 years old) was scanned at approximately 3° nasal from the macula.
The image acquisition speed was 200 kHz and the acquisition time was approximately 0.56 s.
The beam diameter on the cornea was approximately 5 mm (1/e$^2$), which corresponds to an effective NA in air of approximately 0.15.
The illumination power on the cornea was approximately 2.2 mW, which is under the safety limit.
The HOAs from the 2nd to the 4th radial orders (12 coefficients) were taken into account.
The MIP of 8-\um slab at the cone outer segment tips (COST), the inner and outer segment junction (IS/OS), and the deep capillary plexus (DCP) are shown in \figurename~\ref{fig:retina-200k}.
In all projection image regions, the new CAC shows sharper structures when compared with the conventional CAC\@.
The enlarged images of the COST and IS/OS regions [\figs~\ref{fig:retina-200k}(d-f) and \ref{fig:retina-200k}(j-l)] show sharpness improvements more clearly.

The radial spatial frequency profiles of the \enface images and the ratios of the radial frequency spectra obtained when using the new CAC to that acquired using the conventional CAC are shown in \figurename~\ref{fig:retina-sp-cost-200k}.
The results also show that the normalized spectral density of COST and IS/OS images at $\le$ 0.2 \um$^{-1}$ was higher for the new CAC when compared with that of the conventional CAC\@.
The spatial spectrum of the DCP image also shows that the new CAC has higher spatial frequency components when compared with the conventional CAC at around 0.04 \um$^{-1}$.

At the eccentricities of the retinal images obtained, the cone photoreceptor spacing in the human eye may be approximately 8 \um\cite{curcio_human_1990}.
The increased spatial frequency components at $\le$ 0.2 \um$^{-1}$ and the anatomical knowledge of the human retina indicate that the cone photoreceptor structure has been restored well with the new CAC\@.
As shown in these results, the new CAC outperforms the conventional CAC in terms of the volumetric aberration correction performance.

\section{Discussion}

\subsection{Comparison with the previous study of phase-error crosstalk}
\label{sec:discuss-crosstalk}

\begin{figure}
    \centering
    \includegraphics[width=9cm]{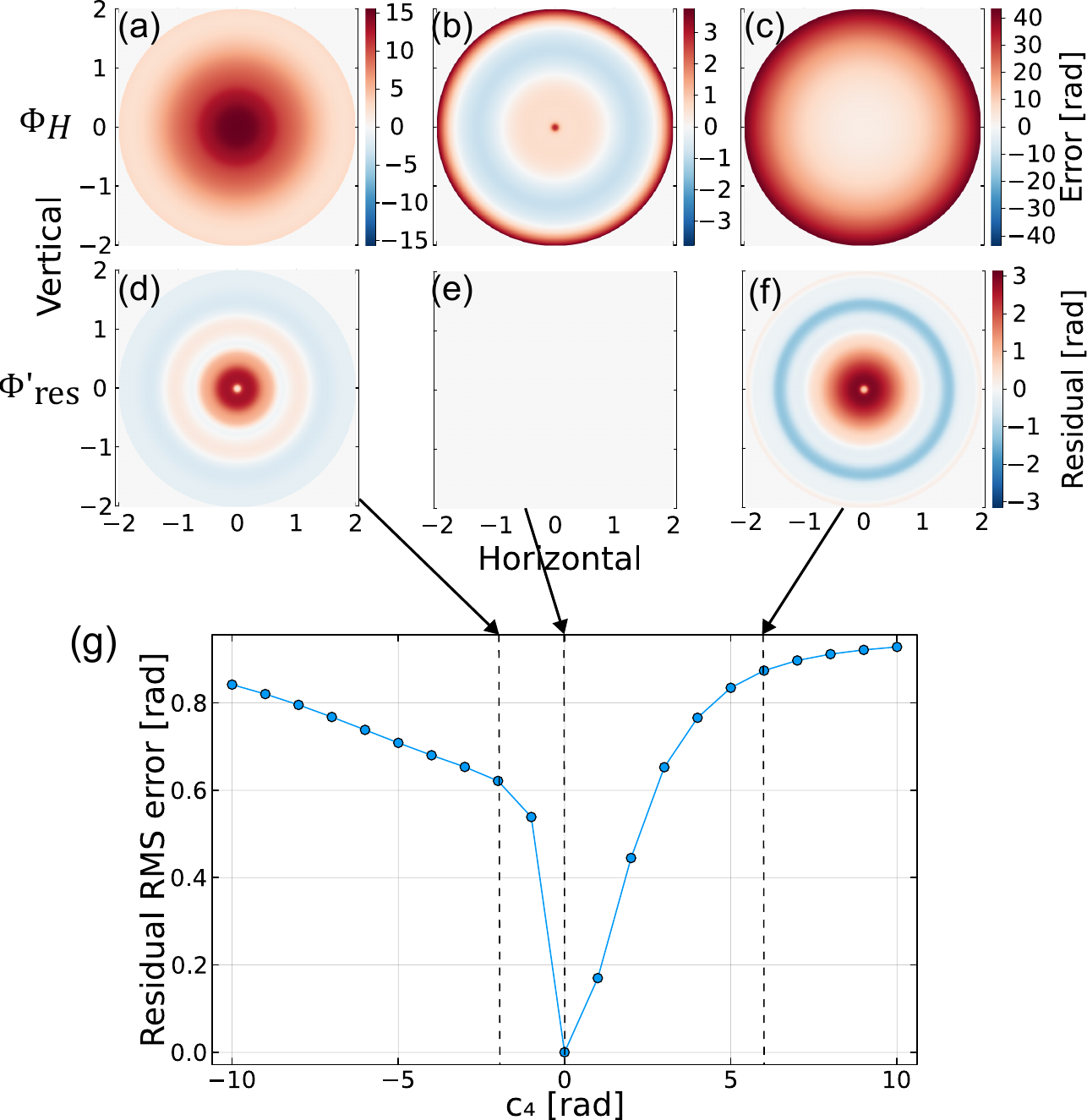}
    \caption{Simple 2D simulation of defocus-dependent multi-Zernike-mode crosstalk.
    The spherical aberration $c_{12}$ = 0.841 rad RMS and various defocus $c_4$ =[-10, 10] rad RMS are applied to the pupil.
    (a,b,c) Phase errors.
    (d,e,f) residual phase errors after subtracting the phase errors of single-Zernike-mode cases of $c_4$ and $c_{12}$.
    (a,d) at $c_4$ = 2 rad RMS, (b,e) at $c_4$ = 0 rad RMS, and (c,f) at $c_4$ = 6 rad RMS.
    (g) The residual RMS phase errors along the defocus.
}\label{fig:crosstalk-sim}
\end{figure}

A previous study by Liu et al.\cite{liu_closed-loop_2021} investigated the phase-error crosstalk in the PSFD-OCT system by numerical simulation and experimental investigation.
They concluded that the crosstalk is negligible in PSFD-OCT\@.
However, our numerical simulation (Section~\ref{sec:CAC_simulation} and Supplement 1 Section~\ref{S-sec:other_example}) suggests that the crosstalk may not be negligible when both defocus and HOAs are present.

The difference between our simulation and Liu et al.'s one lies in the use of 3D convolution of 3D pupils considering broadband wavelengths, as opposed to 2D convolution of 2D pupils using a single wavelength.
However, this point alone cannot be considered a significant reason for the discrepancy.
The significant difference is that Liu et al. investigated several cases of single-Zernike mode aberration.
They numerically simulated the convolution of the pupil with wavefront error expressed by a single Zernike mode and experimentally applied a wavefront error using a deformable mirror to shape the wavefront to match with a single Zernike mode.
Clearly, their investigation cannot evaluate crosstalk in the more general case where multiple Zernike modes are present.

In practice, significant crosstalk and depth (defocus) dependency can be observed using a simple 2D numerical simulation, similar to the approach taken by Liu et al., but involving multiple Zernike modes.
When we assume the defocus $Z_4$ and spherical aberration $Z_{12}$ exist, the self-convolution of the pupil is
\begin{equation}
    \left|H(\bm{\uprho}; c_4, c_{12})\right| e^{\mathrm{i} \Phi_H(\bm{\uprho}; c_4, c_{12})} = 
    \left[P(\bm{\uprho}) e^{\mathrm{i} \Phi_\mathrm{SP}(\bm{\uprho}; c_4, c_{12})}\right]
    \otimes_{\bm{\uprho}}^2
    \left[P(\bm{\uprho}) e^{\mathrm{i} \Phi_\mathrm{SP}(\bm{\uprho}; c_4, c_{12})}\right],
\end{equation}
where $\Phi_\mathrm{SP}(\bm{\uprho}; c_4, c_{12}) = c_4 Z_4(\bm{\uprho}) + c_{12} Z_{12}(\bm{\uprho})$ is the wavefront error in radian and $\bm{\uprho} = [\rho \cos\theta, \rho\sin\theta]$ is the coordinate on a unit disc.
Here, we set $P(\bm{\uprho}) = e^{-\frac{\ln{2}}{2}\frac{\rho^2}{0.4^2}}$ as a Gaussian beam profile with the fill rate of 0.4.
The difference in crosstalk between single-Zernike-mode and multi-Zernike-mode cases can be assessed by subtracting the double-pass phase errors of each single-Zernike-mode aberration case as follows:
\begin{equation}
    e^{\mathrm{i}\Phi_\mathrm{res}(\bm{\uprho}; c_4, c_{12})}
    =
    e^{\mathrm{i}\left[\Phi_H(\bm{\uprho}; c_4, c_{12}) - \Phi_H(\bm{\uprho}; 0, c_{12}) - \Phi_H(\bm{\uprho}; c_4, 0)\right]}
\end{equation}
and removing the piston $e^{\mathrm{i}\Phi'_\mathrm{res}(\bm{\uprho}; c_4, c_{12})} = e^{\mathrm{i}\left[\Phi_\mathrm{res} - \overline{\Phi_\mathrm{res}}\right]}$.
$\Phi'_\mathrm{res}$ implies the impact of considering multi-Zernike-mode aberration instead of multiple single-Zernike-mode aberrations.
The residual RMS phase error is calculated as $\sqrt{\frac{1}{\pi 2^2}\int_{0}^{2}\int_{0}^{2\pi}|\Phi'_\mathrm{res}(\bm{\uprho}; c_4, c_{12})|^2 \rho \mathrm{d}\rho \mathrm{d}\theta}$.
The results of the simulation with various $c_4$ values are shown in \figurename~\ref{fig:crosstalk-sim}.
The simulated $c_4$ range of [-10, 10] radians is approximately [-210, 210] \um defocus when the cut-off NA is 0.2$n_\mathrm{BG}$, the central wavelength is 1.05 \um, and the background refractive index is $n_\mathrm{BG} = 1.34$.
The double-pass phase error at in-focus $\Phi_H(\bm{\uprho}; 0, c_{12})$ [\figurename~\ref{fig:crosstalk-sim}(b)] is a single-Zernike mode ($Z_{12}$) case.
However, the distribution of the phase error apparently exhibits higher-order errors, not the same as Liu's simulation [Fig. 1(b), $Z_8$ in Liu et al.].
This is because the crosstalk also depends on the amount of aberrations.
The residual RMS phase error is increased as defocus increases [\figurename~\ref{fig:crosstalk-sim}(g)].
This simulation suggests that crosstalk may become significant in the out-of-focus state because it is amplified by the increased defocus error $c_4$, even though it is negligible in the in-focus state.
And that is why depth (defocus) dependent estimation and correction would be required for the conventional CAC\@.
Hence, it should be taken into account that the investigation with single-Zernike-mode cases, such as Liu et al. did, cannot be applied to general cases.
The assessment of crosstalk needs to be done under the same conditions as the real application.

\subsection{Amplitude distribution on the pupil}
\label{sec:discuss-apodization}

In this manuscript, we fix the amplitude distribution of both illumination and collection pupils to be the same Gaussian because optic-fiber-based PSFD-OCT is assumed.
The design of the conventional computational refocus [Eq.~(\ref{eq:38})] is based on this assumption.
Thus, if the amplitude distributions $P_\mathrm{ill}$ and $P_\mathrm{col}$ do not satisfy the assumption, these filters may not work well.

If the detection process is performed through a free-space and infinitesimally small pinhole, then the collection pupil may be approximated as unity, i.e., $A_\mathrm{col} = 1$.
Then, Eq.~(\ref{eq:24}) becomes
\begin{equation}\label{eq:40}
    \tilde{\Gamma}(\bm{\upnu}_\parallel, z_0 - z_\mathrm{s}; n_{\mathrm{BG}}k_\mathrm{c})
    =
    - \frac{4\uppi^{2}}{k_\mathrm{s}^{2}}
    \frac{\uppi \mathrm{e}^{-\frac{\Delta f_\mathrm{ill}^2 \phi_\mathrm{def}^2 (z_0 - z_\mathrm{s})^2}{1+4\Delta f_\mathrm{ill}^4 \phi_\mathrm{def}^2 (z_0 - z_\mathrm{s})^2} |\bm{\upnu}_\parallel|^2}}
    {1/(\Delta f_\mathrm{ill}^2 )-\mathrm{i}2\phi_\mathrm{def} (z_0 - z_\mathrm{s})}
    \mathrm{e}^{\mathrm{i}
        \frac{1 + 2\Delta f_\mathrm{ill}^4 \phi_\mathrm{def}^2 (z_0 - z_\mathrm{s})^2}
        {1 + 4\Delta f_\mathrm{ill}^4 \phi_\mathrm{def}^2 (z_0 - z_\mathrm{s})^2}
        \phi_\mathrm{def} (z_0 - z_\mathrm{s}) |\bm{\upnu}_\parallel|^2},
\end{equation}
where $A_\mathrm{ill} = \mathrm{e}^{-\frac{|\bm{\upnu}_\parallel |^2}{\Delta f_\mathrm{ill}^2}}$ and $\Delta f_\mathrm{ill} = \text{NA}_\mathrm{ill}^{(\mathrm{e}^{-1})} / \lambda_\mathrm{c}$.
The phase error due to the defocus no longer varies linearly with the amount of defocus.
When $\Delta f_\mathrm{ill}^4 \phi_\mathrm{def}^2 z^2 \ll 1$ $\left[|z| \ll \frac{1}{\Delta f_\mathrm{ill}^2 \phi_\mathrm{def}}\right]$, the phase error is then close to that in the plane wave illumination case [Eq.~(\ref{S-eq:S3})].
In the opposite case where $\Delta f_\mathrm{ill}^4 \phi_\mathrm{def}^2 z^2 \gg 1$ $\left[|z| \gg \frac{1}{\Delta f_\mathrm{ill}^2 \phi_\mathrm{def}}\right]$, the phase error approaches that of Eq.~(\ref{eq:36}).
It should be noted that the quadratic phase error does not vary linearly with the defocus distance $z$.

The phase-only refocus filter according to the above approximations is given by:
\begin{equation}\label{eq:41}
    \tilde{\Gamma}^{-1}_{\mathrm{CR}, \mathrm{NAp}} (\bm{\upnu}_\parallel, l_\mathrm{s}; n_{\mathrm{BG}}k_\mathrm{c})
    = \mathrm{e}^{
        \mathrm{i}
        \frac{1 + 2\Delta f_\mathrm{ill}^4 \phi_\mathrm{def}^2 \delta z_\mathrm{s}^2(l_\mathrm{s})}
        {1 + 4\Delta f_\mathrm{ill}^4 \phi_\mathrm{def}^2 \delta z_\mathrm{s}^2(l_\mathrm{s})}
        \phi_\mathrm{def} |\bm{\upnu}_\parallel|^2
        \delta z_\mathrm{s}(l_\mathrm{s})
    }.
\end{equation}

The issue of the distributions of $P_\mathrm{ill}$ and $P_\mathrm{col}$ is not only important for the phase-based method.
The shape of the cTF is dependent on these distributions, and thus the optimal resampling positions for ISAM will also vary when $P_\mathrm{ill}$ and $P_\mathrm{col}$ have different distributions\cite{coquoz_interferometric_2017}.
Further examples of CR filters for other illumination and collection configurations are presented in Supplement 1 (Section~\ref{S-sec:CR_other_OCT}).

The design of the conventional CAC [Eqs.~(\ref{eq:47})-(\ref{eq:48})] is also based on the assumption that both pupils have the same Gaussian distributions.
Thus, the performance of the conventional CAC should degrade if the actual amplitude distributions differ significantly from this assumption.
The crosstalk of the illumination and collection aberrations would be more complicated.
On the other hand, the proposed method obtained by simulating the both pupils can be adapted by modifying the amplitude distributions $A'$ in Eq.~(\ref{eq:47}) to match the actual distributions of each pupil.

\subsection{Selection of computational aberration correction methods}
\label{sec:discuss-select}

The numerical simulation results suggested brief guidelines for the selection processes of the CR and CAC methods.
Although ISAM for PSFD-OCTs does not perform a complete correction (Section~\ref{sec:OCTrefocusing}), ISAM will work quite well at a high effective NA of 0.536 in fiber-based PSFD-OCT if there is only defocus (i.e., no HOAs) to be addressed [\figs~\ref{S-fig:PSFD-HighNA} and \ref{S-fig:PSFD_each-HighNA}].
Because the Gaussian illumination and collection pupils do not contain a high fraction of high-frequency components, the limit of the approximation of ISAM for confocal OCT is perhaps mitigated.
The limitation of filtering refocusing methods will be restricted compared with ISAM because the ISAM relies on only paraxial approximation, while filtering methods rely on paraxial and narrow-band approximations (Section~\ref{sec:OCTrefocusing}).

The filtering method also restores the lateral resolution well in high NA cases; however, the axial resolution under out-of-focus conditions does not compare well to the ideal case.
This finding will be particularly important when judging the CR performance in high NA cases.
Almost all previous CR works were assessed based on the lateral resolution, and not on the axial resolution, to evaluate the CR performance.
When only the lateral resolution is assessed, then degradation of the axial resolution may be overlooked.

In the cases of medium and low NA values around and lower than 0.2, the filtering method works well [\figs~\ref{fig:PSFD} and \ref{fig:PSFD_each}].
Because the filtering approach is rather simple and fast, it may be preferable for use in low-to-medium NA cases.

The situation becomes more complex when both defocus and HOAs are present.
These effects interact with each other via the convolution of the illumination and collection pupils [Eq.~(\ref{eq:12})].
The filtering method designed in this study appears to work more effectively than the conventional filtering method and the ISAM + HOPE correction approach.
The interaction of the HOAs and the defocus may cause the depth-dependent phase errors in the OCT signal.
This would be the reason why the conventional filtering method underperformed when compared with the filtering method based on the OCT image formation theory for correction at all depths.
In addition, the cTF shape is also modulated irregularly by this interaction.
This may be the issue in the case where ISAM is used with HOAs because the ISAM is based on the shape of the cTF\@.
Furthermore, ISAM requires knowledge of the focus depth position to remove the phase shift caused by the offset of the focus (Supplement 1, Section~\ref{S-sec:ISAM_implementation}).
Focus depth estimation or iterative optimization of the ISAM process with respect to the focus depth may be required.
However, this becomes more difficult in cases where significant HOAs occur.

\subsection{Limitations of the current theory and numerical simulation}
\label{sec:discuss-limits-theory}

The current theory and numerical simulation are based on the fact that the light suffers only a single scattering event in tissues.
The propagation from the surface of the tissue to the point of the scattering event and vice versa is assumed to be in a transparent medium.
The multiple scattering events are not considered in the theory.
The effects of computational methods on the images suffering from multiply scattered light thus cannot be predicted nor simulated with the presented theory and simulator.
The theoretical model of the signal with significant multiple scattering is an important future work for enabling the investigation of the effects of the multiple scattering on imaging performance and the computational methods to overcome these effects.

\subsection{Limitations of the current computational aberration correction method}
\label{sec:discuss-limits}

The computational aberration correction performance is limited by several factors.
The wavefront errors not only affect the phase of the OCT spatial frequency signal, but also affect its amplitude.
The convolution [Eq.~(\ref{eq:12})] of the aberrated pupils could cause the irregular shape of the spatial frequency gain of the OCT signal, $\left|\tilde{h}_\mathrm{OCT}\right|$, as shown in \figurename~\ref{S-fig:SystemFunction_Amp} and could thus lead to the irregular shape and broadening of the PSF even if the phase errors are well corrected (Section~\ref{sec:CAC_simulation} and Supplement 1 Section~\ref{S-sec:other_example}).
The amplitude modulation is also shown to be defocus dependent by the simulation.

Although the theoretical model of OCT image formation predicts the amplitude modulation if the pupil amplitude distribution is known, phase error and amplitude modulation correction need to be considered more carefully.
Because the direct inversion of the system function is like an inversion filter, this will enhance the low-amplitude frequency components.
That will enhance the noise and degrade the signal-to-noise ratio.
When the correction amplifies the frequency components outside of the system's cut-off frequency, it will generate artificial structure in images.
The apriori knowledge of the pupil amplitude distribution should be crucial for effective correction.
As shown in the simulation, the amplitude modulation is defocus dependent.
Even slight discrepancies between the estimated pupil amplitude distribution and the actual distribution can cause significant changes in the corrected image when the focus is significantly off.
Thus, future developments for the correction of both phase and amplitude are required.
These would include proper regularization and estimation of the pupil amplitude distribution.

In addition, the amplitudes of the pupils $P_\mathrm{ill}$ and $P_\mathrm{col}$ also affect both the amplitude and the phase of the OCT spatial frequency components.
In the application of CAC to PSFD-OCT (Section~\ref{sec:PSFD-CAC}), the amplitudes of the pupils are assumed to follow a Gaussian distribution.
If these amplitude distributions deviate from the Gaussian, then the estimation of the aberrations should contain errors and the correction performance will be decreased.
A numerical simulation of the mismatch in the pupil amplitude distribution affects the refocusing [\figurename~\ref{S-fig:LFFD-lNA}].
It seems the effect is marginal, but a comprehensive investigation on aberration correction methods will be required to conclude this issue.

The current CAC design is based on the paraxial and narrow-band approximations.
The assumptions of the methods appear to be valid for PSFD-OCT with a moderate NA, as in the \invivo retina case.
In the case of optical coherence microscopy (OCM), however, a higher NA and a broader spectral bandwidth will be used, and these assumptions would then not be valid.
However, if the application is not \invivo retinal imaging, then the systematic optical aberrations could be suppressed via careful optical design or by using static correction optics.
The current method may already be effective for correction of small residual systematic aberrations.

\section{Conclusion}
\label{sec:conclusion}

In this paper, we reformulated the image formation theory of OCT to include systematic aberrations.
Numerical investigation of computational refocusing in OCT shows that the refocusing of PSFD-OCT works well with the CR filter up to at least moderately high NAs (~0.2).
In addition, the study showed that model-based design of the CR filter that accounts for the pupil amplitude distributions is important.
A new CAC filter for PSFD-OCT was designed based on the theory, and it demonstrated a better aberration correction on volumetric OCT signals, i.e., simultaneous multi-depth correction of systematic aberrations, than the conventional CAC filter for numerical simulations, phantom imaging, and \invivo retinal imaging.
The numerical simulation study shows that the proposed method can obtain the Strehl ratios of more than 0.8 over $\pm$ 100 \um defocus range, while the conventional method cannot achieve this under the simulated conditions.
Analysis of retinal images revealed that the proposed method improved the frequency component corresponding to the density of cone photoreceptors in OCT photoreceptor images by 1.2 to 1.4 times.
The same design approach may be used for other types of OCT systems to improve the performance of defocus and aberration correction.
The basic theory and the numerical simulation method will play important roles in the development of computational aberration correction for OCT\@.

\begin{backmatter}
\bmsection{Funding}
Core Research for Evolutional Science and Technology (JPMJCR2105); Japan Society for the Promotion of Science (21H01836, 21K09684, 22K04962).

\bmsection{Acknowledgments}
We thank to Dr. Kazuhiro Kurokawa (Legacy Research Institute) for fruitful discussion.

\bmsection{Disclosures}

\noindent SM, YY: Topcon (F), Sky Technology (F), Nikon (F), Kao Corp. (F), Panasonic (F), Santec (F), Nidek (F). NF: Nikon (E). LZ: Topcon (F), Sky Technology (F), Nikon (F), Kao Corp. (F), Panasonic (F), Santec (F). L. Zhu is currently employed by Santec.

\bmsection{Data availability} Data underlying the results presented in this paper are not publicly available at this time but may be obtained from the authors upon reasonable request.
An open source simulator of the PSF of OCT and cTF of reflection imaging can be found at \url{https://github.com/ComputationalOpticsGroup/COG-OCTPSF-simulator}\cite{OCTPSF_simulator}.

\bmsection{Supplemental document}
See Supplement 1 for supporting content. 

\end{backmatter}

\end{document}


\maketitle

\subsection{Rigorous OCT image formation theory}\label{sec:OCT_theory}

We began by formulating the imaging theory for the PSFD-OCT system because it is the most common type among the OCT systems in current use.
Other OCT system types can be obtained by modifying the system illumination and collection configurations, as shown in Sections~\ref{sec:CR_other_OCT} and Ref.~\cite{fukutake_four-dimensional_2025-1}.
In the PSFD-OCT system, the collected light from a reflection confocal optics are interfered with the reference light and acquired at multiple optical frequencies.
The light detected via 2D transverse scanning of the focused spot can be expressed as:

\begin{equation}\label{eq:1}
    \begin{split}
        I(\mathbf{r}_{0\parallel}, \omega; z_0 ) =&
        \left\langle \left|
            \sqrt{p_{\mathrm r} S(\omega)} U_{\mathrm r}
            \mathrm{e}^{\mathrm{i} k_{\mathrm r}(\omega) 2 z_\mathrm{r}}
        \right. \right.\\
        & + \left. \left.
            \sqrt{p S(\omega)}
            \mathrm{e}^{\mathrm{i} k_\mathrm{s}(\omega) 2 z_0}
            \iiint
                h_{\mathrm{RCI}} (\mathbf{r}_{0\parallel} - \mathbf{r}_\parallel, z_0 - z, k_{\mathbf{s}}(\omega))
                \eta(\mathbf{r}_\parallel, z, \omega)
            \mathrm{d}\mathbf{r}_\parallel \mathrm{d} z
        \right|^2 \right\rangle_{D,t},
    \end{split}
\end{equation}
where the time-dependent term $\mathrm{e}^{-\mathrm{i} \omega t}$ is omitted, and where the bracket $\langle\cdot\rangle_{D,t}$ means temporal and 2D spatial integration on the detection plane $D$, which will be omitted hereafter.
The unit of $I$ is [J$\cdot$Hz$^{-1}$].
$h_{\mathrm{RCI}}$ is the 3D complex point spread function (cPSF) for reflection confocal imaging (RCI).
$\eta$ is the scattering potential of the sample.
$k_\mathrm{s} = n_\mathrm{BG} k$ is the wavenumber in the sample with background refractive index $n_\mathrm{BG}$.
The second term inside on the right-hand side of Eq.~(\ref{eq:1}) represents 3D spatial integration along $\mathbf{r}_\parallel$ and $z$.
$\mathbf{r}_{0\parallel} = (x_0, y_0)$ is the transversal focal scanning location, $z_0$ is the axial location of the focus, $k$ is the optical wavenumber, and $S$ is the normalized spectral density of the light source [Hz$^{-1}$].
In addition, $p$ and $p_\mathrm{r}$ are the light powers of the sample and reference arms [W], respectively, and $z_\mathrm{r}$ is the single-trip path length of the reference arm.
$U_\mathrm{r}$ is the monochromatic wave function of the reference light at the detector plane [m$^{-1}$], where a wave function is a solution of the Helmholtz equation.
Because both the reference light and the collected backscattered light are coupled into the same single-mode optical fiber, $U_\mathrm{r}$ has the same mode as the signal light; therefore, it can be considered to be a constant complex number where $|U_\mathrm{r}|^2 = 1$.
The wavenumber in a perfect dielectric medium $k(\omega) = \frac{\omega}{v_\mathrm{p}(\omega)}$ is a function of $\omega$, where $v_\mathrm{p}$ is the phase velocity of light.
$k_\mathrm{r}$ is the wavenumber in the reference path.
Hereafter, the variable $\omega$ is dropped for simplicity.
Base on these definitions, the cPSF $h_\mathrm{RCI}$ should have the unit [m$^{-2}$].
Because the cPSF $h_\mathrm{RCI}$ treats the field propagating away from and coming back to the focus point $(\mathbf{r}_{0\parallel}, z_0)$, there is a phase term, $\mathrm{e}^{\mathrm{i} k_\mathrm{s}(\omega) 2 z_0}$, that accounts for the round-trip propagation to and from the focus point with respect to the origin $z = 0$ to balance the optical path length (OPL) with respect to the reference arm length.

The scattering potential of the sample $\eta$ can be expressed as\cite{sung_optical_2009}:
\begin{equation}\label{eq:2}
    \eta(\mathbf{r}_\parallel, z, \omega) = \frac{k^2(\omega)}{4\uppi} \left[n^2(\mathbf{r}_\parallel, z, \omega) - n_\mathrm{BG}^2(\omega)\right].
\end{equation}
By introducing the relative electric susceptibility\cite{lauer_new_2002}, denoted by $\psi = \frac{n^2 - n_\mathrm{BG}^2}{n_\mathrm{BG}^2}$, and then describing it as $\psi = \psi_\mathrm{p} N$, where $\psi_\mathrm{p}$ is the molar relative electric susceptibility [m$^3$] and $N$ is the density distribution of the molecules that contribute to the backscattering [m$^{-3}$], Eq.~(\ref{eq:2}) can then be expressed as:
\begin{equation}
    \eta(\mathbf{r}_\parallel, z, \omega) = \frac{k_\mathrm{s}^2(\omega)}{4\uppi} \psi_\mathrm{p}(\omega) N(\mathbf{r}_\parallel, z).
\end{equation}
Here, the molar relative electric susceptibility $\psi_{\mathrm{p}}$ is assumed to be the same for all of scatterer molecules.

We assume that each optical frequency channel detects a pure monochromatic wave.
The interference signal is the cross-term on the right-hand side of Eq.~(\ref{eq:1}) and is given as:
\begin{equation}\label{eq:3}
    \begin{split}
        I^{'} (\mathbf{r}_{0\parallel}, \omega; z_0)
        =& \sqrt{p p_\mathrm{r}} S(\omega)
\mathrm{e}^{\mathrm{i} 2 [k_\mathrm{s}(\omega) z_0 - k_\mathrm{r}(\omega) z_\mathrm{r}]} \\
        & \times U_\mathrm{r}^*
        \iiint
            h_\mathrm{RCI} [\mathbf{r}_{0\parallel} - \mathbf{r}_\parallel, z_0 - z, k_{\mathbf{s}}(\omega)]
            \frac{k_\mathrm{s}^2 (\omega)}{4\uppi} \psi_\mathrm{p} (\omega) N(\mathbf{r}_\parallel, z)
        \mathrm{d}\mathbf{r}_\parallel \mathrm{d} z
    \end{split}
\end{equation}
Here, $I^{'}$ is the complex-valued signal because its complex conjugate is ignored.

\subsection{Complex point-spread function for reflection confocal imaging}
\label{sec:PSF_RCI}

\begin{figure}
    \centering
    \includegraphics[width=0.7\linewidth]{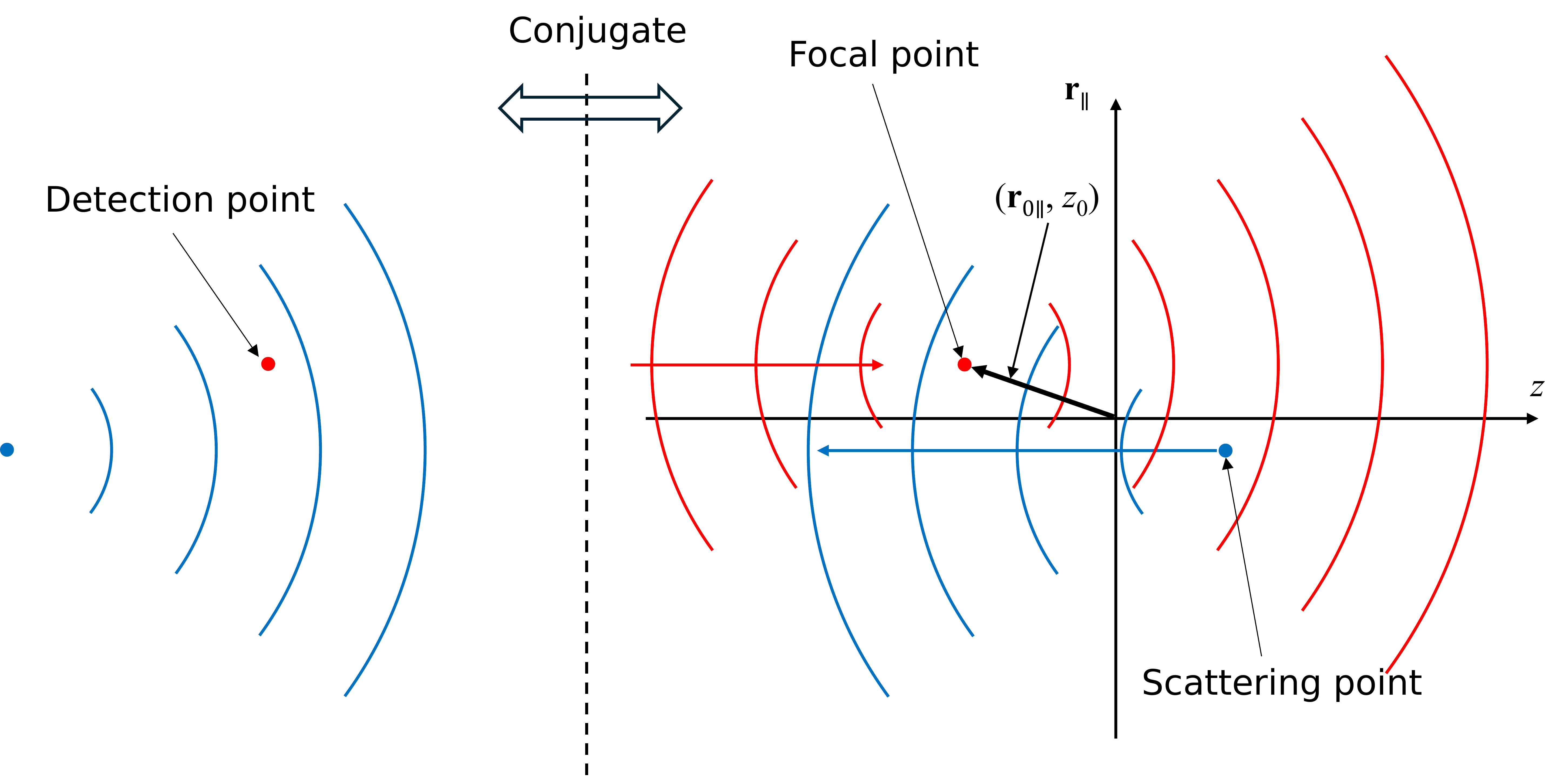}
    \caption{Schematic diagram illustrating the geometry of the illumination beam (red) focused at the red dot and wave backscattered (blue) from the blue dot.}
    \label{fig:scheme}
\end{figure}

The cPSF $h_\mathrm{RCI}$ expresses the collected light field produced by illuminating a point scatterer and collecting the backscattered light via the optics.
Here, $h_\mathrm{RCI}$ is defined using the illumination field to the sample and the collected field mode of the backscattered field from the sample.
Each step is described in the followings.

\subsubsection{Illumination field}

The illumination field in the space $V_\mathrm{ill}$ around the sample [\figurename~\ref{m-fig:fieldspace}(a)] can be expressed using a distribution $P_\mathrm{ill}$ on a sphere centered at the origin of $V_\mathrm{ill}$ with the scalar Debye integral\cite{braat_imaging_2019} as:
\begin{equation}\label{eq:4}
    f_\mathrm{ill} (\mathbf{r}_{\parallel}, z, k)
    =
    \frac{\mathrm{i} k}{2\uppi}
    \iint_{|\bm{\upsigma}_\parallel| \le 1}
        P_\mathrm{ill} (-\bm{\upsigma}_\parallel)
        \mathrm{e}^{\mathrm{i} k(\mathbf{r}_\parallel \cdot \bm{\upsigma}_\parallel + z \sigma_z )}
    \mathrm{d}\Omega
\end{equation}
where $\bm{\upsigma}_\parallel = (\sigma_x, \sigma_y)$ and $\sigma_z$ are directional cosines, and $\sigma_x^2 + \sigma_y^2 + \sigma_{z}^2 = 1$.
The unit vector $\bm{\upsigma} = (\sigma_x, \sigma_y, \sigma_z)$ indicates the propagation direction of a plane wave component.
Equation~(\ref{eq:4}) represents the sum of all plane wave components over the solid angle $\Omega$ on a hemisphere with radius $R$ (described by the blue curve in \figurename~\ref{m-fig:fieldspace}(a)).

\subsubsection{Backscattered field}

The illumination field $f_\mathrm{ill}$ is scattered by a scatterer located at $(\mathbf{r}_\parallel, z)$.
The scattered field at $\left(\mathbf{r}_\parallel^{'}, z^{'}\right)$ can be expressed as:
\begin{equation}\label{eq:5}
    f_\mathrm{s} \left(\mathbf{r}_\parallel^{'}, z^{'}, k\right) = g\left(\mathbf{r}_\parallel^{'} - \mathbf{r}_\parallel, z^{'}-z, k\right) f_\mathrm{ill}  (\mathbf{r}_\parallel, z, k)
\end{equation}
where $g(\mathbf{r})=\mathrm{e}^{\mathrm{i} k|\mathbf{r}|}/|\mathbf{r}|$ is the scalar Green's function [\figurename~\ref{m-fig:fieldspace}(b)].
If we consider only backscattered far-field propagation modes, then the Green's function can be replaced with backward far-field modes\cite{sheppard_rayleighsommerfeld_2013} alone by using Weyl's identity.
\begin{equation}\label{eq:6}
    \begin{split}
        f_\mathrm{bs} \left(\mathbf{r}_\parallel^{'}, z^{'}, k\right)
        &=
        \frac{\mathrm{i} k}{2\uppi}
        \iint_{|\bm{\upsigma}_\parallel| \le 1}
            \mathrm{e}^{\mathrm{i} k\left[\left(\mathbf{r}_\parallel^{'} - \mathbf{r}_\parallel \right) \cdot \bm{\upsigma}_\parallel - \left(z^{'} - z\right) \sigma_z \right]}
        \mathrm{d}\Omega
        f_\mathrm{ill}(\mathbf{r}_\parallel, z, k) \\
        &=
        \frac{\mathrm{i} k}{2\uppi}
        \iint_{|\bm{\upsigma}_\parallel| \le 1}
            \mathrm{e}^{\mathrm{i} k\left(\mathbf{r}_\parallel^{'} \cdot \bm{\upsigma}_\parallel - z^{'} \sigma_z \right)}
            \mathrm{e}^{\mathrm{i} k\left(-\mathbf{r}_\parallel \cdot \bm{\upsigma}_\parallel + z \sigma_z \right)}
        \mathrm{d}\Omega
        f_\mathrm{ill}(\mathbf{r}_\parallel, z, k),
    \end{split}
\end{equation}
where the sign of $\sigma_z$ is negative because the propagation direction is backward.

\subsubsection{Collection of the backscattered field and point-spread function}

The light collected from the single scattering point, i.e., the cPSF of RCI $h_\mathrm{RCI} (\mathbf{r}_{0\parallel} - \mathbf{r}_\parallel, z_0 - z, k)$, is the backscattered field at the location conjugated to the detection point, i.e., $\mathbf{r}^{'} = \mathbf{r}_0$, after each plane wave mode is weighted with the collection function $P_\mathrm{col}$.
Because Eqs.~(\ref{eq:4})-(\ref{eq:6}) are defined using $\mathbf{r}_{0} = \mathbf{0}$,

\begin{equation}\label{eq:7}
    h_\mathrm{RCI}(\mathbf{0}-\mathbf{r}_\parallel, 0-z, k)
    =
    \frac{\mathrm{i} k}{2\uppi}
    \iint_{|\bm{\upsigma}_\parallel| \le 1}
        P_\mathrm{col}\left(\bm{\upsigma}_\parallel\right)
        \mathrm{e}^{\mathrm{i} k(-\mathbf{r}_\parallel\cdot\bm{\upsigma}_\parallel + z\sigma_z )}
    \mathrm{d}\Omega
    f_\mathrm{ill}(\mathbf{r}_\parallel, z, k)
\end{equation}
Therefore, the collected field near the focus can also be expressed using a the combination of plane waves (the Debye diffraction integral) in the same way as the illumination field:
\begin{equation}\label{eq:8}
    f_\mathrm{col}(\mathbf{r}_\parallel, z, k)
    =
    \frac{\mathrm{i} k}{2\uppi}
    \iint_{|\bm{\upsigma}_\parallel| \le 1}
        P_\mathrm{col}\left(\bm{\upsigma}_\parallel\right)
        \mathrm{e}^{\mathrm{i} k(-\mathbf{r}_\parallel \cdot \bm{\upsigma}_\parallel + z \sigma_z )}
    \mathrm{d}\Omega.
\end{equation}

Then, the complex point-spread function, $h_\mathrm{RCI}$ is
\begin{equation}\label{eq:9}
    \begin{split}
        h_\mathrm{RCI}(\mathbf{r}_\parallel, z, k) =& f_\mathrm{col}(-\mathbf{r}_\parallel, -z, k) f_\mathrm{ill}(-\mathbf{r}_\parallel, -z, k) \\
        =&
        \frac{-k^2}{4\uppi^2}
        \iint_{|\bm{\upsigma}_\parallel| \le 1}
            P_\mathrm{col} (\bm{\upsigma}_\parallel) \mathrm{e}^{\mathrm{i} k(\mathbf{r}_\parallel\cdot\bm{\upsigma}_\parallel - z \sigma_z)}
        \mathrm{d}\Omega \\
        & \times
        \iint_{|\bm{\upsigma}_\parallel| \le 1}
            P_\mathrm{ill}(-\bm{\upsigma}_\parallel) \mathrm{e}^{-\mathrm{i} k(\mathbf{r}_\parallel\cdot\bm{\upsigma}_\parallel + z \sigma_z)}
        \mathrm{d}\Omega.
    \end{split}
\end{equation}

\subsubsection{Spatial frequency representation}

The diffraction integrals in illumination field and collection mode [Eqs.~(\ref{eq:4}) and (\ref{eq:8})] can be treated as 2D spatial Fourier transforms (see Section~\ref{sec:2DFT_focused_field}).
The lateral 2D Fourier transform of the illumination and collection fields, $\tilde{f}_\mathrm{ill}$ and $\tilde{f}_\mathrm{col}$, can be expressed as:
\begin{equation}\label{eq:10}
    \begin{split}
        \tilde{f}_\mathrm{ill}(\bm{\upnu}_{\parallel}, z, k) &= \mathcal{F}_{\mathbf{r}_\parallel} \left[f_\mathrm{ill} (\mathbf{r}_{\parallel}, z, k)\right] (\bm{\upnu})\\
        &=
            \mathrm{i} \frac{2\uppi}{k} \frac{P_\mathrm{ill}\left(-\frac{2\uppi}{k} \bm{\upnu}_\parallel\right)}{\sigma_z \left(\frac{2\uppi}{k} \bm{\upnu}_\parallel\right)} \mathrm{e}^{\mathrm{i} k z\sigma_z \left(\frac{2\uppi}{k} \bm{\upnu}_\parallel\right)}\\
\tilde{f}_\mathrm{col}(\bm{\upnu}_{\parallel}, z, k) &= \mathcal{F}_{\mathbf{r}_\parallel} \left[f_\mathrm{col}(\mathbf{r}_\parallel, z, k)\right] (\bm{\upnu})\\
        &=
            \mathrm{i}\frac{2\uppi}{k} \frac{P_\mathrm{col}\left(-\frac{2\uppi}{k} \bm{\upnu}_\parallel \right)}{\sigma_z \left(-\frac{2\uppi}{k} \bm{\upnu}_\parallel\right)} \mathrm{e}^{\mathrm{i} k z\sigma_z \left(- \frac{2\uppi}{k} \bm{\upnu}_\parallel\right)},\\
\end{split}
\end{equation}
where $\mathcal{F}_{x}[f](\nu)$ is the Fourier transform of the function $f$ from the $x$-axis to the $\nu$-axis, and
\begin{equation}\label{eq:11}
    \sigma_z (\bm{\upsigma}_\parallel) = \sqrt{1 - |\bm{\upsigma}_\parallel|^2}.
\end{equation}
Therefore, $\sigma_z(-\bm{\upsigma}_\parallel) = \sigma_z(\bm{\upsigma}_\parallel)$.
The lateral spatial frequency $\bm{\upnu}_\parallel$ represents the Fourier transform pair of $\mathbf{r}_\parallel$.

The 2D spatial Fourier transform of $h_\mathrm{RCI}$ is
\begin{equation}
    \tilde{h}_\mathrm{RCI}(\bm{\upnu}_\parallel, z, k_\mathrm{s})
    =
\left[
        \tilde{f}_\mathrm{ill}(-\bm{\upnu}_\parallel, -z, k_\mathrm{s})
        \otimes_{\bm{\upnu}_\parallel}^2
        \tilde{f}_\mathrm{col}(-\bm{\upnu}_\parallel, -z, k_\mathrm{s})
    \right](\bm{\upnu}_\parallel, z, k_\mathrm{s}),
\end{equation}

\subsection{Analytical expression for some cTFs}\label{sec:ctf_anal}

\begin{figure}
    \centering
    \includegraphics[width=13cm]{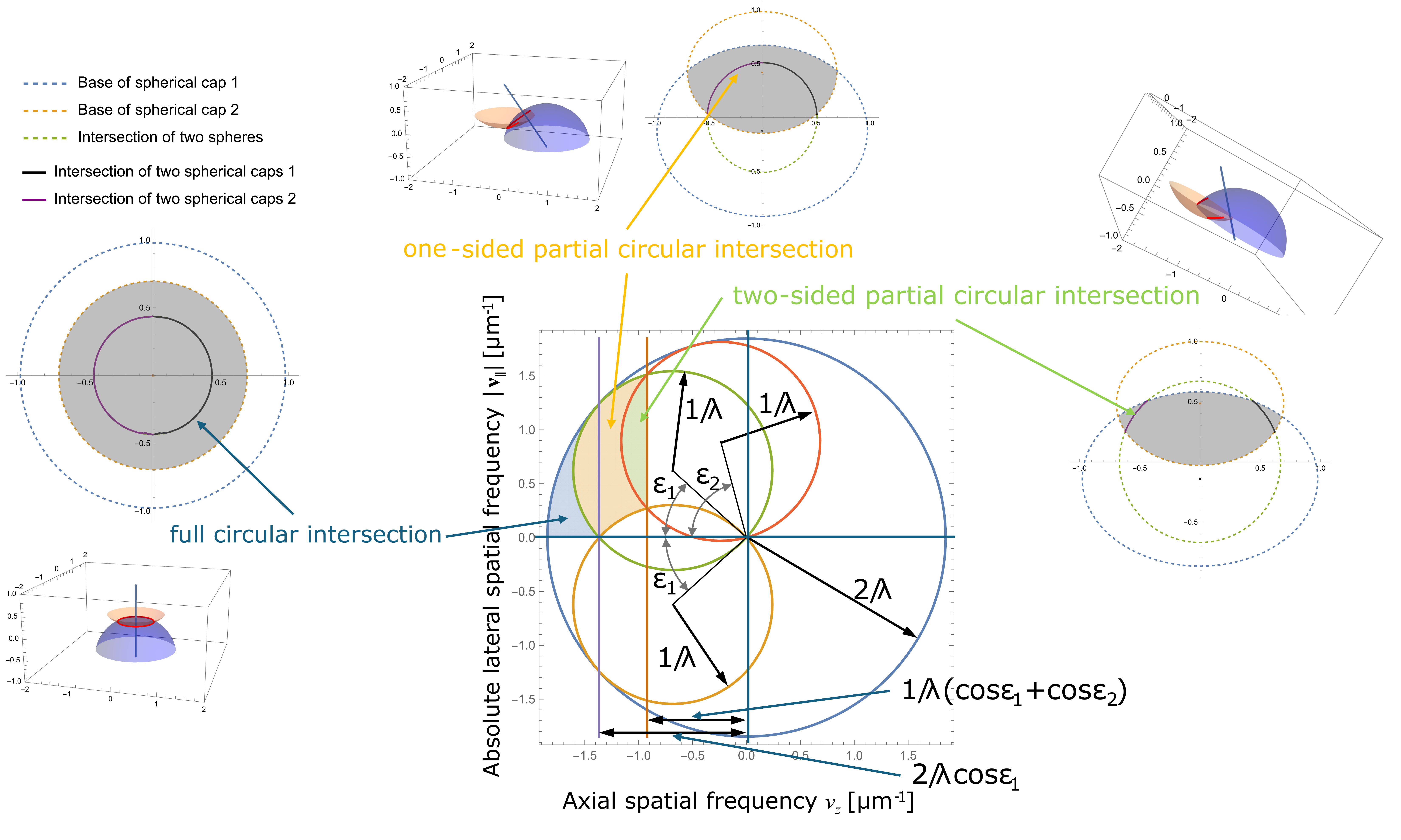}
    \caption{Intersections of pupils and the cTF}\label{fig:convolution_pupils}
\end{figure}

The 3D Fourier transform of $h_\mathrm{RCI}(\mathbf{r}_\parallel, z, k)$ (which is expressed using Eq.~(\ref{eq:9})) is called the coherent transfer function (cTF) for RCI\cite{sheppard_three-dimensional_1994,gu_image_1991}.
\begin{equation}\label{eq:S1}
    H_\mathrm{RCI}(\bm{\upnu}_\parallel, \nu_z, k_\mathrm{s})
    = \iiint h_\mathrm{RCI}(\mathbf{r}_\parallel, z, k_\mathrm{s})
    \mathrm{e}^{-2\uppi\mathrm{i} (\bm{\upnu}_\parallel \cdot \mathbf{r}_\parallel + \nu_z z)}
    \mathrm{d}x \mathrm{d}y \mathrm{d}z.
\end{equation}
From Eqs.~(\ref{m-eq:13}) and (\ref{eq:10}) in the main text, the cTF is obtained as the 3D convolution of the 3D Fourier transforms of $f_\mathrm{ill}$ and $f_\mathrm{col}$.
\begin{equation}\label{eq:S2}
    H_\mathrm{RCI}(\bm{\upnu}_\parallel, \nu_z, k_\mathrm{s})
    =
    - \frac{4\uppi^2}{k_\mathrm{s}^2}
    \begin{aligned}[t]
        & \left[
\frac{P_\mathrm{ill}\left(\frac{2\uppi}{k_\mathrm{s}} \bm{\upnu}_\parallel\right)}
            {\sigma_z\left(\frac{2\uppi}{k_\mathrm{s}} \bm{\upnu}_\parallel\right)}
            \delta\left(\nu_z + \frac{k_\mathrm{s}}{2\uppi} \sigma_z \left(\frac{2\uppi}{k_\mathrm{s}} \bm{\upnu}_\parallel\right)\right)
        \right.\\
        &\quad \left. \otimes_{\bm{\upnu}}^3\ 
\frac{P_\mathrm{col}\left(\frac{2\uppi}{k_\mathrm{s}} \bm{\upnu}_\parallel\right)}
            {\sigma_z\left(\frac{2\uppi}{k_\mathrm{s}} \bm{\upnu}_\parallel\right)}
            \delta\left(\nu_z + \frac{k_\mathrm{s}}{2\uppi} \sigma_z \left(\frac{2\uppi}{k_\mathrm{s}} \bm{\upnu}_\parallel\right)\right)
        \right](\bm{\upnu}_\parallel, \nu_z, k_\mathrm{s}).
    \end{aligned}
\end{equation}
This represents the 3D convolution of spherical shell caps within the spatial frequency domain\cite{sheppard_three-dimensional_1994,villiger_image_2010}.
Threfore, when the pupils have vales of unity, i.e., when $P_\mathrm{ill} = P_\mathrm{col} = 1$ for all $\bm{\upsigma}_\parallel$, the 3D Fourier transforms of the illumination/collection fields are hemispherical shells:
\begin{equation}
    \mathcal{F}_{\mathbf{r}}[f_\mathrm{ill/col}(-\mathbf{r}, k_\mathrm{s})|_{P_{\mathrm{ill}/\mathrm{ill}}=1}](\bm{\upnu}, k_\mathrm{s}) =
    \begin{cases}
        2\mathrm{i}\delta\left(|\bm{\upnu}|^2 - \frac{k_\mathrm{s}^2}{4\uppi^2}\right), & \text{if } \nu_z \le 0\\
        0, & \text{otherwise}
    \end{cases}.
\end{equation}
The cTF is then the 3D convolution of two hemispherical shells:
\begin{equation}
    H_\mathrm{RCI}(\bm{\upnu}_\parallel, \nu_z, k_\mathrm{s})|_{P_\mathrm{ill} = P_\mathrm{col} = 1}
    =\begin{cases}
        \left[
            2\mathrm{i}\delta\left(|\bm{\upnu}|^2 - \frac{k_\mathrm{s}^2}{4\uppi^2}\right)
            \otimes_{\bm{\upnu}}^3
            2\mathrm{i}\delta\left(|\bm{\upnu}|^2 - \frac{k_\mathrm{s}^2}{4\uppi^2}\right)
        \right](\bm{\upnu}_\parallel, \nu_z, k_\mathrm{s}), & \text{if } \nu_z \le 0\\
        0, & \text{otherwise}
    \end{cases}.
\end{equation}
This can be calculated as the 3D convolution of spherically symmetrical functions [see Eq.~(48) in Ref.~\cite{baddour_operational_2010}] as follows:
\begin{equation}\label{eq:anal-ctf-hemisphere}
    H_\mathrm{RCI}(\bm{\upnu}_\parallel, \nu_z, k_\mathrm{s})|_{P_\mathrm{ill} = P_\mathrm{col} = 1}
    = \begin{cases}
        -\frac{4\uppi^2}{k_\mathrm{s}|\bm{\upnu}|}, & \text{if } |\bm{\upnu}| \le \frac{k_\mathrm{s}}{\uppi} \text{ and } \nu_z \le 0\\
        0, & \text{otherwise}
    \end{cases}.
\end{equation}

Because the intersection of two shifted spherical shells forms a circle, the 3D convolution of the spherical shells can be re-formulated using a line integral along the circular intersection.
The cTF is proportional to the length of this circular intersection of the two spherical shell caps at the shift $\bm{\upnu}$ between them.
When the two pupils are cylindrically symmetrical with respect to the $\nu_z$ axis, there are three possible intersection types: $D_1$: full circular intersection; $D_2$: a one-sided partial circular intersection; and $D_3$: a two-sided partial circular intersection, as shown in \figurename~\ref{fig:convolution_pupils}.
Then, the cTF in the case of limited pupils can be calculated as follows:
\begin{equation}
    H_\mathrm{RCI}(\bm{\upnu}_\parallel, \nu_z, k_\mathrm{s}) =
    - \frac{2\uppi}{k_\mathrm{s}|\bm{\upnu}|} \times \begin{cases}
        \begin{aligned}
            & \left\{
            \int_{-\beta_2}^{\beta_1}
                P_{\mathrm{ill}} \left[\bm{\upsigma}_{\parallel, \mathrm{ill}}^{\circ} (\beta) \right]
                P_{\mathrm{col}} \left[\bm{\upsigma}_{\parallel, \mathrm{col}}^{\circ} (\beta) \right]
                \mathrm{d}\beta \right. \\
            & \ \ \left. +
            \int_{\pi-\beta_1}^{\pi + \beta_2}
                P_{\mathrm{ill}} \left[\bm{\upsigma}_{\parallel, \mathrm{ill}}^{\circ} (\beta) \right]
                P_{\mathrm{col}} \left[\bm{\upsigma}_{\parallel, \mathrm{col}}^{\circ} (\beta) \right]
                \mathrm{d}\beta
            \right\}
        \end{aligned}, & \text {if } (|\bm{\upnu}_\parallel|, \nu_z) \in D\\
        0, & \text{otherwise}
    \end{cases},
\end{equation}
where $D = [D_1 \veebar D_2 \veebar D_3]$, and
\begin{equation}
    \bm{\upsigma}_{\parallel, \mathrm{ill}/\mathrm{col}}^{\circ} (\beta; \bm{\upnu}, k_\mathrm{s}) = \frac{2\uppi}{k_\mathrm{s}} \left[\frac{\bm{\upnu}_{\parallel}}{2} \pm \mathbf{R}(\varphi) \cdot \mathbf{m}(\bm{\upnu}, \beta, k_\mathrm{s})\right]
\end{equation}
represents the 2D coordinates of each pupil corresponding to the points on the circular intersection.
Here, $\mathbf{R}$ is the 2D rotation matrix, $\varphi = \tan^{-1}\frac{\nu_y}{\nu_x}$, and
\begin{equation}
    \mathbf{m}(\bm{\upnu}, \beta, k_\mathrm{s}) = \begin{bmatrix}
        \zeta(\bm{\upnu}, k_\mathrm{s})\cos(\beta) \\
        \frac{|\nu_z|}{|\bm{\upnu}|}\zeta(\bm{\upnu}, k_\mathrm{s})\sin(\beta)
    \end{bmatrix}.
\end{equation}
$\mathbf{m}$ represents the 2D spatial frequency coordinates of the point on the circular intersection.
Because the circular intersection is delineated to the $(\nu_x, \nu_z)$-plane by the angle of $\tan^{-1}\frac{|\bm{\upnu}_\parallel|}{\nu_z}$, $\mathbf{m}$ represents the coordinates of an ellipse.
$\zeta = \sqrt{\frac{k_\mathrm{s}^2}{4\uppi^2} - \frac{|\bm{\upnu}|^2}{4}}$ is the radius of the circular intersection.
$\beta$ is the angle along the circular intersection with respect to $\varphi - \frac{\uppi}{2}$.

When $P_\mathrm{ill} = P_\mathrm{col} = 1$, the cTF can then be calculated as:
\begin{equation}\label{eq:anal-ctf-unity}
    H_\mathrm{RCI}(\bm{\upnu}_\parallel, \nu_z, k_\mathrm{s})|_{P_\mathrm{ill} = P_\mathrm{col} = 1}
    = -\frac{4\uppi}{k_\mathrm{s}|\bm{\upnu}|} \times
    \begin{cases}
        \beta_1 + \beta_2, & \text{if } (|\bm{\upnu}_\parallel|, \nu_z) \in D\\
0
        , & \text{otherwise}
    \end{cases},
\end{equation}
where
\begin{equation}
\beta_{1/2}
    = \begin{cases}
        \frac{\uppi}{2}, 
        & \text{if } \frac{1}{|\bm{\upnu}|} \begin{bmatrix}
            |\bm{\upnu}_\parallel|\\
            |\nu_z|
        \end{bmatrix} \cdot \begin{bmatrix}
            \frac{k_\mathrm{s}}{\uppi}\sin\epsilon_{\mathrm{ill}/\mathrm{col}}\\
            \frac{k_\mathrm{s}}{\uppi}\cos\epsilon_{\mathrm{ill}/\mathrm{col}}
        \end{bmatrix} < |\bm{\upnu}|\\
\sin^{-1}\frac{|\bm{\upnu}|}{2\zeta|\bm{\upnu}_\parallel|} \left[|\nu_z| - \frac{k_\mathrm{s}}{\uppi}\cos\epsilon_{\mathrm{ill}/\mathrm{col}}\right], & \text{otherwise}
    \end{cases}.
\end{equation}
$\epsilon$ is the angle of the spherical caps, and $\sin\epsilon_{\mathrm{ill}/\mathrm{col}}$ represents the maximum length of $\bm{\upsigma}_\parallel$ for each of the illumination and collection pupils.

When $P_\mathrm{ill}$ and $P_\mathrm{col}$ are 2D Gaussian distributions with the same 1/e width NA$_w$, the cTF can then be obtained as:
\begin{equation}\label{eq:anal-ctf-gaussian}
    H_\mathrm{RCI}^{(\mathrm{G,G})}(\bm{\upnu}_\parallel, \nu_z, k_\mathrm{s}) = 
    -\frac{4\uppi}{k_\mathrm{s}|\bm{\upnu}|} \times
    \begin{cases}
        \xi(\beta_1) + \xi(\beta_2), & \text{if } (|\bm{\upnu}_\parallel|, \nu_z) \in D\\
0
        , & \text{otherwise}
    \end{cases},
\end{equation}
where
\begin{equation}
    \begin{split}
        \xi(\beta_{1/2})
        =& \mathrm{e}^{-\frac{|\bm{\upnu}|^2 + \nu_z^2 + \frac{\uppi^2}{k^2} \left[|\bm{\upnu}_\parallel|^4 - \nu_z^2 \left(|\bm{\upnu}|^2 + \nu_z^2\right) \right]}{\mathrm{NA}_w^2 |\bm{\upnu}|^2}} \\
        & \times \begin{cases}
            \frac{\uppi}{2}
            I_0 \left[ \frac{|\bm{\upnu}_\parallel|^2}{\mathrm{NA}_w^2} \left(\frac{\uppi^2}{k^2} - \frac{1}{|\bm{\upnu}|^2}\right) \right],
            & \text{if } \frac{1}{|\bm{\upnu}|} \begin{bmatrix}
                |\bm{\upnu}_\parallel|\\
                |\nu_z|
            \end{bmatrix} \cdot \begin{bmatrix}
                \frac{k_\mathrm{s}}{\uppi}\sin\epsilon_{\mathrm{ill}/\mathrm{col}}\\
                \frac{k_\mathrm{s}}{\uppi}\cos\epsilon_{\mathrm{ill}/\mathrm{col}}
            \end{bmatrix} < |\bm{\upnu}|\\
            \int_0^{\beta_{1/2}}
                \mathrm{e}^{\frac{|\bm{\upnu}_\parallel|^2}{\mathrm{NA}_w^2}
                \left(\frac{\uppi^2}{k^2} - \frac{1}{|\bm{\upnu}|^2}\right)\cos2\beta}
            \mathrm{d}\beta, & \text{otherwise}
        \end{cases},
    \end{split}
\end{equation}
where $I_0$ is the modified Bessel function of the first kind and of order zero.

\subsection{OCT signal and cTF}
\label{sec:oct_signal_ctf}

From Eq.~(\ref{eq:3}), the \enface 2D Fourier transform of the OCT interference signal can be rewritten using the cTF $H$ as follows:
\begin{equation}\label{eq:34}
    \begin{split}
        \tilde{I}^{'} (\bm{\upnu}_{\parallel}, \omega; z_0)
        =&
        \sqrt{pp_\mathrm{r}} U_\mathrm{r}^*
        \frac{k_\mathrm{s}^2(\omega)}{4\uppi}
        \psi_\mathrm{p}(\omega)
        S(\omega)
\mathrm{e}^{\mathrm{i} 2 [k_\mathrm{s}(\omega) z_0 - k_\mathrm{r}(\omega) z_\mathrm{r}]} \\
        & \times
        \int
            H(\bm{\upnu}_\parallel, \nu_z, k_\mathrm{s}(\omega)) \tilde{N}(\bm{\upnu}_\parallel, \nu_z)
            \mathrm{e}^{\mathrm{i}2\uppi z_0 \nu_z}
        \mathrm{d}\nu_z,
    \end{split}
\end{equation}
where $H$ and $\tilde{N}$ represent the 3D spatial Fourier transforms of $h$ and the object structure $N$, respectively.
The integration over $z$ in Eq.~(\ref{eq:3}) is converted into an inverse Fourier transform from $\nu_z$ to $z_0$ in Eq.~(\ref{eq:34}).

\subsection{OCT PSF with paraxial and narrow-band approximations}\label{sec:oct_psf_paraxial_narrowband}

The paraxial OCT PSF is derived as Eq.~(\ref{m-eq:28}).
The peak magnitude of $h_\mathrm{RCI}$ is approximately proportional to the squared wavenumber, $\sim k^{2}$: this is caused by focusing.
In the current model, the same solid angle for the focusing field is assumed for all wavenumbers, and thus the spot size of the illumination beam is dependents on the wavenumber.
A higher wavenumber, equates to a smaller spot size and a higher the peak value, $\sim k$.
This also holds the same for the collection field, $\sim k$.
As a result, the order of the wavenumber dependence on the peak signal amplitude is system-dependent.

We assume that $\psi_\mathrm{p}$ is independent of $\omega$ as $\psi_\mathrm{p} (\omega) \sim \psi_\mathrm{p} (\omega_\mathrm{c})$; this assumption will be reasonable when the frequency dependence of the refractive indices of the scatterers and the surrounding medium are similar, i.e., when $n(\omega) \propto n_\mathrm{BG}(\omega)$.
Additionally, the distribution shape of $\tilde{\Gamma}$ is assumed to be independent of $\omega$ as $\tilde{\Gamma}\left(\bm{\upnu}_\parallel, z, k(\omega)\right) \sim K(\omega) \tilde{\Gamma}^{'}\left(\bm{\upnu}_\parallel, z; k(\omega_\mathrm{c})\right)$ (the narrow-band and no-chromatic aberration approximation) where $K(\omega)\approx k^2(\omega)$ is the frequency-dependent scaling factor of $\tilde{\Gamma}^{'}$, and then Eq.~(\ref{m-eq:28}) becomes
\begin{equation}\label{eq:30}
    h_\mathrm{OCT} (\mathbf{r}_\parallel, z, \tau)
    \approx
    \psi_\mathrm{p} (\omega_\mathrm{c})
    \Gamma^{'}\left(\mathbf{r}_\parallel, z; k_\mathrm{s}(\omega_\mathrm{c} )\right)
    G(\tau, z).
\end{equation}
The function $G$ relates the axial position $z$ to the optical delay $\tau$ and is given by
\begin{equation}\label{eq:29}
    G(\tau, z) = 
    \int
        k_\mathrm{s}^{2}(\omega) K(\omega) S(\omega)
        \mathrm{e}^{- \mathrm{i} [\tau\omega + 2 (z - z_0) k_\mathrm{s}(\omega) + 2 z_\mathrm{r} k_\mathrm{r}(\omega)]}
    \mathrm{d}\omega.
\end{equation}
Equation~(\ref{eq:30}) is the frequently used form of the OCT's point spread function (PSF) where the lateral and axial (delay) resolutions are separable.
This represents a somehow reasonable approximation when there are no strong absorption peaks for the sample over the light source's spectrum.

The OCT cPSF without high-order dispersion mismatch can now be described using Eq.~\ref{eq:apsf} as:
\begin{equation}\label{eq:parapsf}
h_\mathrm{OCT} (\mathbf{r}_\parallel, z, \tau)
        \propto \psi_\mathrm{p} (\omega_\mathrm{c})
\Gamma^{'}\left(\mathbf{r}_\parallel, z; k_\mathrm{s}(\omega_\mathrm{c} )\right)
        \gamma\left\{\frac{2}{c} \left[
            l + n_\mathrm{g,BG} (\omega_\mathrm{c}) (z - z_0) + n_\mathrm{g,r} (\omega_\mathrm{c}) z_\mathrm{r}
        \right]\right\},
\end{equation}
where $\gamma$ is the complex temporal coherence function, $l = \frac{c\tau}{2}$ is the single-pass optical path length (OPL) corresponding to the optical delay $\tau$, $c$ is the speed of light in vacuum, $C$ is a phase term [Eq.~\ref{eq:C}], and $n_{g}$ is the group index.

\subsection{Axial (delay) point-spread function of OCT under the paraxial approximation}
\label{sec:axial_psf}

If no high-order ($\ge$ 2nd order) dispersion occurs in either the medium or the reference arm, the wavenumber around the optical frequency $\omega_\mathrm{c}$ can then be considered to be:
\begin{equation}
    \begin{split}
        k_\mathrm{s} (\omega) =& \frac{\omega_\mathrm{c}}{v_\mathrm{p,BG} (\omega_\mathrm{c})} + \frac{1}{v_\mathrm{g,BG} (\omega_\mathrm{c})} (\omega - \omega_\mathrm{c})\\
        =& n_\mathrm{g,BG} (\omega_\mathrm{c})\frac{\omega}{c} + \left[  n_\mathrm{BG} (\omega_\mathrm{c}) - n_\mathrm{g,BG} (\omega_\mathrm{c}) \right] \frac{\omega_\mathrm{c}}{c},
    \end{split}
\end{equation}
and
\begin{equation}
    \begin{split}
        k_\mathrm{r} (\omega) =& \frac{\omega_\mathrm{c}}{v_\mathrm{p,r} (\omega_\mathrm{c})} + \frac{1}{v_\mathrm{g,r} (\omega_\mathrm{c})} (\omega - \omega_\mathrm{c})\\
        =& n_\mathrm{g,r} (\omega_\mathrm{c}) \frac{\omega}{c} + \left[n_\mathrm{r} (\omega_\mathrm{c}) - n_\mathrm{g,r} (\omega_\mathrm{c}) \right] \frac{\omega_\mathrm{c}}{c},
    \end{split}
\end{equation}
where $v_\mathrm{p} = \frac{c}{n}$, $v_\mathrm{g} = \frac{1}{\partial k / \partial \omega}$, and $n_\mathrm{g} = \frac{c}{v_\mathrm{g}}$ are the phase velocity, the group velocity, and the group index, respectively.
$c$ is the speed of light in vacuum.
When $q$ is an integer, $q \in \mathbb{Z}$, the Eq.~(\ref{eq:29}) is
\begin{equation}\label{eq:33}
    G(\tau, z)
    =
    C \sum_{m=0}^{2+q} \left\{R_m B_m \gamma\left[\frac{2}{c} \left(l + n_\mathrm{g,BG} (\omega_\mathrm{c}) (z - z_0) + n_\mathrm{g,r} (\omega_\mathrm{c}) z_\mathrm{r}\right)\right] \right\}
\end{equation}
where $\gamma(\tau) \propto \mathcal{F}_\omega [S(\omega)](\tau)$ represents the complex temporal coherence function.
The operator $B_m$ is given by
\begin{equation}\label{eq:B}
    B_m = \left(\frac{\omega_\mathrm{c}}{c}\right)^{(2 + q - m)} \frac{\partial^m}{\partial l^m},
\end{equation}
the constant $R_m$ is given by
\begin{equation}\label{eq:R}
    R_m = \binom{2+q}{m}
    \left[n_\mathrm{BG} (\omega_\mathrm{c}) - n_\mathrm{g,BG} (\omega_\mathrm{c}) \right]^{(2 + q - m)}
    \left[-\mathrm{i} n_\mathrm{g,BG} (\omega_\mathrm{c}) \right]^m
\end{equation}
and the phase term $C$ is given by
\begin{equation}\label{eq:C}
    C = \mathrm{e}^{-\mathrm{i} 2\frac{\omega_\mathrm{c}}{c}
    \left\{
        \left[n_\mathrm{BG} (\omega_\mathrm{c}) - n_\mathrm{g,BG} (\omega_\mathrm{c})\right] (z - z_0)
        + \left[n_\mathrm{r} (\omega_\mathrm{c}) - n_\mathrm{g,r} (\omega_\mathrm{c}) \right] z_\mathrm{r} 
    \right\}}.
\end{equation}
The single-pass OPL corresponding to the optical delay $\tau$, i.e., $l = \frac{c\tau}{2}$, is the axial direction coordinate in OCT images.
Usually, the spreading width of $\Gamma$ along $z$ (the confocal gate) is considerably broader than that of $\gamma$ (the temporal coherence gate) under conditions where the paraxial approximation holds.
Therefore, Eq.~(\ref{eq:33}) can be treated as the axial PSF of OCT\@.
At this stage, we can make the simple assumption that the lateral and axial resolutions can be separated.
This assumption thus requires both the paraxial and narrow-band approximations.

In practical OCT applications, the light source spectrum has a narrow bandwidth when compared with its central frequency $\Delta\omega \ll \omega_\mathrm{c}$.
Then, the derivative in $B_m$ does not change the shape of the axial PSF signficiantly.
In this case, the operation of Eq.~(\ref{eq:B}) on $\gamma(l)$ can be approximated as $B_m \gamma(l) \approx B_0 \gamma(l) = \left(\frac{\omega_\mathrm{c}}{c}\right)^{(2 + q)} \gamma(l)$.
When $n_\mathrm{BG} \approx n_\mathrm{g,BG}$, $\sum_{m=0}^{2+q} R_m \approx R_{2+q} = \left[-\mathrm{i} n_\mathrm{g,BG} (\omega_\mathrm{c}) \right]^{2+q}$.
Then, the axial PSF of OCT [i.e., Eq.~(\ref{eq:33})] is approximately proportional to the temporal coherence function $\gamma$ as follows:
\begin{equation}\label{eq:apsf}
    G(\tau, z)
    \approx
    \left[-\mathrm{i} k_\mathrm{s}(\omega_\mathrm{c}) \right]^{2+q}
    C
    \gamma\left[\frac{2}{c} \left(
        l + n_\mathrm{g,BG} (\omega_\mathrm{c}) (z - z_0) + n_\mathrm{g,r} (\omega_\mathrm{c}) z_\mathrm{r}
    \right)\right].
\end{equation}

The axial PSF is different from that of Villiger and Lasser\cite{villiger_image_2010}, which they reported as $\mathcal{F}_\omega [k(\omega)S(\omega)](\tau)$.
The difference may come from the fact that we took the $\omega$-dependent cTF size into account by assuming that $P_{\mathrm{ill}/\mathrm{col}}(\bm{\upsigma}_\parallel)$ is $\omega$-independent.
This $\omega$-dependent cTF size assumption provides the factor of $k^q$ (see Eq.~(\ref{eq:anal-ctf-hemisphere})).

\subsection{Numerical simulation method of cTF}\label{sec:numerical_simulation-ctf}

For the cTF simulation, the 3D data are calculated for the cTF for either one or several optical frequencies $\omega_{i}$.
As similar to Sec.~\ref{m-sec:simulations}, a 2D convolution is then calculated between $\tilde{f}_\mathrm{ill}$ and $\tilde{f}_\mathrm{col}$.
This procedure is iterated for the desired optical frequencies.
A sufficiently high sampling density along $z$ is required to accurately simulate the phase change along $z$ direction.
The axial sampling step is set to be less than $\lambda/4$.
The 1D Fourier transform for the axial location $z$ is calculated for the cTF simulation.
These last-stage Fourier transforms are calculated using a zoom FFT\cite{leutenegger_fast_2006} to avoid calculation of the blank regions.

\subsection{Validation of numerical simulation}\label{sec:validation-cTFsim}

\begin{figure}
    \centering
    \includegraphics[width=12cm]{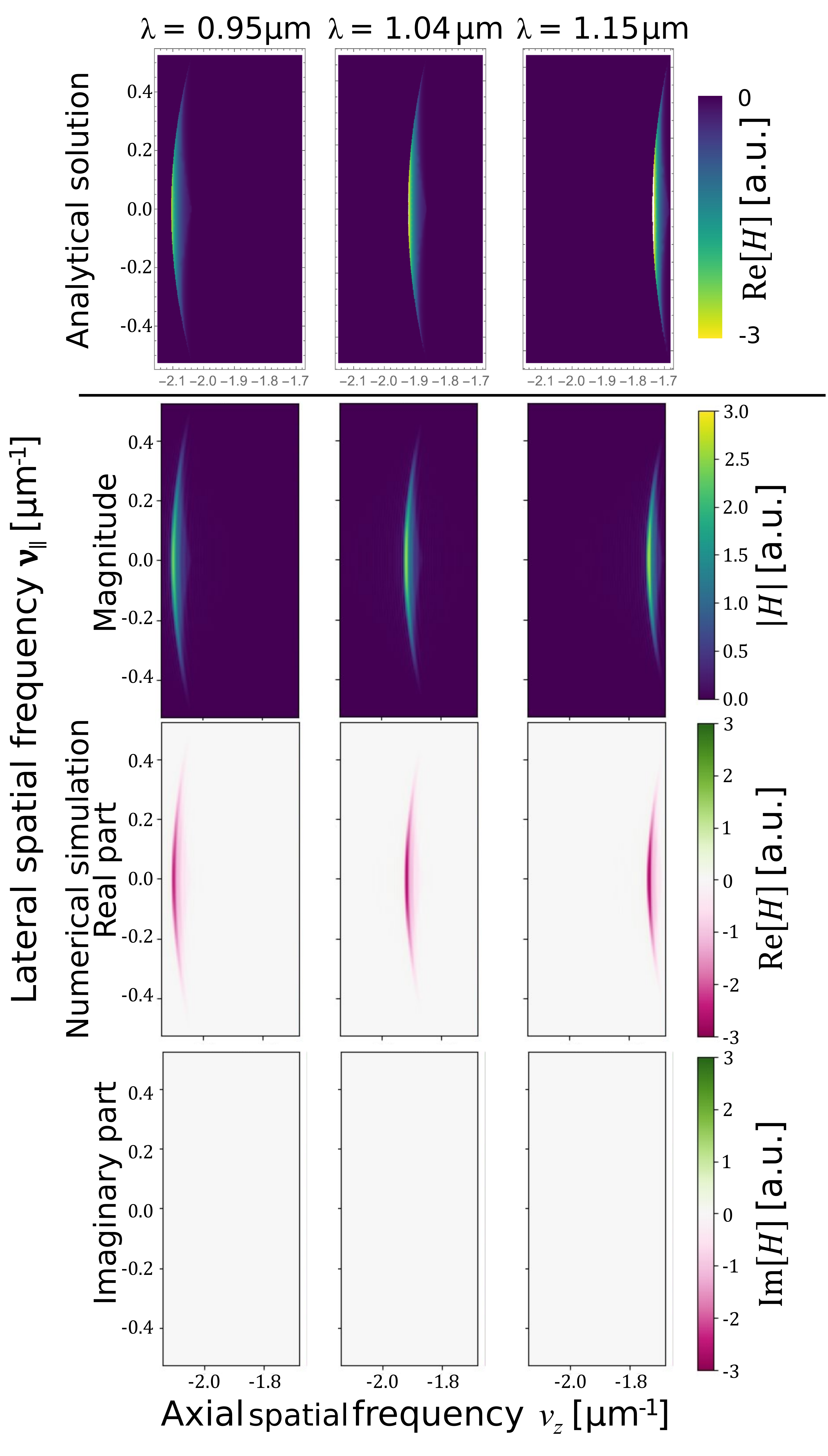}
    \caption{Comparison of numerical simulation method with analytical method.
    Both the illumination and collection pupils represent Gaussian distributions with 0.2 1/e widths in the NA, and their cut-off NAs are both 0.25.
    The analytical result was calculated using Eq.~(\ref{eq:anal-ctf-gaussian}).}\label{fig:Validation-sim}
\end{figure}

\begin{figure}
    \centering
    \includegraphics[width=12cm]{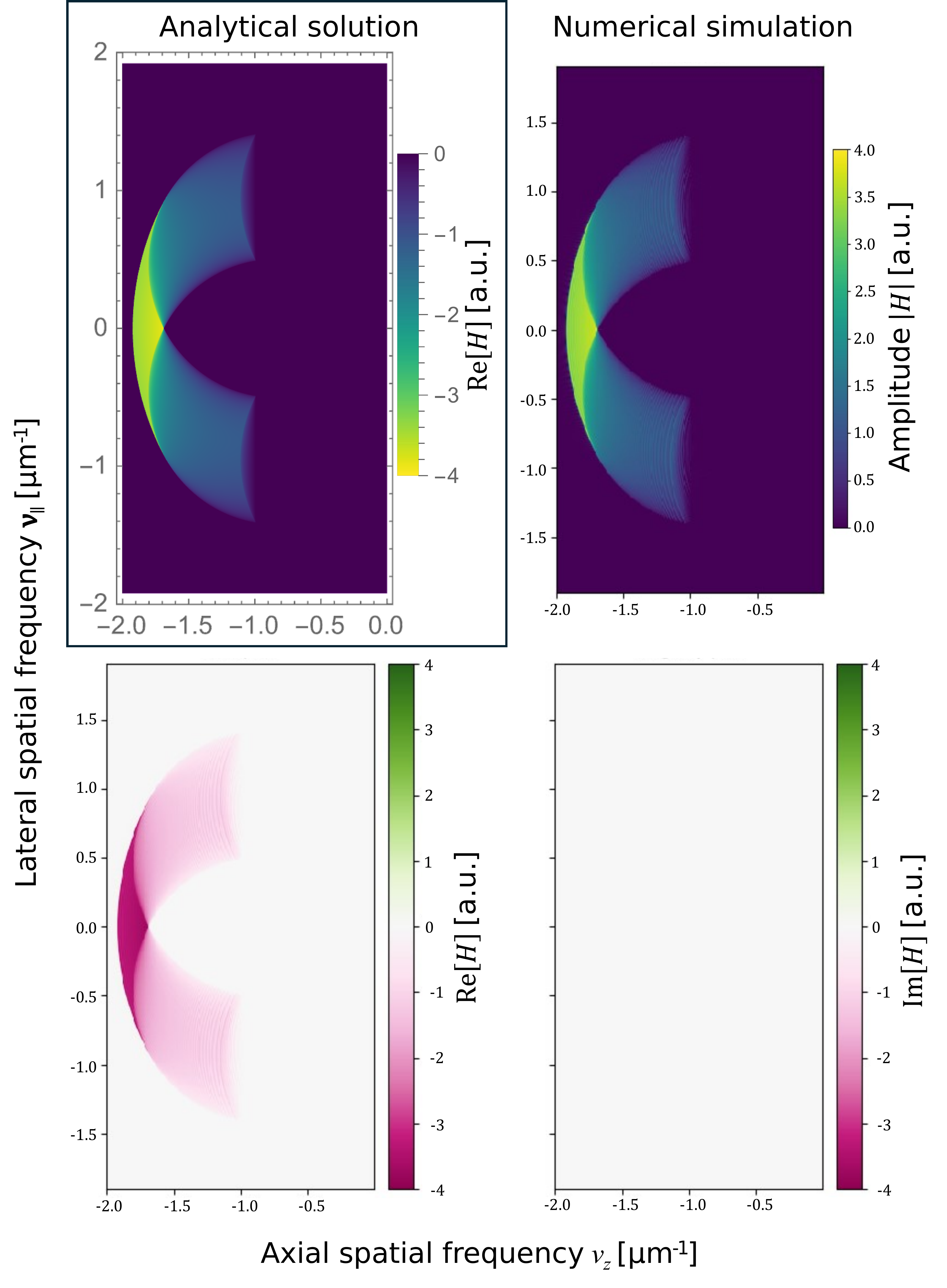}
    \caption{Comparison of numerical simulation method with analytical method.
    Both the illumination and collection pupils have values of unity.
    Their cut-off NAs are 0.988 and 0.479, respectively.
    The analytical results were obtained using Eq.~(\ref{eq:anal-ctf-unity}.)}\label{fig:Validation-sim-2}
\end{figure}

The cTF $H$ of a specific case has been simulated numerically as shown in \figs~\ref{fig:Validation-sim} and \ref{fig:Validation-sim-2} using Eq.~(\ref{m-eq:12}) and a 1D digital Fourier transform along $z$.
The analytical results for Section~\ref{sec:ctf_anal} are also shown.
In \figurename~\ref{fig:Validation-sim}, the calculated results for the three wavelength cases with truncated Gaussian illumination and collection pupils are given.
The cut-off NA was 0.25, and the 1/e width of the Gaussian pupils was 0.2 in the NA\@.
In \figurename~\ref{fig:Validation-sim-2}, illumination and collection pupils with unity values and cut-off NAs of 0.988 and 0.479, respectively, were used at the wavelength of 1.04 \um.
Both cases used $n_{\mathrm{BG}} = 1.0$.
There are no significant differences in shape and amplitude between the numerical results and the analytical results.
This indicates that, our numerical simulation procedure follows the image formation theory well.

\subsection{CTF with aberrations}
\label{sec:CTF_aberrations}

\begin{figure}
    \centering
    \includegraphics[width=11cm]{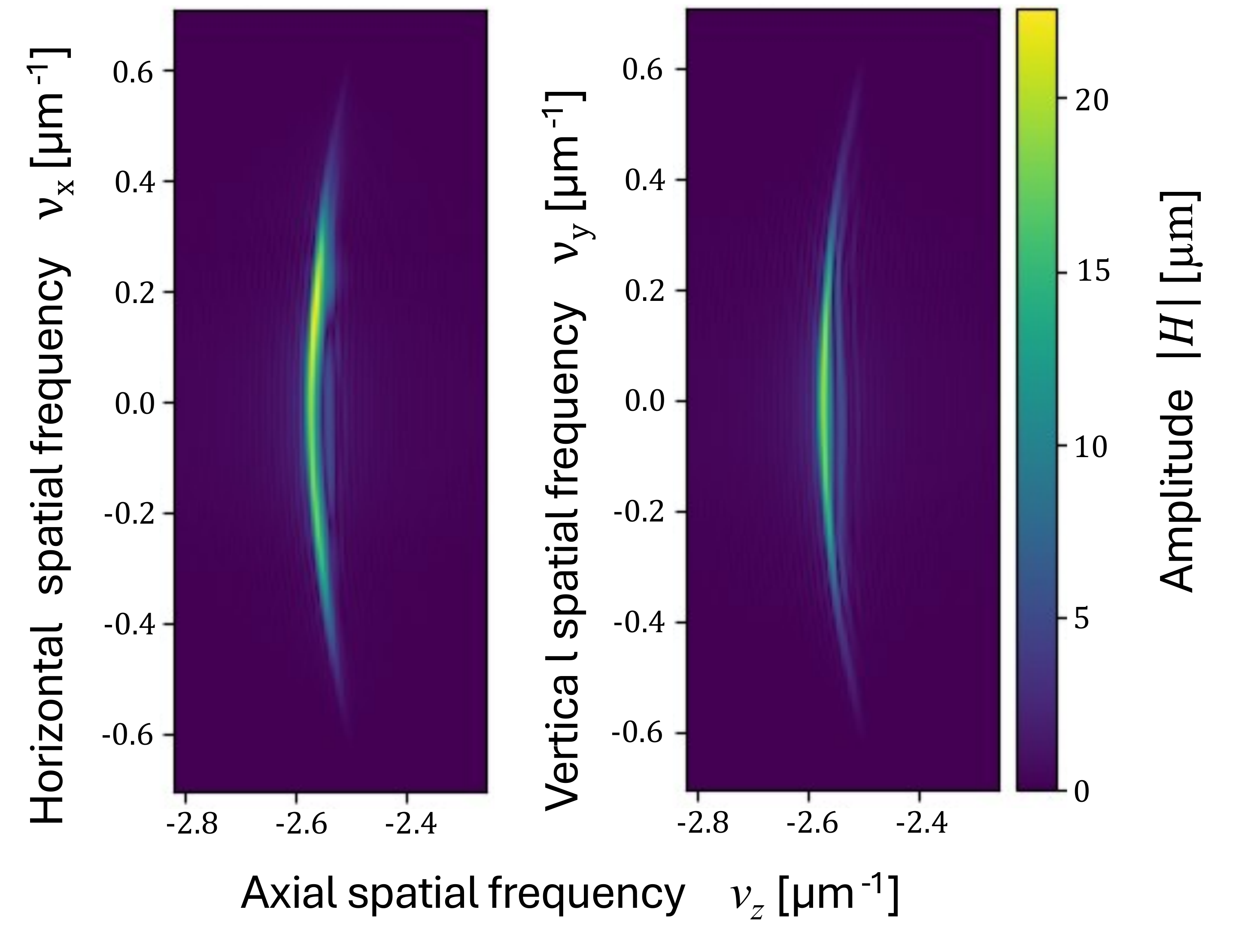}
    \caption{An example of simulated cTF with aberrations.
    }\label{fig:CTF_aberrations}
\end{figure}

A simulated result of a cTF with aberrations is shown in \figurename~\ref{fig:CTF_aberrations}.
Both the illumination and collection pupils represent Gaussian distributions with 0.2 $n_{\mathrm{BG}}$ $e^{-1}$ widths in the NA, and their cut-off NAs are both 0.25 $n_{\mathrm{BG}}$.
The wavelength in vacuum is 1.04 nm, and $n_{\mathrm{BG}} = 1.34$
High-order aberrations set as the same to Section~\ref{m-sec:CAC_simulation}

\subsection{Interpretation of the aberrations estimated by computational aberration correction}

\begin{figure}
    \centering
    \includegraphics[width=13cm]{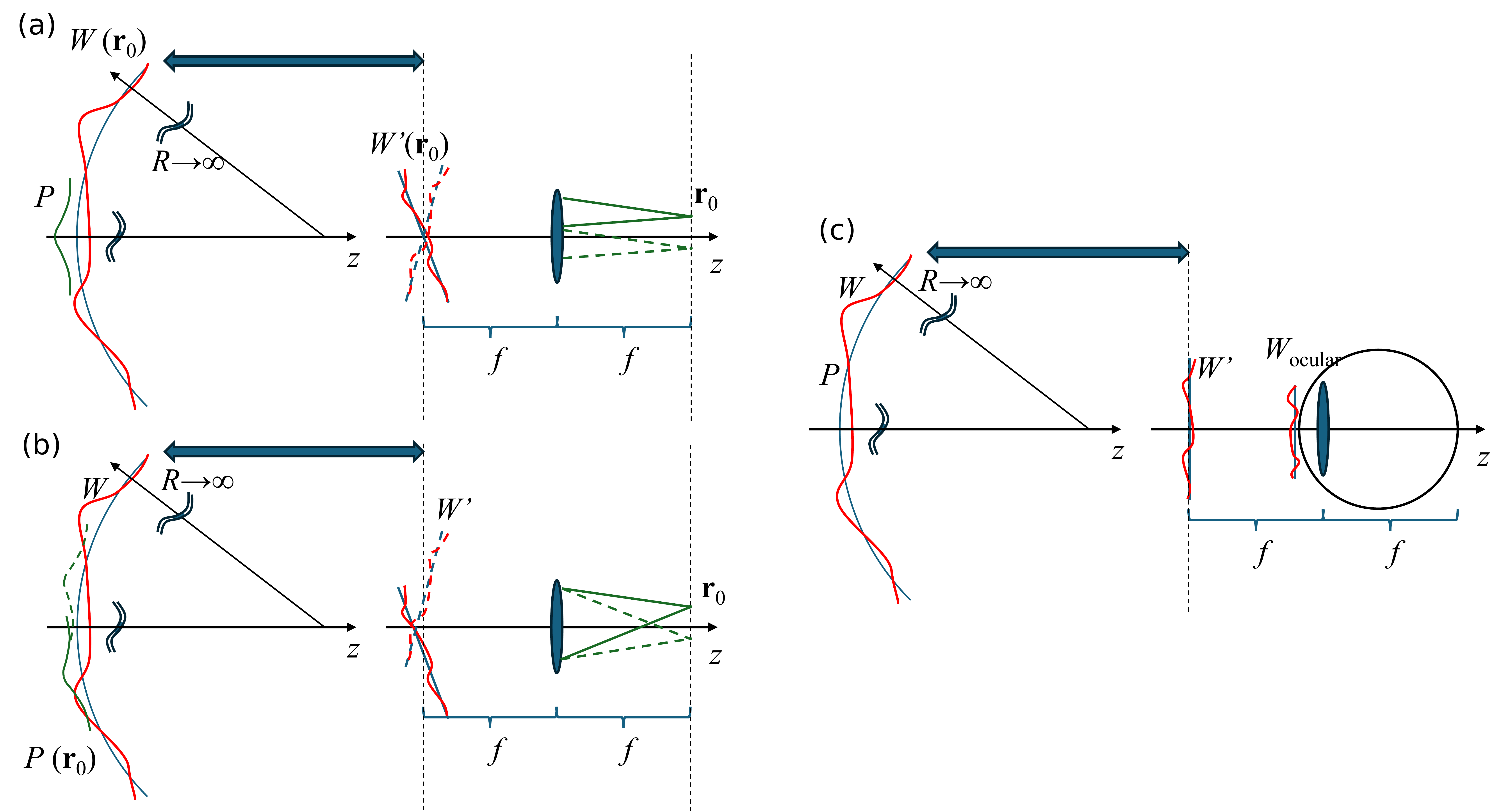}
    \caption{Relationships between estimated aberrations and imaging optics configurations.
    (a) Relationship between the pupil and the estimated wavefront aberrations of the translational scanning system.
    $f$: focal length of the objective.
    (b) Relationship between the pupil and the estimated wavefront aberrations of the fan-beam scanning system.
    $f$: focal length of the objective.
    (c) Relationship between the estimated wavefront aberrations based on our computational aberration correction (CAC) $W$ and the ocular aberration $W_\mathrm{ocular}$.
    $f$: focal length of the eye.}\label{fig:Interpret_Aberrations}
\end{figure}

Strictly speaking, the pupil $P$ or the wavefront errors $W$ will be dependent on the scanning location $\mathbf{r}_0$, according to the beam scanning mechanism.
In the translational beam scanning case [\figurename~\ref{fig:Interpret_Aberrations}(a)], the displacement of the scanning beam on the physical aperture is equivalent to the displacement of the aperture of the optical system.
The same plane wave component will suffer from different wavefront aberrations.
Therefore, $W$ becomes a function of $\mathbf{r}_0$ as $W(\mathbf{r}_0)$.

In contrast, fan-beam scanning [\figurename~\ref{fig:Interpret_Aberrations}(b)] tilts the beam wavefront at the physical aperture.
As a result, $P$ becomes a function of $\mathbf{r}_0$ as $P(\mathbf{r}_0)$.

In both cases, the cTF becomes a function of $\mathbf{r}_0$ as $h_\mathrm{RCI}(\mathbf{r}_\parallel, z, k, \mathbf{r}_0)$.
To be able to assume that the image is constructed via a convolution of the object structure and $h_\mathrm{RCI}$, the scanning range should be small to allow $h_\mathrm{RCI}$ to be regarded as being constant over the scanning range.
The maximum of this range may define the acceptable scanning range for the CAC\@.

Note here that the ocular aberrations $W_\mathrm{ocular}$ are usually defined as wavefront errors at the anterior part of the eye\cite{thibos_standards_2002}.
This corresponds to the plane located immediately before the objective.
This definition differs from that of $W$, as shown in \figurename~\ref{fig:Interpret_Aberrations}(c).
Many studies of retinal imaging methods, including adaptive optics imaging and aberrometers for the eye, have reported the aberrations as being defined at this plane.
Therefore, the different definitions for $W$ may make it difficult to directly compare the aberrations estimated using CAC with PSFD-OCT with the ocular aberrations reported by other studies.

\subsection{Computational refocusing of several other OCT types}
\label{sec:CR_other_OCT}

\subsubsection{Design of computational refocusing filters}

\paragraph{Full-field plane wave illumination}

When the illumination is in the form of a plane wave, $A_\mathrm{ill}$ is a delta function $\delta(\bm{\upnu}_\parallel - \frac{n_{\mathrm{BG}}k_\mathrm{c}}{2\uppi}\bm{\upsigma}_{\mathrm{ill}\parallel})$, where $\bm{\upsigma}_{\mathrm{ill}\parallel}$ represents the propagation direction of the illumination plane wave in the \enface plane.
Threfore:
\begin{equation}\label{eq:S3}
    \tilde{\Gamma}(\bm{\upnu}_\parallel, z_0 - z_\mathrm{s}; n_{\mathrm{BG}} k_\mathrm{c})
    \propto A_\mathrm{col}\left(\bm{\upnu}_\parallel - \frac{n_{\mathrm{BG}}k_\mathrm{c}}{2\uppi}\bm{\upsigma}_{\mathrm{ill}\parallel}\right)
    \mathrm{e}^{\mathrm{i} \phi_\mathrm{def} (z_0 - z_\mathrm{s}) \left|\bm{\upnu}_\parallel - \frac{n_{\mathrm{BG}}k_\mathrm{c}}{2\uppi}\bm{\upsigma}_{\mathrm{ill}\parallel}\right|^2}.
\end{equation}
This corresponds to full-field coherent illumination, such as that in full-field SS-OCT\@.
The refocusing filter is given by
\begin{equation}\label{eq:S4}
    \tilde{\Gamma}^{-1}_{\mathrm{CR, FF}} (\bm{\upnu}_\parallel, l_\mathrm{s}; n_{\mathrm{BG}}k_\mathrm{c})
    = \mathrm{e}^{\mathrm{i} \phi_\mathrm{def} \delta z_\mathrm{s}(l_\mathrm{s}) \left|\bm{\upnu}_\parallel - \frac{n_{\mathrm{BG}}k_\mathrm{c}}{2\uppi}\bm{\upsigma}_{\mathrm{ill}\parallel}\right|^2}.
\end{equation}
A refocus filter that simply invert the defocus phase $\phi_\mathrm{def}$ can be also used:
\begin{equation}\label{eq:S5}
    \tilde{\Gamma}^{-1}_{\mathrm{CR, FF'}} (\bm{\upnu}_\parallel, l_\mathrm{s}; n_{\mathrm{BG}}k_\mathrm{c}) = \mathrm{e}^{\mathrm{i} \phi_\mathrm{def} \delta z_\mathrm{s}(l_\mathrm{s}) |\bm{\upnu}_\parallel |^2}.
\end{equation}
This filter is frequently used for full-field OCT with plane wave illumination.
Note that this refocus will cause a depth-dependent lateral shift because of the frequency offset of $\frac{n_{\mathrm{BG}}k_\mathrm{c}}{2\uppi}\bm{\upsigma}_{\mathrm{ill}\parallel}$.
If the illumination angle of the plane wave is small, this shift is negligible.

\paragraph{Line-field illumination with no apodized detection}

In the case of line-field OCT (LF-OCT), $A_\mathrm{ill}$  can be defined as a 1D Gaussian and 1D delta function as:
\begin{equation}\label{eq:S6}
    A_\mathrm{ill} = \mathrm{e}^{-\frac{\nu_x^2}{\Delta f_\mathrm{ill}^2}} \delta(\nu_y),
\end{equation}
which represents that a illumination beam is focusing along the $x$ direction and plane wave along the $y$ direction without a tilt.
When the size of a slit (i.e., the height of the detector of the sensor array) is significantly smaller than the focusing optical spot size on the sensor array, $A_\mathrm{col}$ can then be approximated as unity along the $x$ direction.
Then,
\begin{equation}\label{eq:S7}
    \begin{split}
        \tilde{\Gamma}(\bm{\upnu}_\parallel, z_0 - z_\mathrm{s}; n_{\mathrm{BG}}k_\mathrm{c})
        \propto& \sqrt{\frac{\uppi}{\frac{1}{\Delta f_\mathrm{ill}^2} - 2\mathrm{i} \phi_\mathrm{def} (z_0 - z_\mathrm{s})}}\\
        \times& \mathrm{e}^{-\frac{\Delta f_\mathrm{ill}^2 \phi_\mathrm{def}^2 (z_0 - z_\mathrm{s})^2}
            {1 + 4\Delta f_\mathrm{ill}^4 \phi_\mathrm{def}^2 z^2} \nu_x^2}
        \mathrm{e}^{\mathrm{i} \frac{1+2\Delta f_\mathrm{ill}^4 \phi_\mathrm{def}^2 (z_0 - z_\mathrm{s})^2}
            {1 + 4\Delta f_\mathrm{ill}^4 \phi_\mathrm{def}^2 z^2} \phi_\mathrm{def} (z_0 - z_\mathrm{s}) \nu_x^2}\\
        \times& A_\mathrm{col}^{(y)} (\nu_y) \mathrm{e}^{\mathrm{i} \phi_\mathrm{def} (z_0 - z_\mathrm{s}) \nu_y^2}.
    \end{split}
\end{equation}
In this case, the phase-only refocusing filter would be given by:
\begin{equation}\label{eq:S8}
    \tilde{\Gamma}^{-1}_{\mathrm{CR}, \mathrm{LF}} (\bm{\upnu}_\parallel, l_\mathrm{s}; n_{\mathrm{BG}}k_\mathrm{c})
    = \mathrm{e}^{\mathrm{i} \phi_\mathrm{def} \delta z_\mathrm{s}(l_\mathrm{s})
        \left(\frac{1 + 2\Delta f_\mathrm{ill}^4 \phi_\mathrm{def}^2 \delta z_\mathrm{s}^2(l_\mathrm{s})}
            {1 + 4\Delta f_\mathrm{ill}^4 \phi_\mathrm{def}^2 \delta z_\mathrm{s}^2(l_\mathrm{s})} \nu_x^2
        + \nu_y^2
        \right)}.
\end{equation}

\paragraph{Line-field illumination and Gaussian apodized detection}

When a mask is present at the collection pupil, the phase errors caused by defocus will be different.
If we assume that $A_\mathrm{col}$ has a 2D Gaussian distribution, then the phase-only refocusing filter in this case would be
\begin{equation}\label{eq:S13}
    \tilde{\Gamma}^{-1}_{\mathrm{CR}, \mathrm{LF-GC}} (\bm{\upnu}_\parallel, l_\mathrm{s}; n_{\mathrm{BG}}k_\mathrm{c})
    = \mathrm{e}^{\mathrm{i} \phi_\mathrm{def} \delta z_\mathrm{s}(l_\mathrm{s})
        \left(\frac{1}{2} \nu_x^2 + \nu_y^2\right)}.
\end{equation}

\subsubsection{Numerical simulations of computational refocusing for several OCT types}

\begin{table}
    \centering
    \caption{Simulation parameters used for each configuration.}\label{table:CR-configs}
    \begin{tabular}{l c c c c c}
    Config. & \makecell{PSFD,\\mid-NA\\(\figs~\ref{m-fig:PSFD}-\ref{m-fig:PSFD_each})} & \makecell{PSFD,\\high-NA\\(\figs~\ref{fig:PSFD-HighNA}-\ref{fig:PSFD_each-HighNA})} & \makecell{FF-SS,\\mid-NA\\(\figs~\ref{fig:FFSS}-\ref{fig:FFSS_each})} & 
\makecell{LF-FD,\\low-NA\\(\figurename~\ref{fig:LFFD-lNA})}\\
    \midrule\\
    \makecell[l]{Illumination NA \\(cut-off)} & 0.25$n_\mathrm{BG}$ & 0.5$n_\mathrm{BG}$ & 0 & 
0.25$n_\mathrm{BG}$ ($x$)\\
    Illumination NA (eff.) & 0.15$n_\mathrm{BG}$ & 0.4$n_\mathrm{BG}$ & N/A & 
0.05$n_\mathrm{BG}$ ($x$)\\
    Collection NA (cut-off) & 0.25$n_\mathrm{BG}$ & 0.5$n_\mathrm{BG}$ & 0.25$n_\mathrm{BG}$ & 
0.25$n_\mathrm{BG}$\\
    Collection NA (eff.) & 0.15$n_\mathrm{BG}$ & 0.4$n_\mathrm{BG}$ & N/A & 
N/A\\
    \makecell[l]{Central wavelength\\ in vacuum [nm]} & 1050 & 1050 & 830 &
    830\\
    \makecell[l]{Wavelength bandwidth\\ in vacuum [nm]} & 100 & 100 & 80 &
    80\\
    \makecell[l]{Refractive index $n_\mathrm{BG}$} & 1.34 & 1.34 & 1.34 &
    1.34\\
    \end{tabular}
\end{table}

Each simulation configuration used for generating results in the main document and this supplementary material is summarized in \tablename~\ref{table:CR-configs}.

\paragraph{PSFD-OCT with mid-NA}

\begin{figure}
    \centering
    \includegraphics[width=8cm]{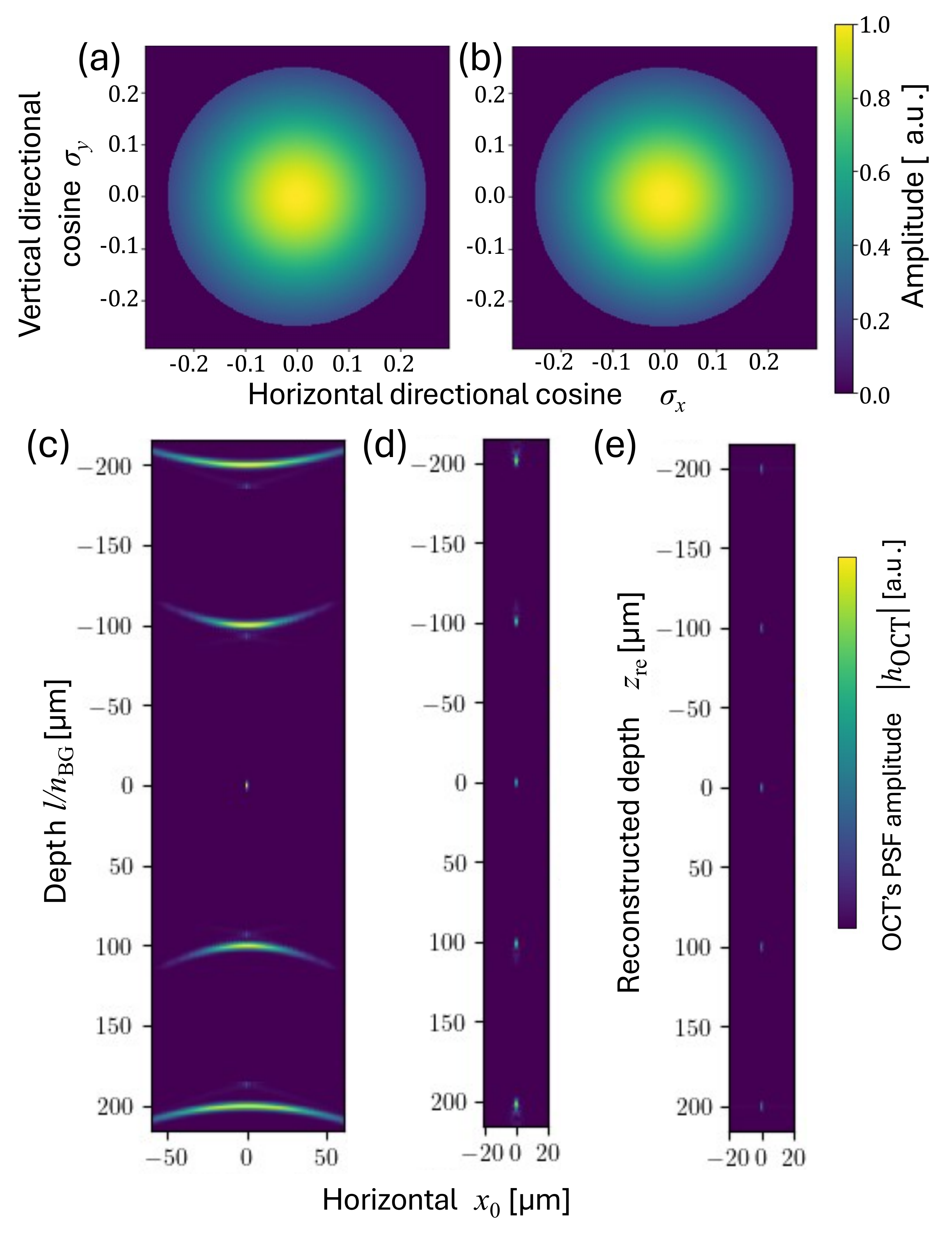}
    \caption{
        Numerical simulations of PSFD-OCT PSFs with/without computational defocus corrections.
        The cut-off NA is 0.5 $n_\mathrm{BG}$, and the effective NA of is 0.4 $n_\mathrm{BG}$, where $n_\mathrm{BG}$ = 1.34.
        The simulated (a) illumination and (b) collection pupil distributions are used to calculate (c) the OCT PSFs.
        (d) The computational refocusing filter [Eq.~(\ref{m-eq:38})] is applied to the simulated OCT signals.
        (e) The PSFs after the ISAM for PSFD-OCT (Section~\ref{sec:ISAM_PSFD}) applied.
    }\label{fig:PSFD-HighNA}
\end{figure}

\begin{figure}
    \centering
    \includegraphics[width=13cm]{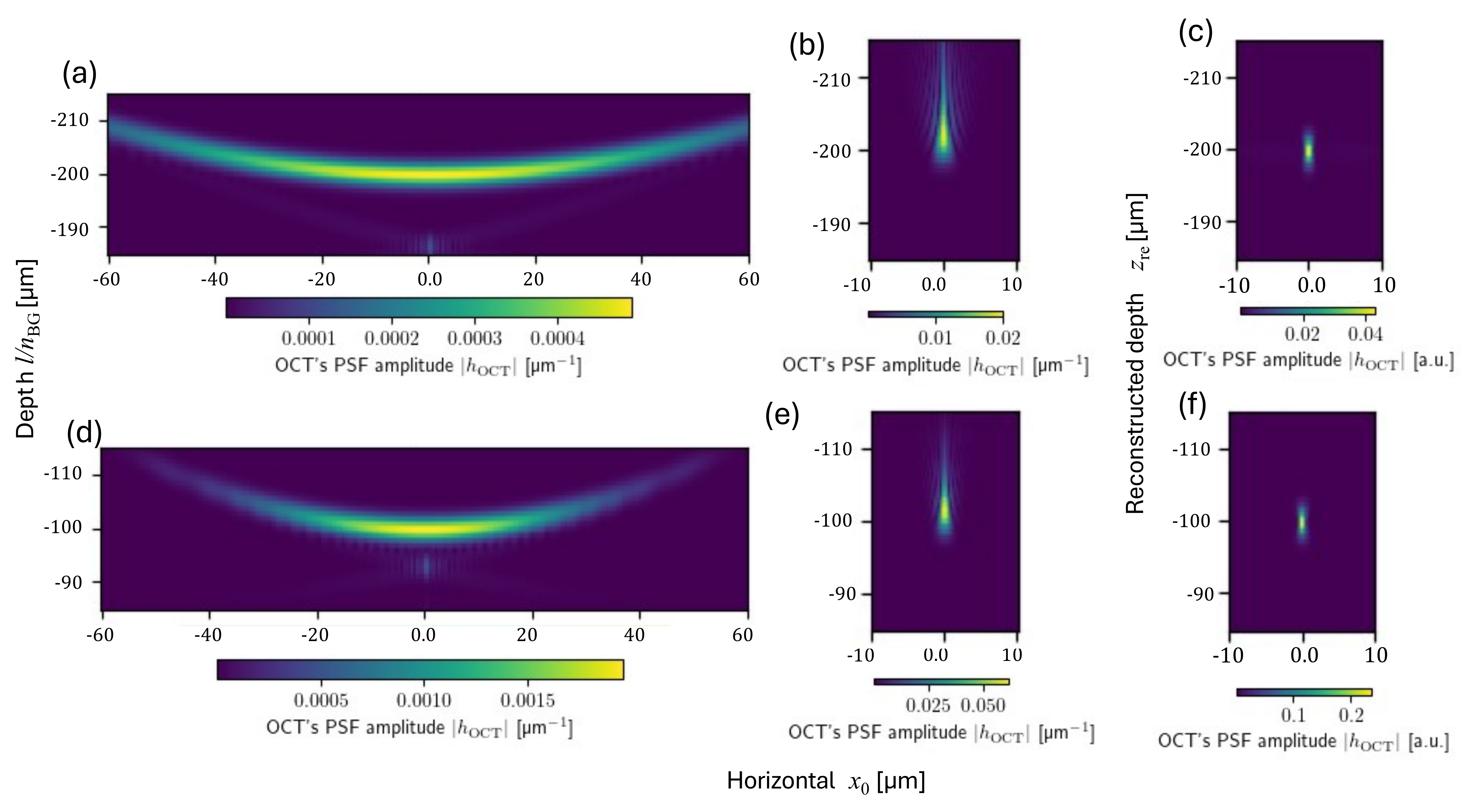}
    \caption{
        Numerically simulated PSFD-OCT PSFs at (a-c) -200 \um defocus and (d-f) -100 \um defocus.
        (d,e) The computational refocusing filter [Eq.~(\ref{m-eq:38})] is applied.
        (c,f) The PSFs after the ISAM for PSFD-OCT (Section~\ref{sec:ISAM_PSFD}) applied.
    }\label{fig:PSFD_each-HighNA}
\end{figure}

PSFD-OCT signals with high NA were simulated.
The simulated pupil amplitudes $P_\mathrm{ill}$ and $P_\mathrm{col}$ and PSFs with several defocus values are shown in \figurename~\ref{fig:PSFD-HighNA}.
The illumination pupil $P_\mathrm{ill}$ and the collection pupil $P_\mathrm{col}$ are identical Gaussian distributions with a cut-off NA of 0.5$n_\mathrm{BG}$ and an effective ($\mathrm{e}^{-1}$) NA of 0.4$n_\mathrm{BG}$.
The central wavelength was 1050 nm, the full-width at half maximum (FWHM) bandwidth was 100 nm, and the background refractive index was 1.34.

The CR filter [from Eq.~(\ref{m-eq:38})] and the simple ISAM implementation [Section~\ref{sec:ISAM_PSFD}] are applied to the simulated OCT signals.
The enlarged PSFs with the defocus values of 100 and 200 \um are shown in \figurename~\ref{fig:PSFD_each-HighNA}.

\paragraph{Full-field SS-OCT}

\begin{figure}
    \centering
    \includegraphics[width=8cm]{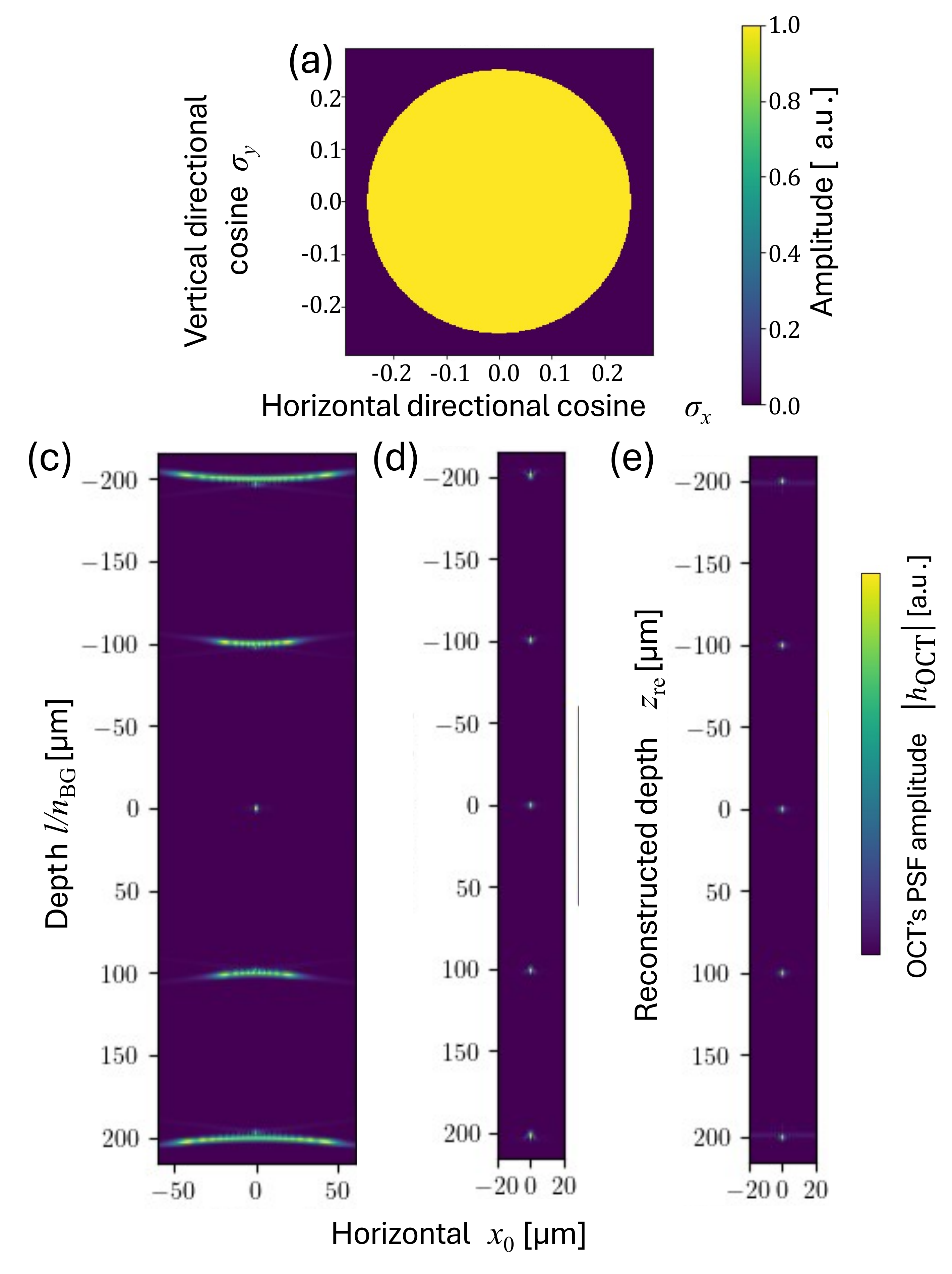}
    \caption{
        Numerical simulations of full-field (FF)-SS-OCT PSFs with/without computational defocus corrections.
        The cut-off NA of collection is 0.25 $n_\mathrm{BG}$, where $n_\mathrm{BG}$ = 1.34.
        The simulated (a) collection pupil distribution is used to calculate (b) the OCT PSFs.
        (c) The computational refocusing filter [Eq.~(\ref{eq:S5})] is applied to the simulated OCT signals.
        (d) The PSFs after the ISAM for PSFD-OCT (Section~\ref{sec:ISAM_FFSS}) applied.
    }\label{fig:FFSS}
\end{figure}

\begin{figure}
    \centering
    \includegraphics[width=13cm]{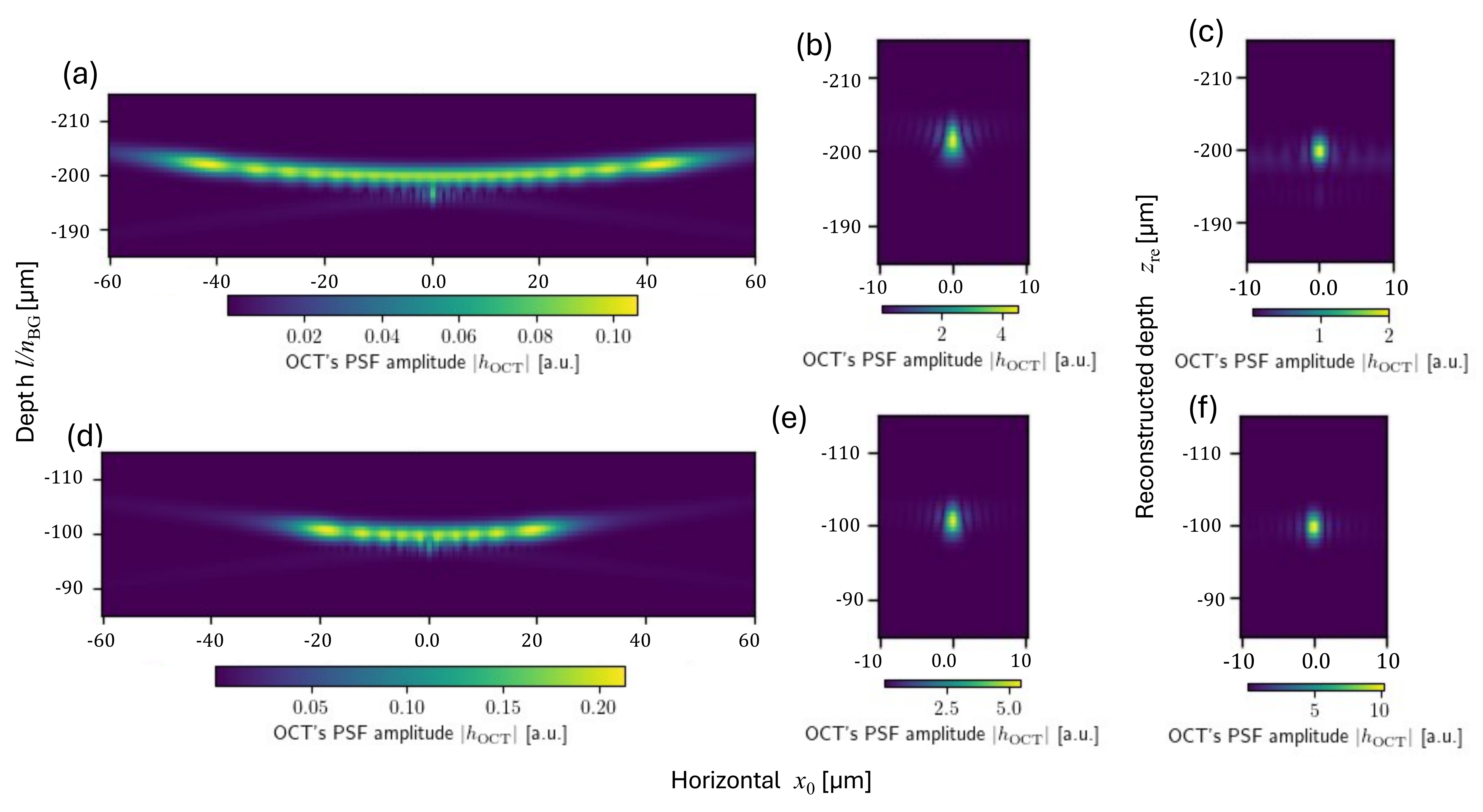}
    \caption{
        Numerical simulation of the PSFs of FF-SS-OCT with/without CR\@.
        Numerically simulated FF-SS-OCT PSFs at (a-c) -200 \um defocus and (d-f) -100 \um defocus.
        (d,e) The computational refocusing filter [Eq.~(\ref{eq:S5})] is applied.
        (c,f) The PSFs after the ISAM for PSFD-OCT (Section~\ref{sec:ISAM_FFSS}) applied.
    }\label{fig:FFSS_each}
\end{figure}

The full-field (FF)-SS-OCT signals were simulated.
The illumination pupil was a delta function with a center at $P_\mathrm{ill} = \delta(|\bm{\upnu}_\parallel|)$, and the collection pupil $P_\mathrm{col}$ was simulated as a circular function with a cut-off NA of 0.25 $n_\mathrm{BG}$.
The central wavelength was 830 nm, the full-width at half maximum (FWHM) bandwidth was 80 nm, and the background refractive index was 1.34.
The PSFs with several defocus errors were corrected with the FF-SS-OCT CR filter [Eq.~(\ref{eq:S4})] and with ISAM for FF-SS-OCT [Section~\ref{sec:ISAM_FFSS}], as shown in \figurename~\ref{fig:FFSS}.

\paragraph{Line-field FD-OCT}

\begin{figure}
    \centering
    \includegraphics[width=11cm]{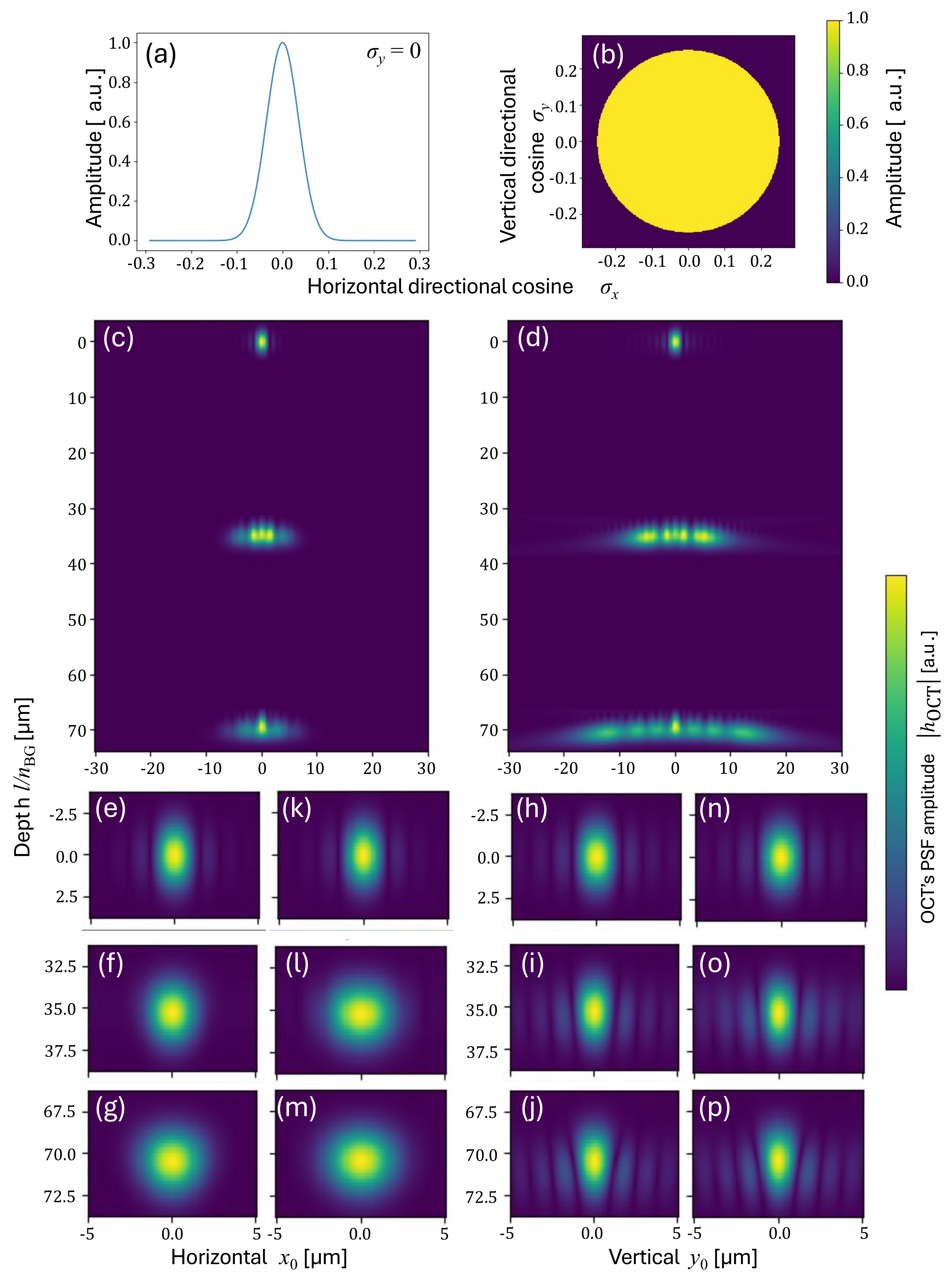}
    \caption{
        Numerical simulations of LF-FD-OCT PSFs with/without computational defocus corrections.
        The cut-off NA is 0.25 $n_\mathrm{BG}$, and the effective NA of is 0.05 $n_\mathrm{BG}$, where $n_\mathrm{BG}$ = 1.34.
        The simulated (a) illumination pupil distribution at $\sigma_y$ = 0 and (b) collection pupil distribution are used to calculate (c,d) the OCT PSFs.
        (e-j) The computational refocusing filter [Eq.~(\ref{eq:S8})] and (k-p) the computational refocusing filter with Gaussian apodization collection [Eq.~(\ref{eq:S13})] are applied to the simulated OCT signals at defocus of (e,k,h,n) 0, (f,l,i,o) 35, and (g,m,j,p) 75 \um.
        Corrected horizontal PSFs via both filters (f,g,l,m) do not reach to that of in-focus.
        The wrong design of the CR filter results in a broader corrected PSF width along the horizontal direction at 35 \um defocus (l) compared with that of a better design (f).
    }\label{fig:LFFD-lNA}
\end{figure}

The LFFD-OCT signals were simulated using a horizontal Gaussian and vertical $\delta$-function distributions for the illumination pupil and a cylinder function for the collection pupil.
The defocused PSFs were corrected by applying the LF-OCT CR filter [Eq.~(\ref{eq:S8})] and the LF-OCT CR filter with the Gaussian collection pupil [Eq.~(\ref{eq:S13})].
The simulated pupils and defocused and corrected PSFs at defocus of 0, 35, and 70 \um are shown in \figurename~\ref{fig:LFFD-lNA}.
Because of the discrepancy between the simulation model and CR filter design in the collection pupil distribution, the corrected PSF at defocus of 35 \um with the Gaussian-collection CR filter [\figurename~\ref{fig:LFFD-lNA}(l)] showed lower resolution along the horizontal direction when compared with that with uniform-collection CR filter [\figurename~\ref{fig:LFFD-lNA}(f)].
Nevertheless, the refocused results obtained with both filters are not as sharp as the in-focus results along the horizontal ($x$-) direction because the two filters are based on different collection pupils than that used in the OCT signal generation.
This implies that the computational refocusing of LF-OCT is particularly difficult when compared with that of other OCT types because of the complex interactions between illumination and collection, and between the horizontal and vertical pupil distributions.

\subsection{Other example of numerical simulation of computational aberration correction}\label{sec:other_example}

\begin{figure}
    \centering
    \includegraphics[width=13cm]{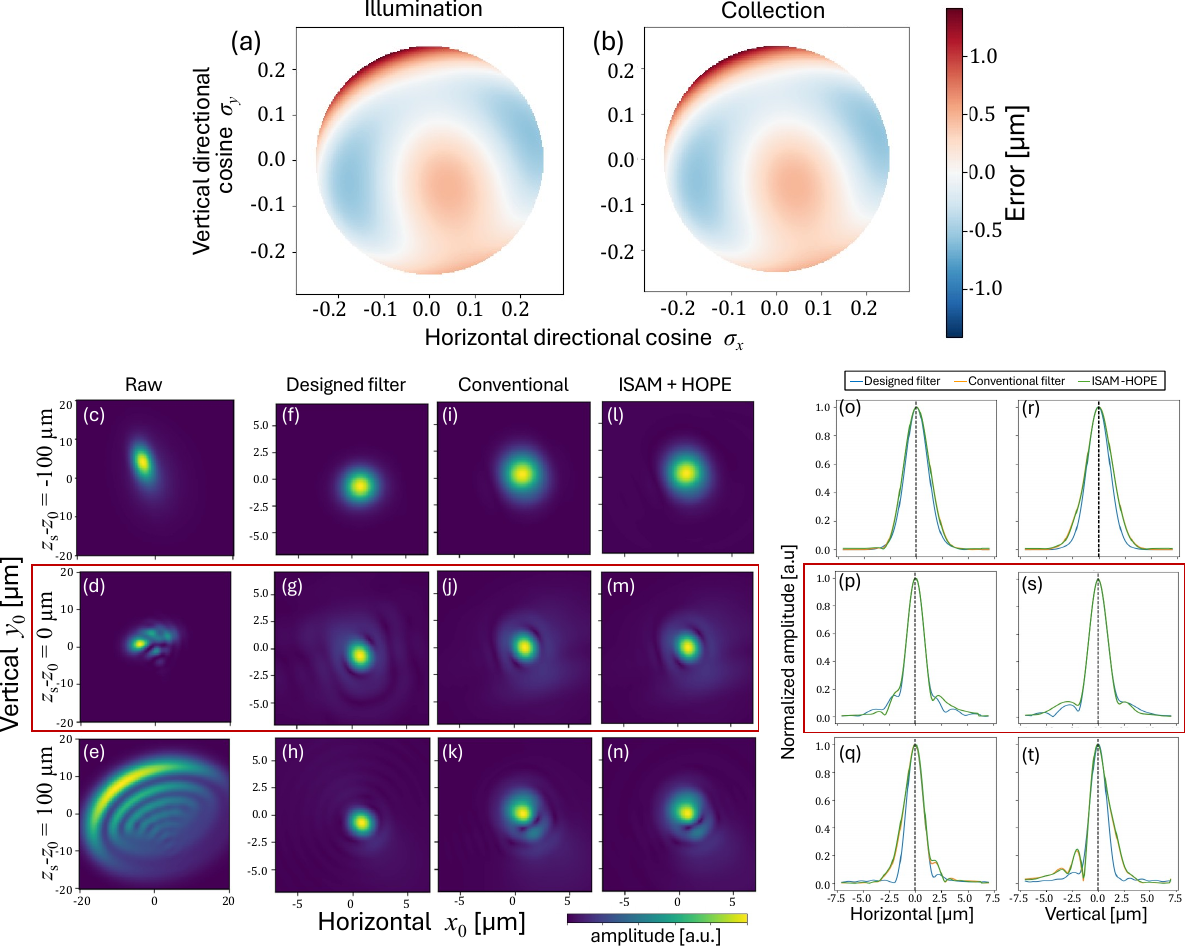}
    \caption{Numerical simulation of aberrated PSFs and corrections.
        (a, b) Numerically simulated wavefront errors on the pupil for PSFD-OCT\@.
        The aberrations for both pupils are assumed to be identical.
        (c,d,e) PSFs with aberrations, (f,g,h) corrected using the designed filter, (i,j,k) corrected using a conventional filter, and (l,m,n) corrected via the ISAM + high-order phase error (HOPE) correction, where its parameters estimated with the conventional filter (i,j,k) are shown.
        The aberration parameters of each method were estimated at the in-focus signal, hence the PSFs are well corrected for all collection methods (red box).
        However, the PSFs with correction using conventional filter exhibit blurring at defocused signals (i,l and k,n).
        (o,p,q) Horizontal and (r,s,t) vertical profiles of corrected PSFs (f--n).
        Profiles are centered at the maximum peak locations and normalized by its maximum values.
        PSFs shape are almost the same at in-focus because correction parameters were optimized with this signal (red box, p, s).
        However, the PSFs corrected by the conventional filter exhibit increased side lobes and broadening with defocus (o, q, r, t).
    }\label{fig:pupil-CAC-PSFD-2}
\end{figure}

\begin{table}
    \centering
    \caption{Strehl ratios of the aberration corrected PSFs [\figurename~\ref{fig:pupil-CAC-PSFD-2}] with different correction methods and defocus amounts.}\label{tab:strehl-ratio}
    \begin{tabular}{r|ccc}
        Defocus & New CAC & Conventional CAC & ISAM + HOPE correction\\
        \midrule
        - 100 \um & 1.000 & 0.860 & 0.846\\
            0 \um & 0.990 & 0.963 & 0.958\\
          100 \um & 0.970 & 0.765 & 0.750\\
    \end{tabular}
\end{table}

Another example of the numerical simulation of inter-depth computational aberration correction for PSFD-OCT is demonstrated.
The HOA coefficients were set as follows: $w_3$: 0.167, $w_5$: -0.193, $w_7$: -0.126, $w_8$: -0.125, and $w_{12}$: 0.148 \um.
The applied wavefront error is shown in \figs~\ref{fig:pupil-CAC-PSFD-2}(a) and \ref{fig:pupil-CAC-PSFD-2}(b).
The RMS wavefront error is 0.344 \um.
The same analysis method of Sec.~\ref{m-sec:CAC_simulation} is applied.
The corrected PSFs distributions and their profiles are shown as \figurename~\ref{fig:pupil-CAC-PSFD-2}(f-t).
The Strehl ratios of them are shown in \tablename~\ref{tab:strehl-ratio}.

The corrected PSFs at 0 \um defocus [\figs~\ref{fig:pupil-CAC-PSFD-2}(g,j,m)] and their Strehl ratios show that the optimization of the coefficients works well.
However, the PSFs at 100 \um defocus with conventional aberration correction exhibit increased side lobes and broadening compared to the designed filter [\figs~\ref{fig:pupil-CAC-PSFD-2}(k,n,q,t)].
The PSFs at -100 \um defocus with conventional aberration correction also exhibit broadening compared to the designed filter along diagonal direction [\figs~\ref{fig:pupil-CAC-PSFD-2}(i) and \ref{fig:pupil-CAC-PSFD-2}(l)].
This additional example also demonstrates the effectiveness of the proposed method in correcting aberrations across different depths.

\begin{figure}
    \centering
    \includegraphics[width=12cm]{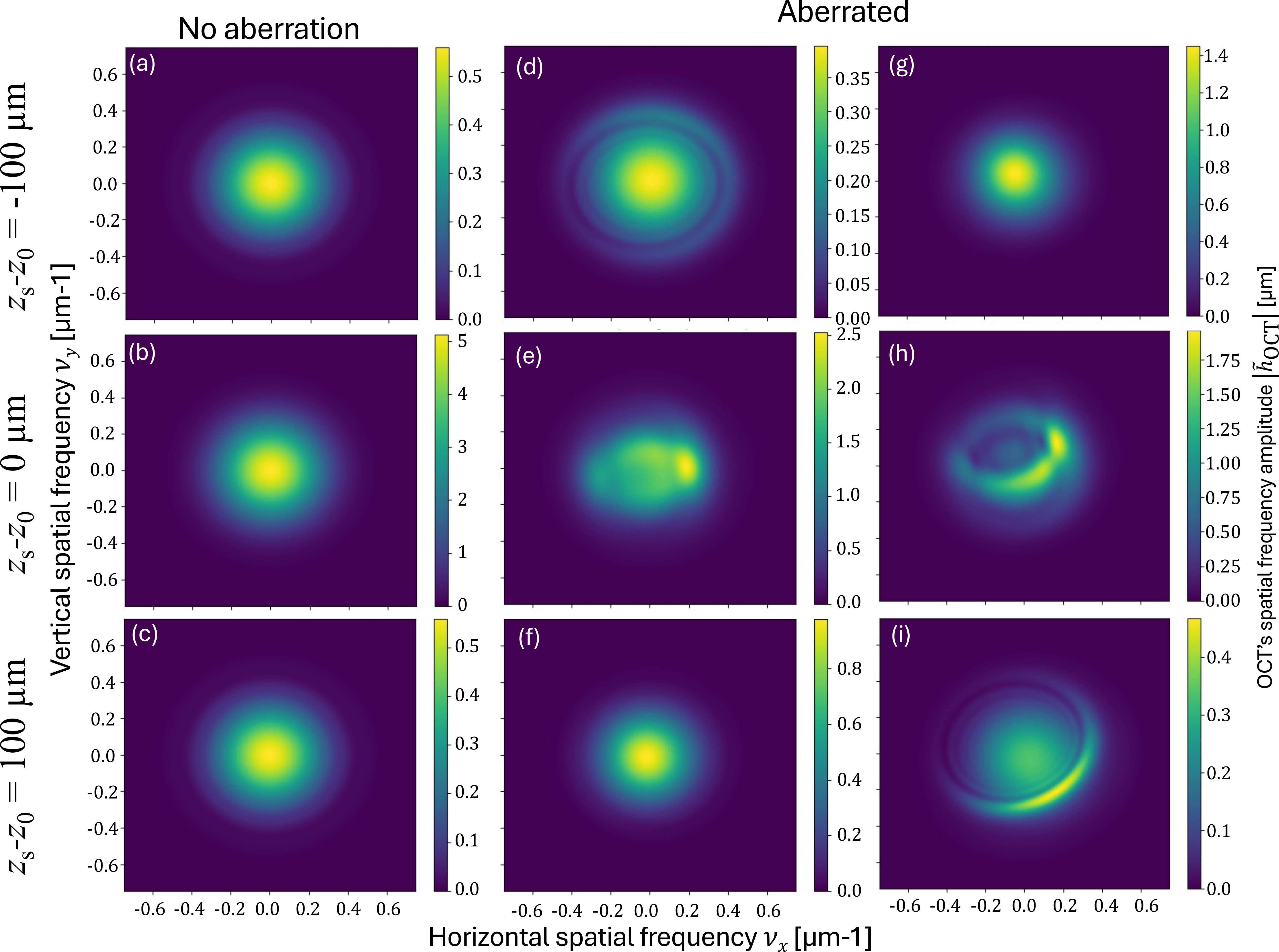}
    \caption{Simulated amplitude of \enface OCT signal's spatial frequency component of PSFD-OCT with different defocus amounts.
    (a,b,c) Without aberrations, (d,e,f) with aberration defined in Sec.~\ref{m-sec:CAC_simulation} (g,h,i) with aberration defined in Sec.~\ref{sec:other_example}.}\label{fig:SystemFunction_Amp}
\end{figure}

\subsection{Implementation details of the computational aberration correction method}
\label{sec:CAC_implementation}

\begin{figure}
    \centering
    \includegraphics[width=13cm]{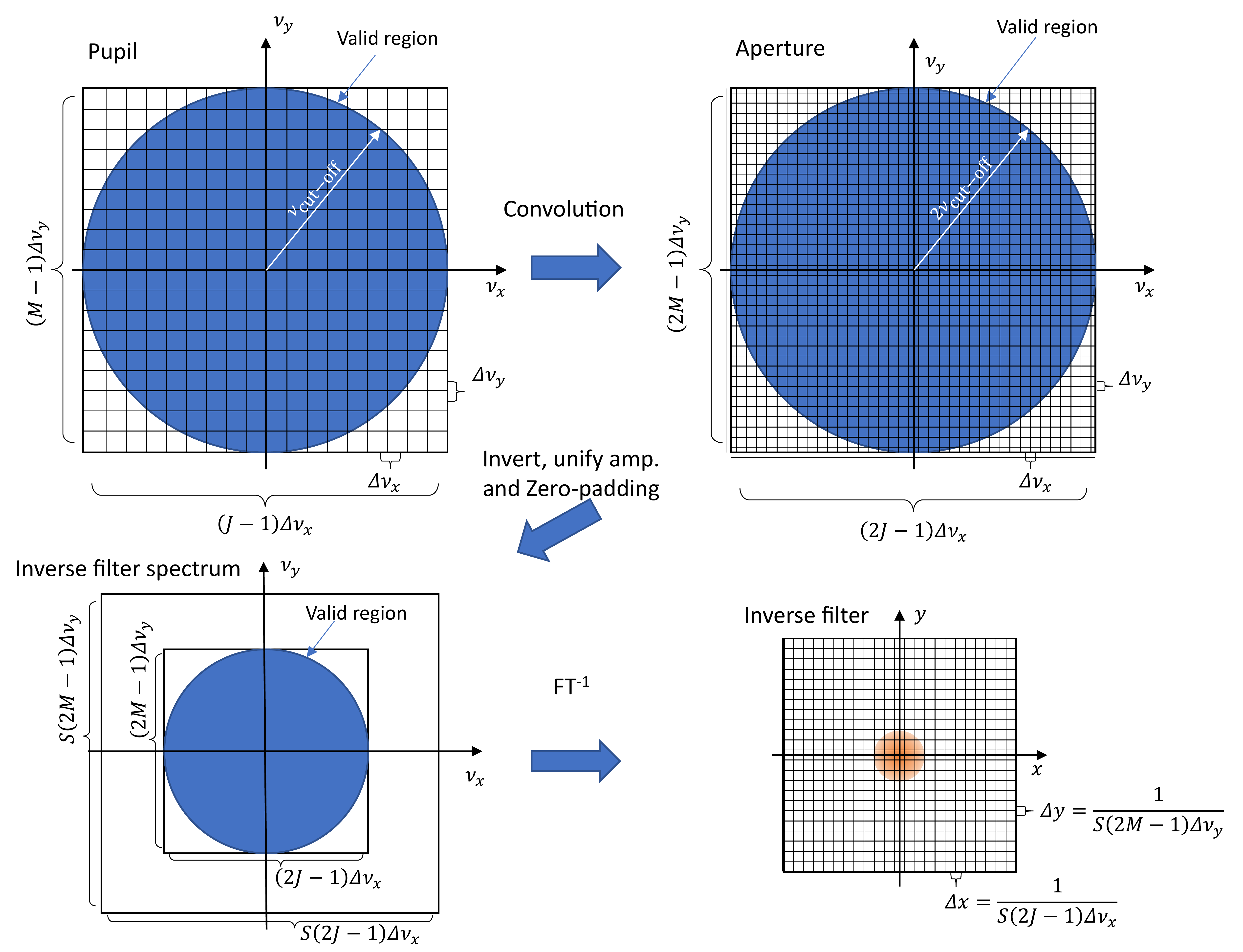}
    \caption{Calculation of the CAC filter.}\label{fig:CAC_filter_calc}
\end{figure}

The aberration correction filter can be obtained with an amplitude $A'$ and the coefficients of the Zernike polynomials $\mathbf{c} = {c_n}$ for wavefront aberrations in the pupil plane as Eqs.~(\ref{m-eq:42}) and (\ref{m-eq:43}).
This filter will be convolved with the complex OCT signal.
The amplitude distribution of the pupil could be approximated using a Gaussian function.
Additionally, for accurate calculation of the numerical convolution, the pupil must have a value of zero at the edges of the calculation grid.
The following truncated Gaussian function is thus used for the pupil amplitude:
\begin{equation}
    A^{'} (\bm{\upnu}_\parallel)
    = \begin{cases}
        \mathrm{e}^{-2 \frac{|\bm{\upnu}_\parallel|^2}{\Delta f^2}} & |\bm{\upnu}_\parallel| < f_\mathrm{co} \\
        0 & |\bm{\upnu}_\parallel| \ge f_\mathrm{co}
    \end{cases},
\end{equation}
where $\Delta f$ represents the the Gaussian function width and $f_\mathrm{co}$ is the cut-off spatial frequency.

In \figurename~\ref{fig:CAC_filter_calc}, the steps and the calculation grid parameters for the inverse filter calculation are summarized.
The spatial frequency grid range is determined by the number of grids (J, M) used and the grid spacing ($\Delta \nu_x$, $\Delta \nu_y$).
In the case of large defocus and high-order aberrations, the wavefront $W_\mathrm{pupil}$ will change rapidly, and thus the small grid spacing is required.
The grid range should be greater than the spatial frequency range of the diffraction-limited OCT signal to achieve the full CAC performance and to preserve the spatial resolution when no aberrations are present.
These limit gives the maximum spatial frequency components of the OCT's diffraction-limited point-spread-function (PSF).
These requirements can be written as $2 f_\mathrm{co} = (J - 0.5) \Delta \nu_x = (M - 0.5) \Delta \nu_y \gg 1/\delta x$, where $\delta x$ is the diffraction-limited lateral resolution.
Additionally, the limit also determines the spatial sampling density of the inverse filter ($\Delta \nu_x$, $\Delta \nu_y$).
If the spatial sampling spacing is not small enough, zero padding before the inverse Fourier transform will be required (where $S$ is the ratio of the grid ranges before and after the zero padding).

We predefined the defocus slope parameter as:
\begin{equation}
    \frac{\uppi \lambda_{0}(\omega_\mathrm{c}) f_\mathrm{co}^2}
    {8 \sqrt{3}}0.55.
\end{equation}
However, the sign of this slope depends on the selected direction of the OPL $l$, whether complex conjugate signal is used or not, the wavelength scanning and detection directions, and whether a Fourier or inverse Fourier transform is used to perform OCT signal reconstruction.
In practice, the sign of the slope can be determined based on which correction is better than the other.

\subsection{Implementation of ISAM}
\label{sec:ISAM_implementation}

Simple ISAM implementations that can be applied to simulated OCT signals for comparison are described in this section.
In the numerical simulations, we always set the zero-delay position at the focal palne, $z_0 = z_\mathrm{r}$.
However, in practice, the FD-OCTs are used with $z_0 \ne z_\mathrm{r}$.
In this case, the signal should be shifted to make the focal plane is at the origin of the axial coordinate to remove the phase term $\mathrm{e}^{\mathrm{i} 2 [k_\mathrm{s}(\omega) z_0 - k_\mathrm{r}(\omega) z_\mathrm{r}]}$ from Eq.~(\ref{eq:34}).
Otherwise, the linear phase change along $\omega$ will be converted into a phase error in the ($\nu_x, \nu_y$)-plane because of the resampling process in ISAM.
The requirement for this shifting operation was also noted previously by Kumar et al.\cite{kumar_numerical_2014}.

\subsubsection{PSFD-OCT}
\label{sec:ISAM_PSFD}

ISAM for PSFD-OCT\cite{ralston_inverse_2006} assumes that the cTF is a spherical cap with a radius of $2 \frac{k_\mathrm{s}}{2\uppi}$.
Then, the $i$-th resampling point along the optical frequency $\omega_{i}$ that is obtained from arbitrary equidistant sampling points along the axial frequency $\nu_z$ is given by:
\begin{equation}\label{eq:S9}
    \omega_{i} (\bm{\upnu}_\parallel)
    = \frac{\uppi c}{n_\mathrm{g, BG}(\omega_\mathrm{c})} \sqrt{|\bm{\upnu}_\parallel|^2 + (i \Delta \nu_z + \nu_{z0})^2},
\end{equation}
where $\Delta \nu_z$ represents the axial frequency spacing and $\nu_{z0}$ is the axial frequency offset with respect to the first resampling point.
In this manuscript, a simple linear interpolation was applied to the OCT signal frequency spectrum $\tilde{I}^{'} (\bm{\upnu}_{\parallel}, \omega; 0)$ [Eq.~(\ref{eq:34})] to obtain the resampled signal at $(\bm{\upnu}_\parallel, \omega_{i}(|\bm{\upnu}_\parallel|))$.

\subsubsection{FF-SS-OCT}
\label{sec:ISAM_FFSS}

ISAM for FF-SS-OCT\cite{marks_inverse_2007} assumes that the cTF is a spherical cap and a radius of $\frac{k_\mathrm{s}}{2\uppi}$ with the shifted center along $\nu_z$ with $\frac{k_\mathrm{s}}{2\uppi}$ when the illumination direction $\bm{\upsigma}_{\mathrm{ill}} = [0, 0, 1]^T$.
Then, the $i$-th resampling point along the optical frequency $\omega_{i}$ obtained from the arbitrary equidistant sampling points along the axial frequency $\nu_z$ is given by:
\begin{equation}\label{eq:S10}
    \omega_{i}(|\bm{\upnu}_\parallel |)
    = \frac{\uppi c}{n_\mathrm{g, BG}(\omega_\mathrm{c})}
    \frac{|\bm{\upnu}_\parallel|^2 + (i \Delta \nu_z + \nu_{z0})^2}{i \Delta \nu_z + \nu_{z0}}.
\end{equation}
As in the Section~\ref{sec:ISAM_PSFD}, a simple linear interpolation is then applied to the OCT signal frequency spectrum $\tilde{I}^{'} (\bm{\upnu}_{\parallel}, \omega; 0)$ [Eq.~(\ref{eq:34})] to resample the signal at the location described by Eq.~(\ref{eq:S10}).

\subsection{Converting directional cosines in the diffraction integral into spatial frequencies}
\label{sec:2DFT_focused_field}

For the diffraction integrals in Eqs.~(\ref{eq:4}) and (\ref{eq:8}), 2D lateral directional cosines $\bm{\upsigma}_\parallel$ can be converted into the 2D spatial frequencies $\bm{\upnu}_\parallel$ and the integral can be described using a 2D Fourier transform.
A function $\phi$ of $\bm{\upsigma}_\parallel$ is weighting the phase term $\mathrm{e}^{\mathrm{i} k \mathbf{r}_\parallel \cdot \bm{\upsigma}_\parallel}$ and is integrated over a spherical plane with a solid angle $\Omega$.
By using the relationship between the spatial frequency and the directional cosines of the propagation directions
\begin{equation}\label{eq:S11}
    \bm{\upsigma}_\parallel = \frac{2\uppi}{k}\bm{\upnu}_\parallel,
\end{equation}
and $\mathrm{d}\Omega = \frac{\mathrm{d}\sigma_x\mathrm{d}\sigma_y}{\sigma_z} = \frac{\mathrm{d}\bm{\upsigma}_\parallel}{\sigma_z}$:
\begin{equation}\label{eq:S12}
    \begin{split}
        \iint_{\Pi} \phi(\bm{\upsigma}_\parallel)
        \mathrm{e}^{\mathrm{i} k \mathbf{r}_\parallel \cdot \bm{\upsigma}_\parallel}
\frac{\mathrm{d}\bm{\upsigma}_\parallel}{\sigma_z(\bm{\upsigma}_\parallel)}
        =&
        \frac{4\uppi^2}{k^2}
        \iint_{-\infty}^{\infty}
            \frac{\phi(\frac{2\uppi}{k}\bm{\upnu}_\parallel)}{\sigma_z (\frac{2\uppi}{k}\bm{\upnu}_\parallel)}
            \mathrm{e}^{2\uppi\mathrm{i} \mathbf{r}_\parallel \cdot \bm{\upnu}_\parallel}
        \mathrm{d}\bm{\upnu}_\parallel\\
        =&
        \frac{4\uppi^2}{k^2}
        \mathcal{F}_{\bm{\upnu}_\parallel}^{-1}
        \left[
            \frac{\phi\left(\frac{2\uppi}{k} \bm{\upnu}_\parallel\right)}{\sigma_z \left(\frac{2\uppi}{k} \bm{\upnu}_\parallel\right)}
        \right](\mathbf{r}_\parallel),
    \end{split}
\end{equation}
The range of $\sigma_x$ and $\sigma_y$ should be $\bm{\upsigma}_\parallel \in \Pi$, which is limited by the optical system.
However, the integration range for $\nu_x$ and $\nu_y$ is treated as $-\infty$ to $\infty$.
The conversion would be made valid by defining the function $\phi$ as being zero outside the domain $\Pi$,
$\phi(\bm{\upsigma}_\parallel) \equiv 0, \text{if}\ \bm{\upsigma}_\parallel \notin \Pi$.

\subsection{Optical setup}
\label{sec:Optical_setup}

\begin{figure}
    \centering
    \includegraphics[width=13cm]{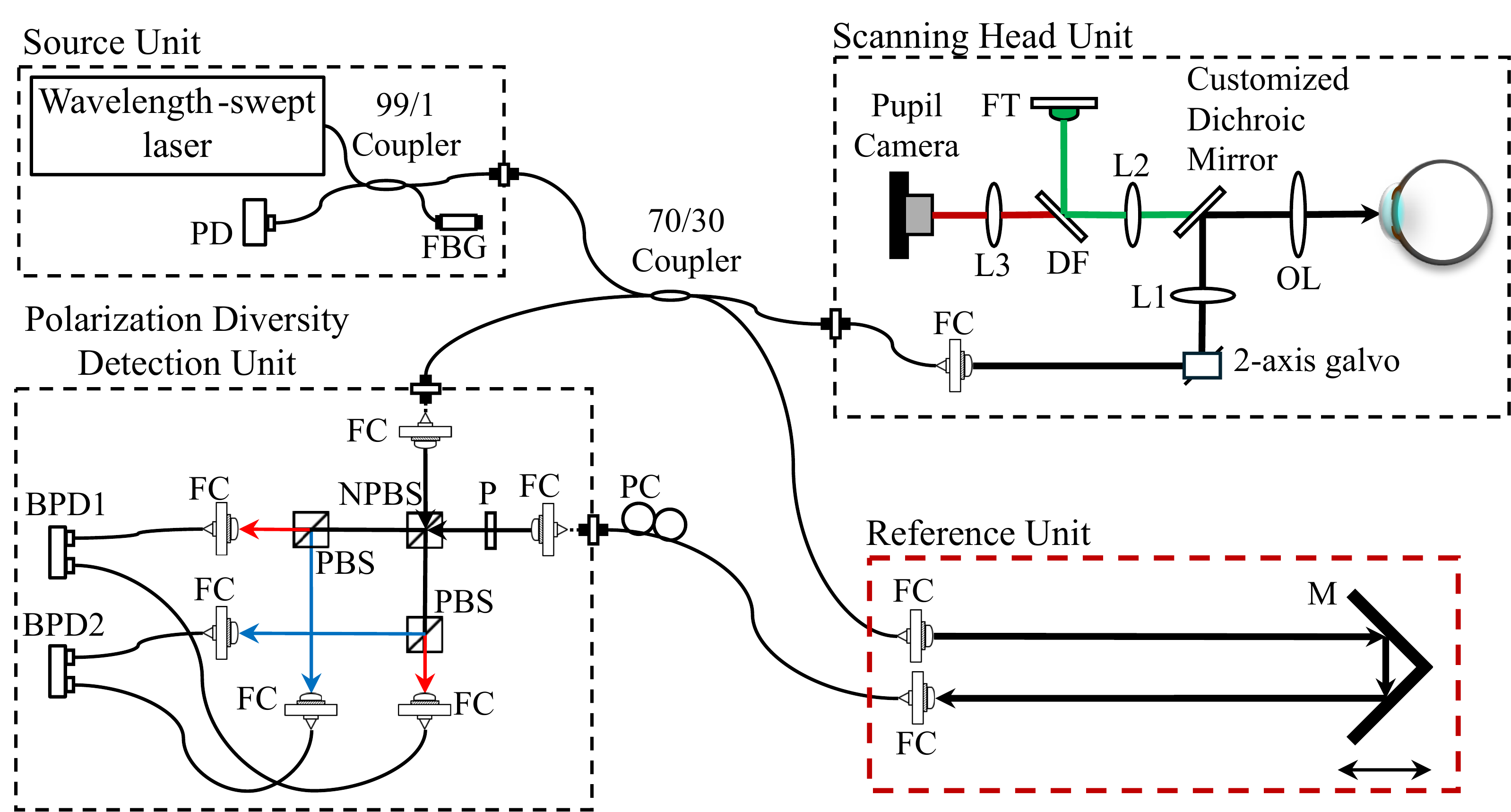}
    \caption{Optical setup of PSFD-OCT\@.
    FT: fixation target, OL: ophthalmic lens, L: lens, M: mirror, PC: polarization controller, NPBS: non-polarization beam splitter, PBS: polarization beam splitter, FC: fiber collimator, DF: dichroic filter, FBG: fiber Bragg grating, P: polarizer, PD: photodiode, and BPD: balanced photodetector.
    }\label{fig:Optical_setup}
\end{figure}

\begin{table}
    \centering
    \caption{Configurations used for the optical setup.}\label{table:Optical_setup}
    \begin{tabular}{c c c}
    Configuration & A & B \\
    \midrule
    Wavelength scanning rate [Hz] & 100,000 & 200,000 \\
    Beam diameter on the cornea (e$^{-2}$) [mm] & $\approx$ 3.4 & $\approx$ 5 \\
    \end{tabular}
\end{table}

Optical setup of the PSFD-OCT system used for experiments are shown in \figurename~\ref{fig:Optical_setup}.
The optical setup is based on a fiber-based interferometer with a polarization-diversity detection.
The reflection from the fiber Bragg grating (FBG) is used as a $k$ clock.
The scanning head has a pupil monitoring path to align the probe entering to the eye.
The subject gated at the fixation target (FT).

Two configurations were used for the optical setup, as summarized in \tablename~\ref{table:Optical_setup}.
Wavelength swept lasers with a wavelength range of 100 nm centered at the 1050 nm were used for both configurations.
For configuration A, the laser had a wavelength scanning rate of 100 kHz, and the beam diameter on the cornea was approximately 3.4 mm.
For configuration B, the laser had a wavelength scanning rate of 200 kHz, and the beam diameter on the cornea was approximately 5 mm.
The data acquisition was performed by ATS9350 (AlazarTech, Canada) for configuration A and ATS9360 (AlazarTech, Canada) for configuration B.